# Towards Refactoring DMARF and GIPSY OSS

Authors

Aaradhna Goyal, Ali Alshamrani, Dhivyaa Nandakumar, Dileep Vanga, Dmitriy Fingerman, Parul Gupta, Riya Ray, Srikanth Suryadevara

Group ID: mes64711

Concordia University, Montreal

## I. ABSTRACT

***We present here an exploratory and investigatory study of the requirements, design, and implementation of two open-source software systems: the Distributed Modular Audio Recognition Framework (DMARF), and the General Intensional Programming System (GIPSY). The inception, development, and evolution of the two systems have overlapped in terms of the involved developers as well as in their applications. DMARF is a platform independent collection of algorithms for pattern recognition, identification and signal processing in audio and natural language text samples. It provides a rich platform for the research community in particular to use, test, and compare various algorithms in the broad field of pattern recognition and machine learning. Intended as a platform for intensional programming (a paradigm mathematically rooted in intensional logic), GIPSY's inception was intended to push the field of intensional programming further, overcoming limitations in the available tools two decades ago.***

***In this study, we present background research into the two systems and elaborate on their motivations and the requirements that drove and shaped their design and implementation. We subsequently elaborate in more depth about various aspects of their architectural design, including the elucidation of some use cases, domain models, and the overall class diagram of the major components. In some contexts, the two systems can be fused and used distributively towards achieving certain tasks, and so we present an architectural view of the DMARF-over-GIPSY run-time architecture. Moreover, we investigate existing design patterns in both systems and provide a detailed view of the involved components in such patterns. Furthermore, we delve deeper into the guts of both systems, identifying code smells and suggesting possible refactorings, some of which we do implement and integrate in both systems, including some design patterns. Implementations of selected refactorings have been collected into patchsets and could be committed into future releases of the two systems, pending a review and approval of the developers and maintainers of DMARF and GIPSY.***

## II. INTRODUCTION

After comprehensive study of two case studies DMARF and GIPSY, their backgrounds are summarized. The source code's quality is measured in terms of number of SLOC, number of Java files, number of classes and methods, using *Eclipse plugin for metrics*.

Actors and stakeholders are identified and two fully-dressed use cases are written, one for each case study. These use cases describe scenarios for application of the case studies. Use case diagrams for both use cases are constructed using Microsoft Visio. Similarly, two domain models corresponding to the use cases are made using Microsoft Visio that show the conceptual classes and their associations involved in the use case. Design class diagrams are created using *ObjectAid UML Explorer* in Eclipse environment. Design class diagrams show the dynamic behavior of the whole system and how classes interact with each other to make the system work as a whole. Next part of the document gives the comparison and mapping of conceptual classes with actual classes. Any kind of discrepancies are also identified and defined for the same. A brief description of the tool used is given at the end of this part. For two of these classes, their relationship with each other is described, complimented by visualization and code snippet for both.

To check the system for any code smells, JDeodorant, Robusta, McCabe and Logiscope tools have been used and identified smells are refactored using JDeodorant .To make the refactoring descriptive, code snippets and class diagrams are included. Refactorings are supplemented by respective test cases, written using JUnit, to validate that refactorings were effective in improving the performance and quality of the code. Four design patterns per case study were identified, defined and explained in detailed. The detail includes code snippets as well as graphically represented classes which implement each pattern.

## III. BACKGROUND

This document focuses on some research papers related to the open-source systems DMARF and GIPSY. Distributed

Modular Audio Recognition Framework (DMARF) is basically the distributed version of MARF. The General Intensional Programming System (GIPSY) is a multi-tier compilation and run-time environment system aimed to provide a platform for intensional programmers that goes beyond mere programming in the standard intensional programming language Lucid.

## A. *OSS Case Studies*

### 1) *DMARF*

Distributed Modular Audio Recognition Framework (DMARF) is based on the classical MARF where the pipeline stages of MARF are made into distributed nodes.[15]

- *Classical MARF:*

MARF (Modular Audio Recognition Framework) is a general open-source research platform associated with various domains like pattern recognition, signal processing and natural language processing (NLP) [15].

The MARF consists of a sequentially organized set of pipelines which communicate with each other for the processing of data. This forms the backbone of MARF. It is comprised of four basic stages as shown in Figure 1 – sample loading, preprocessing, feature extraction and training/ classification.

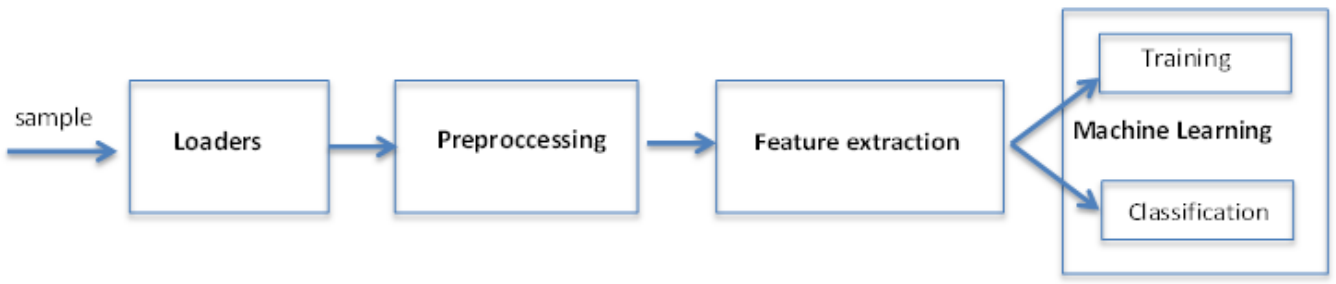

**Figure 1. Abstract view of the Classical MARF**

MARF has been extended to DMARF, the goal of which is to decouple the MARF's stages into distributed services running on different nodes on the network [14, 3]. DMARF aims to facilitate disaster recovery, fault tolerance and high availability of the system as a whole, by having its various stages distributed over the network. Furthermore, the communication between the different components is achieved using Web services (which have extended earlier implementation using RMI and CORBA which imposed limited intra-operability between services and inter-operability with clients [14]), thereby providing services in a uniform fashion independent of the communication technology [9].

To achieve its intended goals as a distributed service provider, the developers of DMARF implemented a general Web service through which clients can seamlessly communicate with the various functionalities in classical MARF, each of which is itself a service [14]. These services expose classical MARF functionalities depicted in Figure 1 each as a service, but also add further services including: natural language processing, speaker identification, language identification, miscellaneous services (e.g. probabilistic parsing [14]).

The requirements for DMARF on these services include [14]:

- Concurrency: lacking in MARF, necessary when processing high volume of voice samples
- Distributed pipeline as an option
- Disaster recovery
- Service replication
- Communication-technology-independence
- Ability to manage, configure and monitor MARF nodes using common network management tools over SNMP protocol [5].

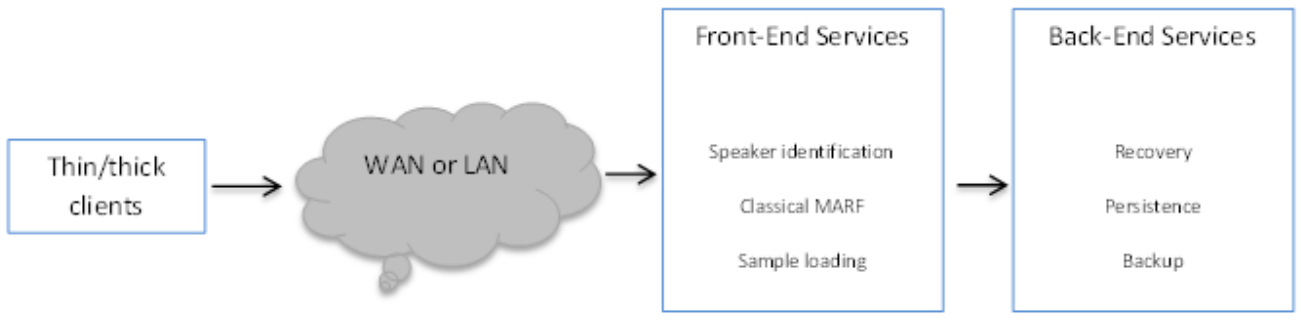

**Figure 2. Abstract View of DMARF**

The DMARF is essentially a complex system composed of multi-level operational layers. The architecture of DMARF is such that it is easy to achieve the desired functionality without any failure. DMARF module runs on different machines as a backup server such that if primary server crashes then the machine will take over the system and provide uninterrupted service by sending the request to the backup server. The division of the DMARF into different layers ensures that there is no loss of data if the system crashes. [9]

DMARF, because of its inclination of use, executes a pipelined or anchored method for joining some of its dispersed hubs after the example distinguish pipeline stages, consequently making it synchronous. While there is synchronicity inside the pipeline, it is shaped for the current subject/ test assessment, and the pipeline association way may be diverse for every new subject or specimen. DMARF without anyone else's input does not display the interest driven model of calculation and permits applications interfacing straightforwardly to a percentage of the hubs staying away from the pipeline on the off chance that they simply oblige the administrations of those specific hubs. DMARF does not manage the execution of unequivocal code, so there is no blunder remedy issue. [11]

In MARF there was the disadvantage of the lack of concurrency when processing bulk inputs, which consequently had a negative impact on performance. Therefore, DMARF had to be developed in order to make the pipeline distributed and to offload heavy processing from the clients, which may be running on lightweight hardware such as a mobile phone, to the backend group of powerful servers, organized into clusters, to perform heavy processing while the client is performing data collection tasks. DMARF was also developed to make MARF independent of communication technology, replicable, disaster recoverable and also to boost its performance. [5]

DMARF components are able to communicate via TCP, RMI, CORBA and XML-RPC. However, sometimes when the DMARF nodes were large, it was difficult to manage them using DMARF proprietary communication protocol. Accordingly, DMARF was extended to support the industry common network management protocol SNMP. The addition of SNMP protocol enabled the integration of DMARF with other industrial network management tools and allowed network administrators to configure, manage, monitor performance and status of nodes, and retrieve statistics from MARF nodes using common familiar tools. [5]

Initially, the DMARF could not be used in autonomous environments due the lack of autonomic design provisions. As a result, DMARF was integrated with Autonomic System Specification Language (ASSL), which assisted the incorporation of the three autonomic features – self-healing, self-optimization and self-protection i.e. the core self-CHOP autonomic properties.

ASSL is an abbreviation for Autonomic System Specification Language. ASSL addresses the issue of formal specification and code generation of autonomic systems (ASs) within a framework. ASSL framework provides a toolset that aids in validation of specifications by checking for syntax and consistency errors against a set of semantic definitions. If the validation checks pass, ASSL generates an operational Java Application skeleton corresponding to the specification. ASSL framework has a multi-tier system architecture where tiers describe the system specification at multiple levels of abstraction thereby addressing the problem of complexity. These aspects are realized by defining ASSL in multiple tiers, which are the AS tier, the ASIP tier and the AE tier.[1]

In the process of making DMARF autonomous with ASSL, for each distributed node a single AE that introduces an autonomic behavior at that node is specified. The novelty in this approach is to safeguard the pipeline which is not possible with common distributed systems. Self-healing in a DMARF-based system refers to the ability to recover with replication such that one route of the pipeline is available. Two type of replication is possible. One is the replication of a service which is increasing the number of nodes per stage and the other is the replication within the node itself. When one of the stages in the DMARF’s pipeline goes offline, the pipeline stalls and to recover it, the following options are considered:

- The use of a replacement node
- Recovery of the failed node and
- Re-routing the pipeline through a different node with the same service functionality until the failed one recovers.

Self-healing algorithm is spread on both the AS-tier and AE-tier levels where events, actions, metrics and special managed element interface functions are used to incorporate the self-healing property in Autonomous DMARF. [7]

To incorporate the self-optimization property in DMARF, two main functional requirements need to be added to the DMARF architecture. The DMARF system handles and computes a lot of data. The main load is handled by the Classification stage which deals with both I/O bound data processing and other computational features including various complex calculations. MARF uses dynamic programming to cache these results, but in DMARF this data is stored across various systems. This can lead to recomputation of already computed values, if the distributed nodes do not communicate with each other. Thus to avoid this, the DMARF Classification nodes need to be equipped with a feature that enables automatic communication between these nodes as soon as the results are cached, to avoid redundant computations. Another feature to be included was the automatic selection of the available most efficient communication protocol i.e. dynamic communication protocol selection. [1]

To equip DMARF with self-managing capabilities, a special automatic manager (AM) is added to each DMARF stage.

This transformed DMARF into Autonomic Elements (AEs) that constitute the Autonomic DMARF (ADMARF). The main reason for using ASSL is ‘security’ i.e. the incoming messages must be secure. This is why, ADMARF systems are more significant in global environment running over the Internet. From an abstract level, the self-protection is achieved by an algorithm which checks for the timing, sender identification and security aspects of the message. DMARF specification defines this algorithm as an ASSL self-protecting tool. [3]

*2) GIPSY*

Intensional programming, also referred to as multidimensional programming, is a programming paradigm that has its mathematical foundation in intensional logic, whereby the outcome of the evaluation of a given logical expression is dependent on the 'context'. The General Intensional Programming System (GIPSY) is a multi-tier compilation and run-time environment system aimed to provide a platform for intensional programmers that goes beyond mere programming in the standard intensional programming language Lucid. GIPSY was conceived of a time (late 90's) where existing intensional programming software tools were becoming outdated and not at pace with the theoretical advancements in the field. The main objective of GIPSY has therefore been to provide a platform for researchers to advance relevant and practical research projects that lend themselves to intensional programming.

GIPSY has various requirements. One of the aims is to create a run-time system for distributed execution of programs written in any variant of Lucid language using an eductive model of computation in order to enhance five quality attributes of architecture i.e. language independence, scalability, flexibility of execution architecture, opacity of run-time considerations, observability [4]. While Lucid is the main programming language to be handled by GIPSY, it is also used for evaluation of Higher Order Intensional Logic (HOIL) expressions [12]. To strengthen Lucid in these areas, Java-Lucid hybrid dialects were later developed [6]. The original GIPSY multi-threaded and distributed architecture which was written in Java RMI is not fully integrated and the detailed work flow need to be clarified. There Are two more separate branches of GIPSY implemented based on Jini and JMS of Demand Migration Framework (DMF) are interoperable and their top interfaces complicating and delaying the execution of Lucid programs in GIPSY run-time system which is General Eduction Engine (GEE) [14]. Later, system got additional requirement to realize the goal-driven self-healing, self-protection, self-optimization and self-configuration which are the aspects of autonomic computing.[8].

The main components of the original GIPSY architecture are depicted in Figure 1. Lucid code is compiled into an intensional data dependency structure (IDS), and it is fed to the General Eduction Engine (GEE), which is the run-time interpreter in GIPSY (where ‘eduction’ implies a demand-driven computation in which each procedural call is executed locally or remotely thus simulating parallel computing [16]). The General Eduction Engine (GEE) was designed to be more flexible than GLU (Granular Lucid).[16] The GEE was also incorporated into the ASSL for the compilation of JOOIP code in order to integrate self-forensics in it. [2]. RIPE is Intensional Run-time Programming Environment. The sequential functions of the GIPSY program are translated into C code using the second stage C compiler syntax, yielding C sequential threads (CST), while “data communication procedures used in a distributed evaluation of the program are also generated by the GIPC according to the data structures definitions written in the Lucid part, yielding a set of intensional communication procedures (ICP)” [16]. GIPC transforms Lucid code into IDS (intensional data dependency structure) which is interpreted at run-time by GEE; or into intensional communication procedures (ICP) [16].

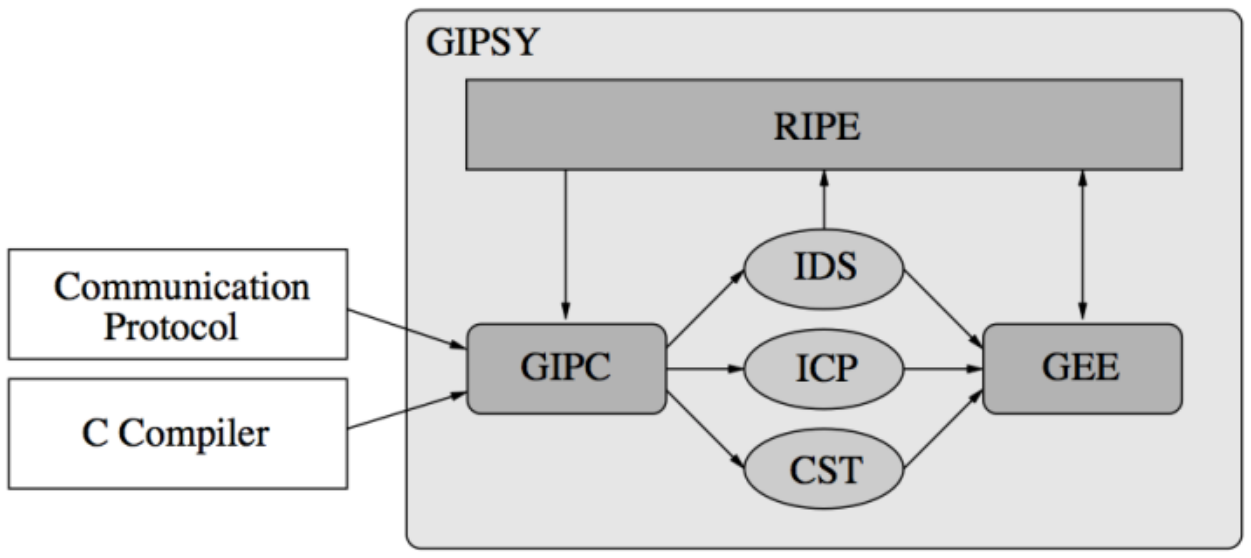

**Figure 3. The GIPSY Architecture**

One of Java-Lucid dialects, JOOIP is bidirectional and allows calling Lucid functions from Java and Java methods from Lucid. Since computation of such hybrid programs with heavy weight Java components is resource-intensive and because Lucid programs are naturally parallel, GEE had to be extended to compute hybrid components distributively. [6, 2] JOOIP, which allowed the mixing of Java and Lucid code by placing the Lucid code anywhere within the Java classes, could be compiled by the GIPSY Eduction Engine i.e. GEE. The JOOIP code, generated by the ASSL toolset, along with the Forensic Lucid specification is passed on to the hybrid General Intensional Programming Compiler (GIPC) of GIPSY for its compilation. The generic nature of GIPL (Generic Intensional Programming Compiler) allows the system to create an intermediate representation called GEER which makes the system language independent, thereby increasing versatility [4]. This is further linked together into an executable code inside the GEE engine resources (GEER) which then has three options for evaluation including traditional eduction model of GEE, Aspect-J based eduction model and probabilistic model checking with the PRISM backend. [2]

To support distributive execution in GIPSY, a Demand Driven Framework (DMF) was defined. DMF defines interfaces for distributed propagation of demands among demand generators and demand workers and two implementations of DMF, called Demand Migration System (DMS) were developed. One DMS was implemented using Jini and another using JMS distributed middleware technologies. JMS-based implementation of DMS was more reliable then Jini because JMS had more flexibility for configuring memory management, but Jini-based implementation provides better throughput and was easier for development and deployment. Both Jini and JMS based implementations of DMS were built with strong dependencies on the specific distributed middleware technologies and weren't able to work with other middleware technologies. In order to perform comparative studies, it was decided to refactor and unify both DMS implementations into one DMS which is able to operate on both Jini and JMS technologies in the same GIPSY runtime environment, and to make it easy to add support for other middleware technologies in the future. As part of the refactoring, Jini-based Transport Agent (JiniTA) and JMS-based Transport Agent (JMSTA) were created. These components implement middleware-specific tasks and are encapsulated from the rest of the system, and other components work with them via an interface. Benefits of doing so made it inexpensive to integrate new Transport Agents possibly developed on other distributive technologies with the rest of DMS [6].

To evaluate the HOIL expressions, the intermediate representation of the compiler is required. The compiler is GEER (Generic Eduction Engine Resource). In GISPY tire concept, execution of GISPY programs will be divided into three different tasks, and those tasks will be assigned to three different tiers. GISPY node is nothing but a computer, and it hosts one or more GISPY tiers. GISPY instance is interconnected by GISPY tiers, which are deployed on GISPY nodes. GISPY instance will execute GISPY programs. The Demand Generator Tier (DGT) will produce the demands according to the program definition and declaration in the GEER instance. Demand Store Tier (DST) migrates the demands among the different tiers. The Demand Work Tier (DWT) executes the functions and methods of the intensional programs. GISPY Instance Manager (GIM) will control the GISPY nodes and tiers. To manage these nodes, GIM allocates nodes to the GISPY instance [12].

A Graph-based automation assistant was created which is perused as a Case study to deal with the GIPSY. The detailed analysis portrays the answer for assessing requests gave at Run-time by utilizing GIPSY Pattern. In GIPSY, the generator chooses an interest while specialist registers the chose request and answer. This system empowers to perform different sorts of projects. In the GIPSY-NODES there is a director node which goes about as a super-node, this node is utilized to deal with the GIPSY Network. The configuration of GIPSY consists of Java Universal Network/Graph (JUNG) library to implement visualization of management aspect and it should be platform tested. [10]

Regardless of the flavor of Lucid, the source code is converted to "generic Lucid" called Generic Intensional Programming Language (GIPL) when the compiler generates an intermediate representation called GEER. Secondly, the ability to declare wrappers to be combined with the procedures enables them to be called by Java Virtual Machine. The run-time GEER instance, possessing both these, is the language independent solution. If needed, at run time, instances of tier are created. The solution provides a provision for scalability of operations by facilitating new tier instances generation once the existing ones are either overloaded or lost. The solution allows new nodes to be added or removed as necessary during run time. Secondly, GIPSY instances are allowed to execute multiple programs concurrently. Hybrid programs can be executed along with pure Lucid programs. All these features increase flexibility. Just like GLU, the type of topology it uses for program execution can be chosen, both before or during the run time. The demands are pooled in DST for use in subsequent run time operations and processed demands are stored in an output buffer called Local Demand Store. This property of observability makes GIPSY eligible to be an experimental base. [4]

To make GIPSY autonomic, AGIPSY architecture was developed. AGIPSY's aim was to realize the goal-driven self-healing, self-protection, self-optimization and self-configuration which are the aspects of autonomic computing. The foundation of this AGIPSY is done with ASSL. AGIPSY is a composition of autonomous GNs. All GNs are considered to be AEs. A GN's control is provided by a node manager (NM) which allows a GM follow its own thread of execution. AGIPSY architecture has all the features of a multi-agent loosely coupled distributed system with decentralized control and data allocation. The following are the autonomic features of AGIPSY. Fault tolerance and recovery: Fault tolerance and recovery is the mechanism where the GNs recover from the failures using ASSL recovery protocol. Self-Management features: The different self-management policies are Self-Configuration, Self-Optimization, Self-Healing, Self-Protection. Dynamic allocation of resources like GIPSY tier instances to a GN is an example for self-configuration. An AGIPSY can optimize its performance by measuring its own with an optimum which is an example for self-optimization. Self-healing is the process of creating resilient GNs which can recover from any faults. AGIPSY should be protected from any malicious or accidental external attacks, which is done by protection filters (Self-Protection). [8]

### B. Summary

In this section we have summarized the measurements that we have calculated from the two open-source systems DMARF and GIPSY. This data is enlisted in Table 1.

**Table 1.** DMARF and GIPSY Measurements

| **Measurements** | **DMARF** | **GIPSY** |
|---|---|---|
| Number of Java Files | 1024 | 592 |
| Number of Classes | 979 | 580 |
| Number of Methods | 6109 | 5746 |
| Lines of Java Code | 77297 | 104073 |

We have used various measurement tools to calculate the size of both the systems which are cited in Table 1. The tools which were used include the Metrics plugin to Eclipse version 1.3.6, SonarQube and also some LINUX commands [Appendix A].

MARF architecture was initially a sequential pipeline. This was later improved to a distributed pipeline i.e. DMARF. These changes were made to incorporate certain features which were not possible with the original stand-alone MARF. Then DMARF architecture was extended to support management of distributed MARF nodes over SNMP [6]. Also, including the self-CHOP autonomic properties enabled the DMARF to become self-adapting [8][2][4]. Thus optimizing the overall system and making it more efficient.

Summarizing the GIPSY case-study, GIPSY is a multi-tier architecture, which has applications in Intensional Programming, Eductive Model of Computation, Lucid Programming, and Hybrid Programming. In order to make the system work on Java RMI, Jini and JMS there was an enhancement to make GLU's Generator-Worker architecture multi-tiered. Also, GIPSY's distributed framework was refactored to work with nodes over JMS and Jini in the same GIPSY runtime environment. In the process of making GIPSY autonomous, ASSL, which has a multi-tier architecture, has been incorporated into GIPSY's architecture, which made GIPSY AGIPSY.

GIPSY's domains are Intensional Programming, Lucid Programming, Hybrid programming and Cloud Computing. AGIPSY's domains are Robotic and Autonomous systems where the human intervention is less.

## IV. Requirements and Design Specifications

The following two use cases are based on the application of DMARF[19] and GIPSY[18],[12] in cyberforensics. The intensional cyberforensics aims at verifying the evidences against the witness claims using forensic lucid specifications[18] which is later used to backtracking the events. To realize this goal, both DMARF and GIPSY are used.

### A. Personas, Actors and Stakeholders

#### 1) DMARF

**Secondary Persona: Police Investigator**

Smith is a 49 year old police investigator who specializes in homicide crimes and his police experience totals 25 years of outstanding service record. He started his police career as a formed patrol officer after completion of his Bachelor's Degree in Criminal Justice. Few years later, after gaining the mandatory initial police experience and knowledge in investigative techniques, he was promoted to the position of a police detective. As a homicide detective, his main duty is investigation of suspicious deaths. On a typical day he interviews suspects, victims and witnesses, to try to build a complete understating of the story of how the crime had occurred. He also works closely with the Crime Scene Investigation Unit, which is responsible for collecting forensic evidences which he needs for working on the cases. He is also often called for testifying at courts using the forensic evidences and information he diligently collected. He has a basic experience as a computer user, as he regularly uses computer for filling reports and managing the information related to cases which he is investigating, however he has no experience with complex software systems. He often works on very complicated cases which involve many forensic evidences, and to assist himself he would like to have a software system which given the founded evidences would help proving or disproving the guilt of the suspect.

**Additional Stakeholders:**

**DMARF Application Expert:**

Upendra is a 27 year old recent graduate in Software Engineering who conducted extensive research within the DMARF ecosystem for his Master's thesis. While looking for a job, he responded to a job advertisement from a regional police department and was hired as a DMARF application expert to help police investigators use the newly acquired DMARF system. He has been working there for the last two years where he had an opportunity to work on different criminal cases.

*2) GIPSY*

**Primary Persona: GIPSY Application Expert**

Ernesto is a 32 year old experienced software engineer. He lives with his wife and two kids in New York city. He likes to travel and to read books. He holds a master's degree in computer science engineering. He is a Java expert who is equally skilled in intensional programming. He has a strong technical sense and a high motivation to experiment with innovative software tools and techniques that aim at helping police detectives to do their work. He has a very good understanding of the General Intensional Programming System. His long-term goal is to become an intensional programmer but currently he is working as a GIPSY application expert dealing with forensic lucid. He is also a part time professor in Columbia university, New York. He teaches the graduate level advanced software architecture course. He has published a few articles on cyber forensics application of GIPSY.

**Additional Stakeholders for DMARF and GIPSY:**

**Research scientist:**
A research student in the broad domain of machine learning and natural language processing has a stake in using, testing, extending or integrating DMARF and/or GIPSY in the contest of pursuing a research agenda under the umbrella of these domains. The scientist can use DMARF and/or GIPSY to train/classify bulk data samples, test their performance against other similar tools, extend the functionalities of some training/classification algorithms, or integrate them as part of a development pipeline having a machine learning/natural language processing step (thru Web services for example).

**Third-party Web services:**
Given the modular design of DMARF and GIPSY, and the accessibility to their various modular components thru various methods, chief among which is Web services, other third-party Web-services can integrate the functionalities of DMARF and GIPSY into the service they provide. Web services have a stake in the availability of DMARF and GIPSY, as well as a stake in their correct implementation of standard protocols of inter-service communication.

## *B. Fully Dressed Use Cases*

*1) DMARF*

**Use case UC1:** Forensic Lucid encoding of forensic evidence samples

**Primary Actor**: DMARF Application Expert
**Secondary Actor:** Police Investigator

**Stakeholders and interests:**

- Police investigator: wants quick processing of evidence sample, consistent output, without omission of any elements in the sample
- DMARF Application Expert: wants to generate Forensic Lucid expressions from audio/ text samples
- Research scientist.
- Third party web services.

**Preconditions:**

- DMARF is up and running
- DMARF sample loaders are functional
- The network is up; distributed components of DMARF are accessible

**Postconditions:**

- Valid forensic Lucid expressions are produced
- DMARF has terminated

**Main success scenario:**

1. The police investigator collects the **audio/ text evidence samples** for analysis and gives it to the **DMARF Application Expert**.
2. The DMARF Application Expert then launches the appropriate DMARF **sample loader**.
3. The samples of evidential statements and witness claims are loaded and pre-processed.
4. The **adapter translator** program translates the MARF's data structures into the **Forensic Lucid-compatible expressions**.

**Extensions:**

2a. If sample loader fails. Indicate error.
2b. If sample loader is offline. Indicate error.

3a. If sample loader rejects sample. Indicate error.

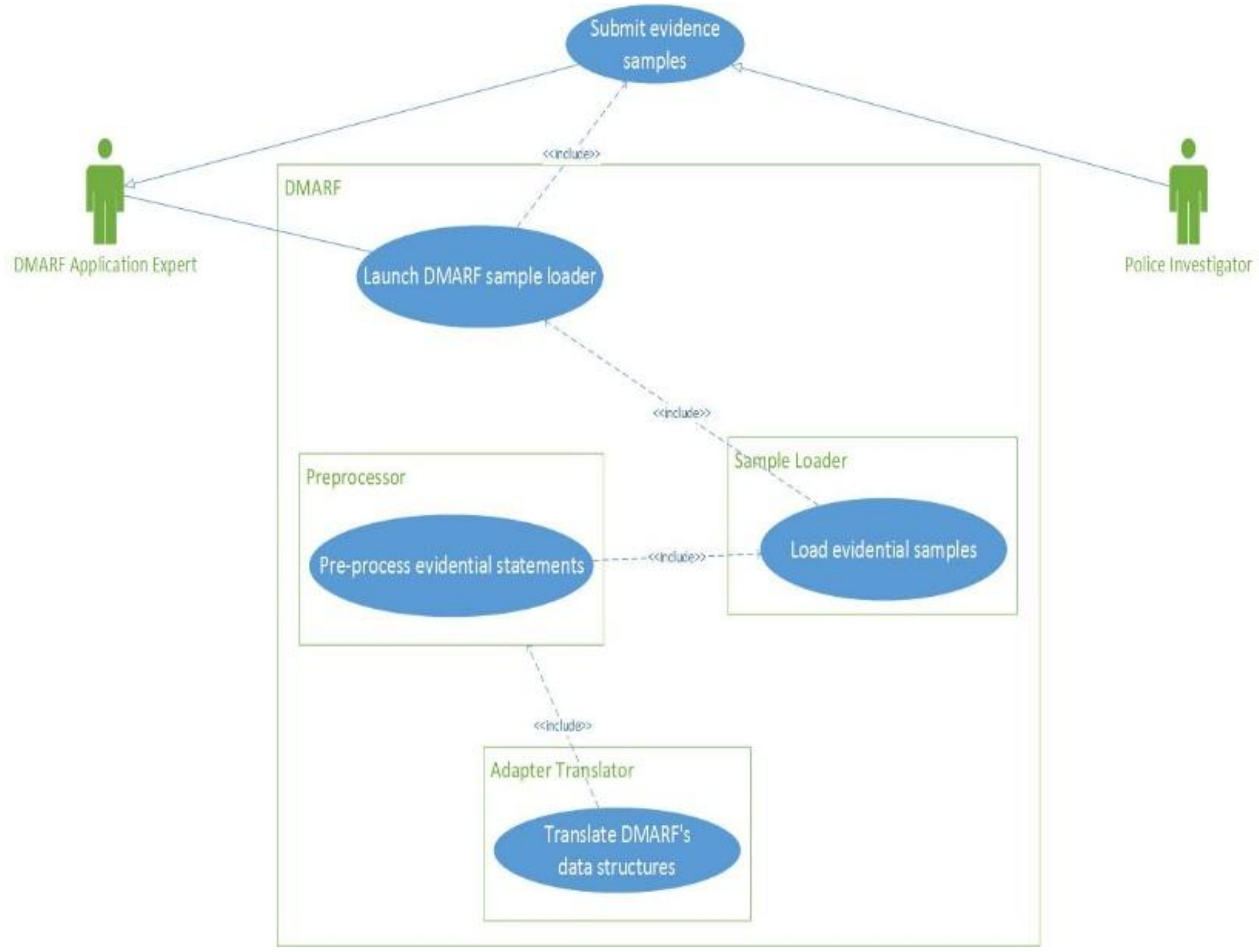

**Figure 4 - Use case diagram for DMARF**

*2) GIPSY*

**Use case UC2:** Identifying conflicting witness testimonies.

**Primary Actors:** GIPSY Application expert
**Secondary Actor:** Police Investigator

**Stakeholders and interests:**

- Police Investigator: wants quick analysis of the sample and generation of the output
- GIPSY Application Expert: wants to process Forensic Lucid expressions using GIPSY.
- Research scientist.
- Third party web services.

**Preconditions:**

- GIPSY is Forensic-Lucid-capable
- GIPSY's Forensic-Lucid components (compiler, parser, semantic analyzer) are functional.
- GIPSY is up and running
- The network is up; distributed components of GIPSY are accessible

**Postconditions:**

- GIPSY has terminated
- A text-based output of conflicting evidences has been produced.

**Main success scenario:**

1. The police investigator brings the **encoded evidences** (Forensic Lucid expressions) and the **crime scene specifications report** to the GIPSY Application expert.

2. The GIPSY Application Expert transforms the crime scene report specifications into **transition functions** using the **Data Flow Graph (DFG) Editor**.
3. The GIPSY Application Expert combines the Forensic Lucid Expressions and the transition functions into a **Forensic Lucid Specification**.
4. The GIPSY Application Expert then runs the GIPSY **Intensional Compiler** (GIPC) with the Forensic Lucid Specification.
5. The GIPSY Application Expert then evaluates the compiled program using the GIPSY **Eduction Engine** (GEE).
6. A **text-based output** of conflicting testimonies and the involved witnesses is produced.

**Extensions:**

3a. Invalid Forensic Lucid syntax in the input expressions. Indicate error.

**Special requirements:**

- Forensic Lucid-aware GIPSY

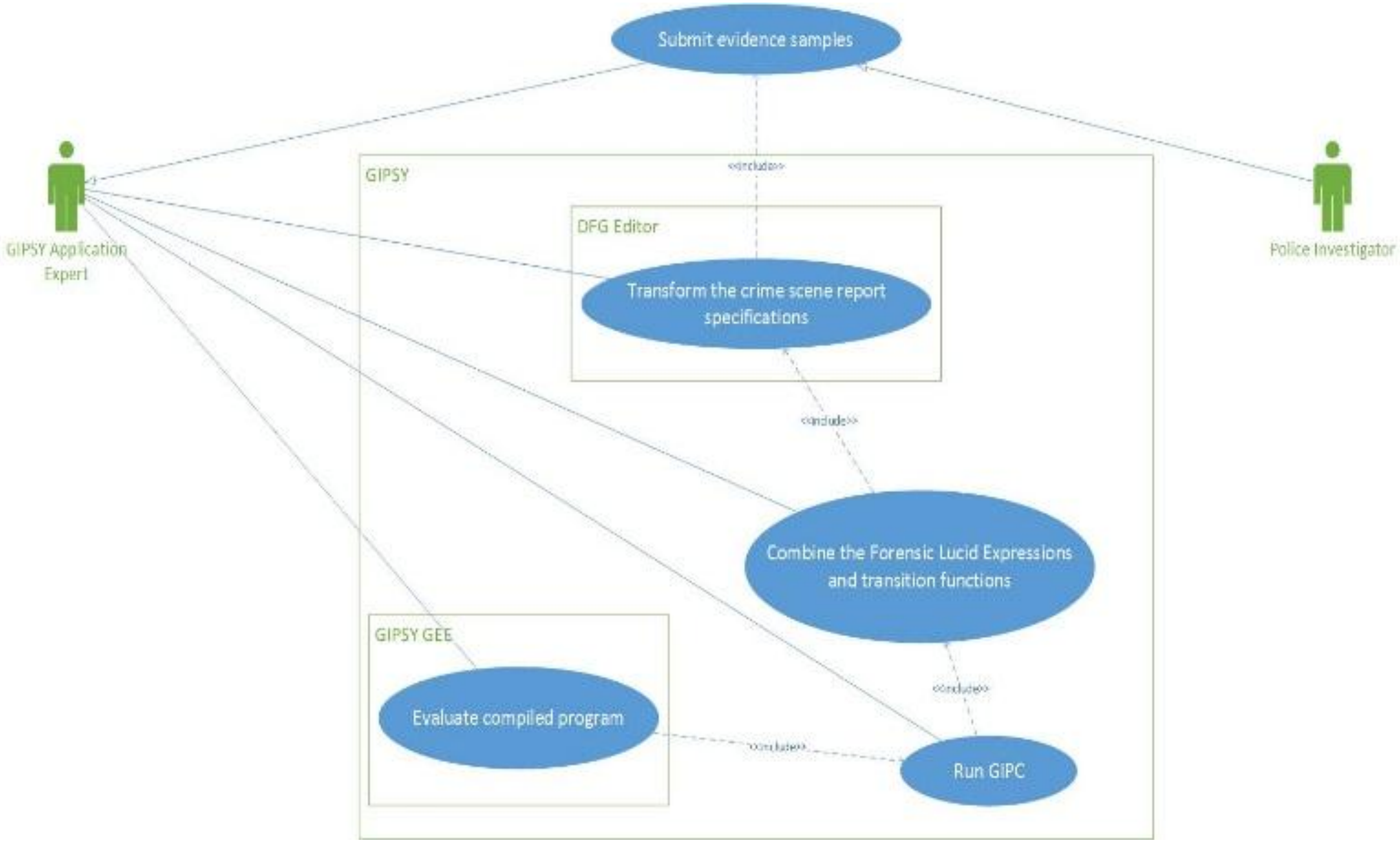

**Figure 5 - Use case diagram for GIPSY**

### C. *Domain Model UML Diagrams*

#### 1) *DMARF*

The Evidence is either an audio or a text version of the claims of witnesses that are loaded into the sample loaders of the DMARF system. There can be different types of sample loaders in the DMARF system. Preprocessing of the loaded evidence is done by the preprocessors. One DMARF system has many preprocessors. There is a composition relation between DMARF system and the sample loaders, preprocessors and adaptor translator. The Adaptor translator is a program and is a part of the DMARF system that generates many contextual forensic expressions depending on the evidential statements. There is one adaptor translator program in the DMARF system for generating many forensic lucid expressions. Forensic lucid expressions are the encoded evidential statements meant for further analysis in a cyberforensic case.

#### 2) *GIPSY*

Forensic lucid expressions are the encoded evidential statements. Crime Scene details are entered into Data Flow

Graph Editor which generates Transition Functions. A single Data Flow Graph Editor can generate many transition functions. One or more contextual Forensic Lucid Expression together with one or more Transition Functions gives the Forensic Lucid specification. Forensic Lucid Specification is compiled by GIPSY Forensic Compiler and is evaluated by the GIPSY Eduction Engine. The Forensic Lucid Compiler and Eduction Engine are part of the GIPSY; therefore there is a Composition relation between Forensic Compiler and GIPSY and between Eduction Engine and GIPSY. Eduction Engine evaluates the compiled Lucid program and produces text based outputs. The text based outputs are the conflicting testimonies and the involved witnesses. Since a single Eduction Engine can produce zero or many text-based outputs, there is a one-to-many relation between them

3) *Fused DMARF-Over-GIPSY Run-time Architecture(DoGRTA)*

The fused domain model shows DMARF using GIPSY's run time which is the GEE's multi-tier architecture for distributed computing. The domain diagram describes the Generator and Worker tiers on the pipeline stages of DMARF inheriting the DGT and DWT of GEE. The tiers in the distributed pipelines work in a way similar to GEE's DWT and DGT while maintaining the semantics of the DMARF's pipeline stages. To explain one pipeline stage, the sample loader generator and Sample loader worker work exactly similar to the Demand generator and worker. The DGT generates the demands, likewise the sample loader generator generates the samples that are to be loaded. The DST stores the demands that are computed and pending. The demand workers select the demands that are in pending state from the storage tier and start computing them. Likewise the sample loader worker picks up a sample that is in pending state and starts working on it. The loaded samples are then stored in the storage tier. The use cases described above for DMARF and GIPSY can still work on this fused architecture successfully as the generator and worker tiers does not break the semantics of the pipeline.

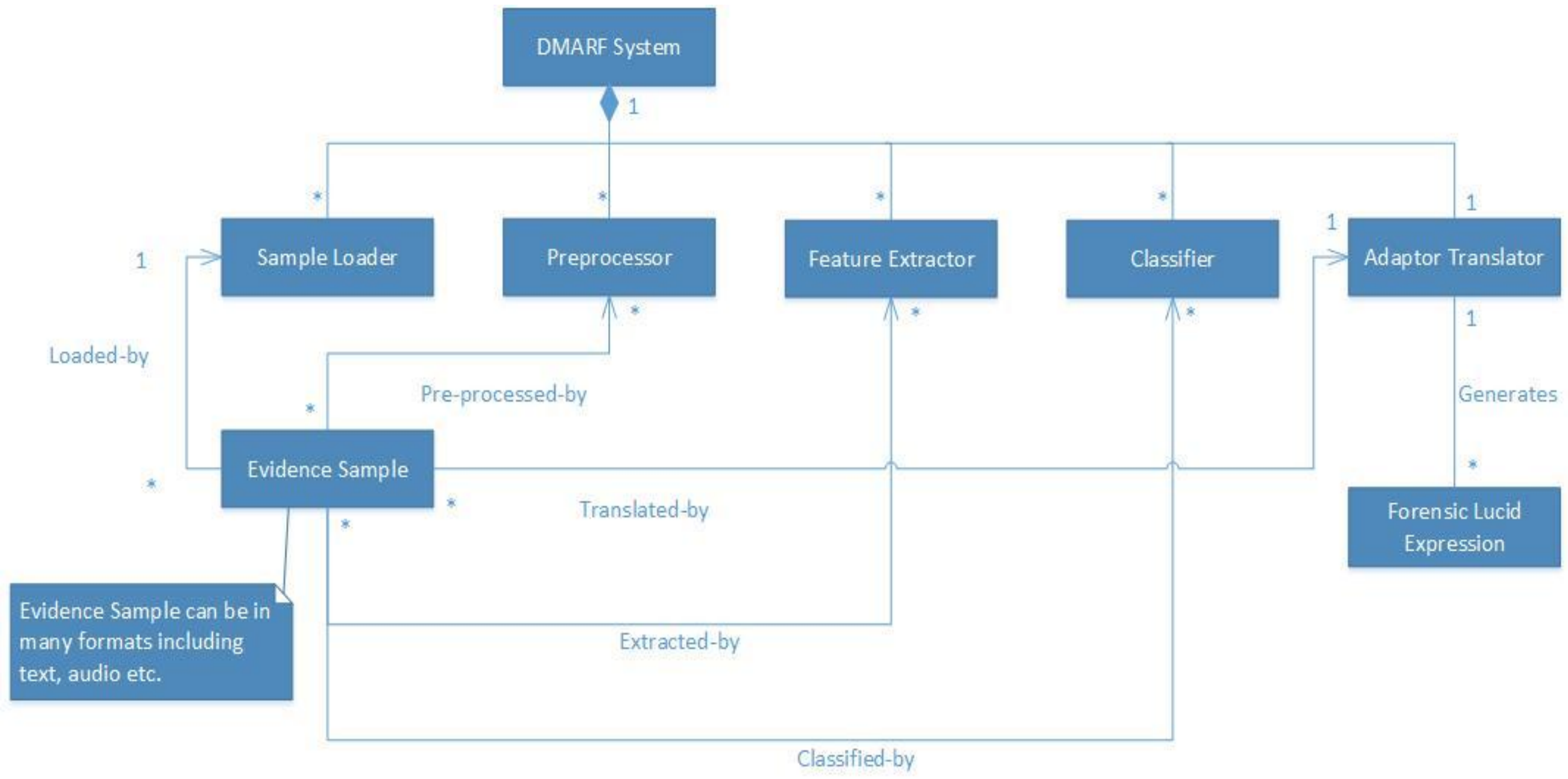

**Figure 6 - DMARF Domain model**

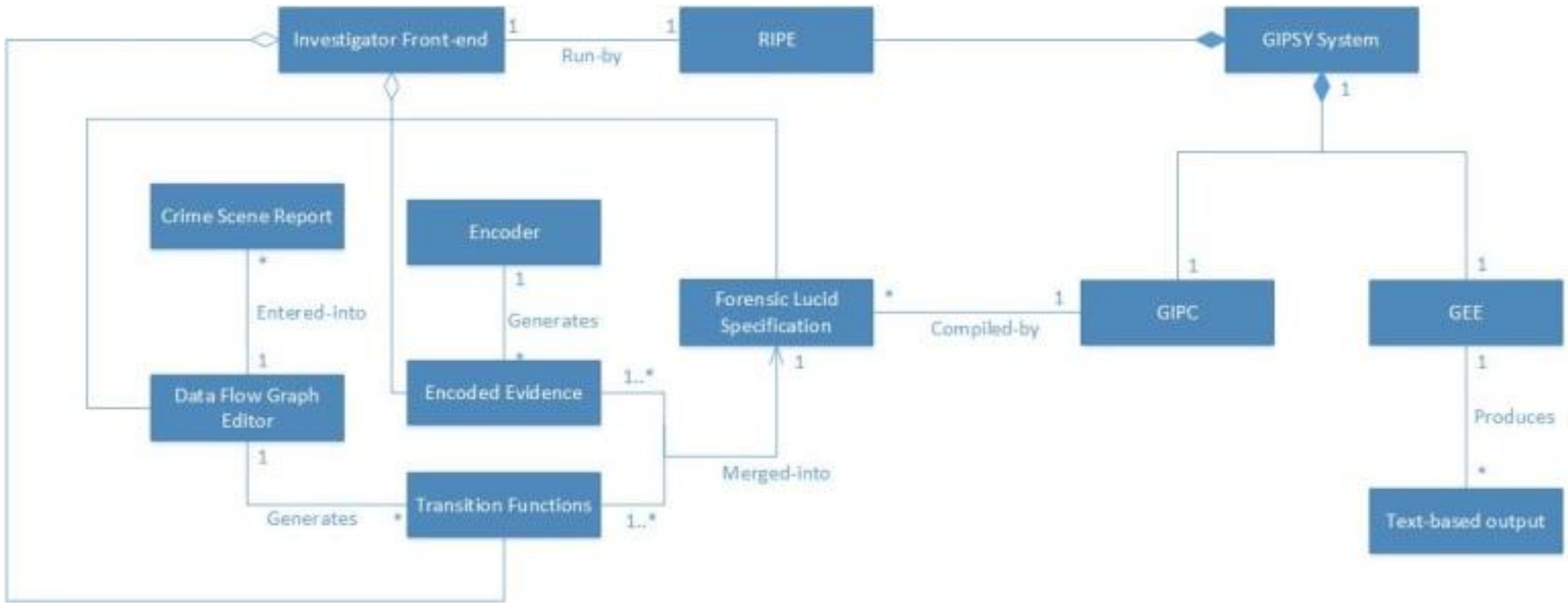

**Figure 7 - GIPSY domain model**

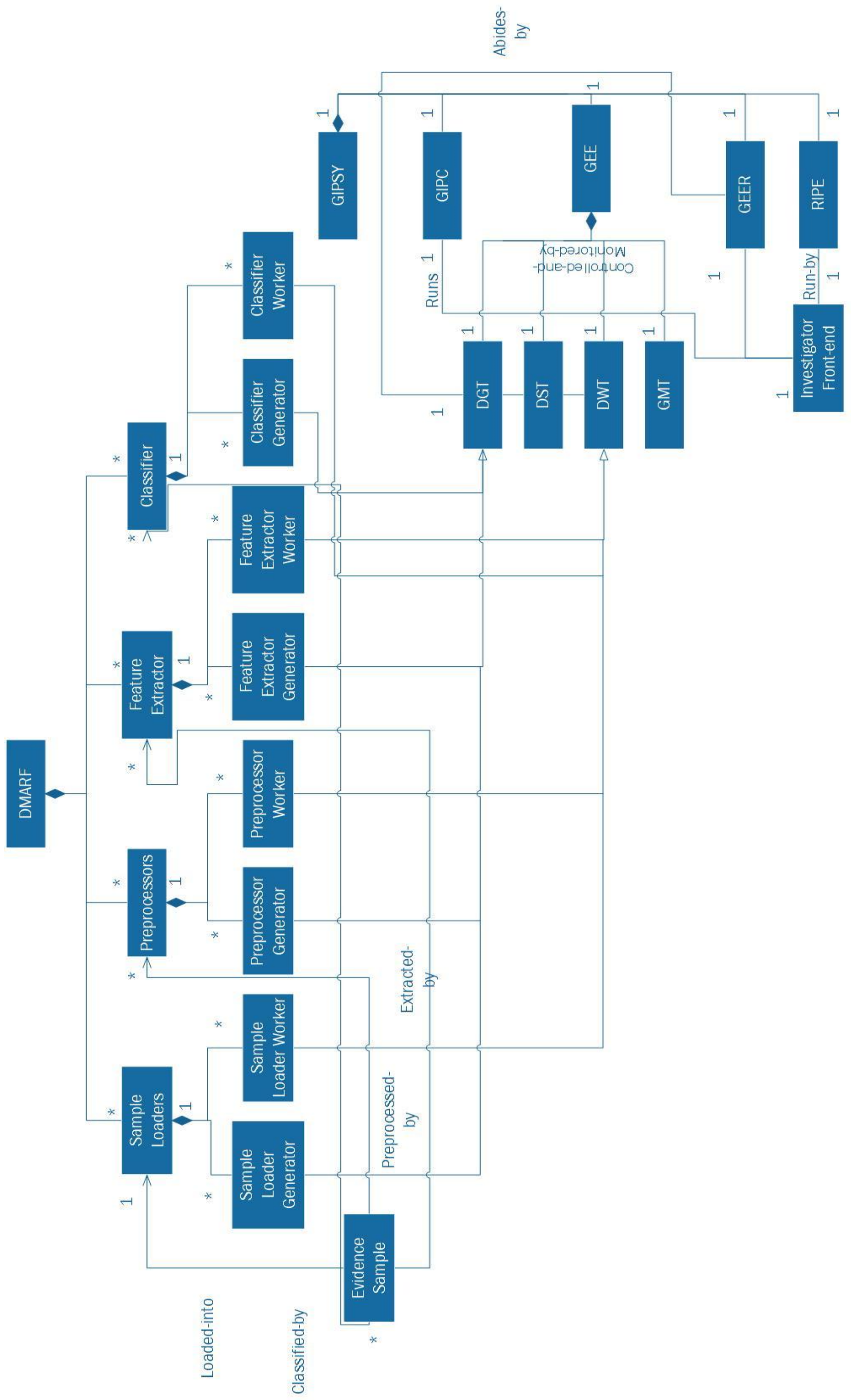

**Figure 8- DoGTRA Domain model**

*D. Actual Architecture UML Diagrams*

*1) DMARF*

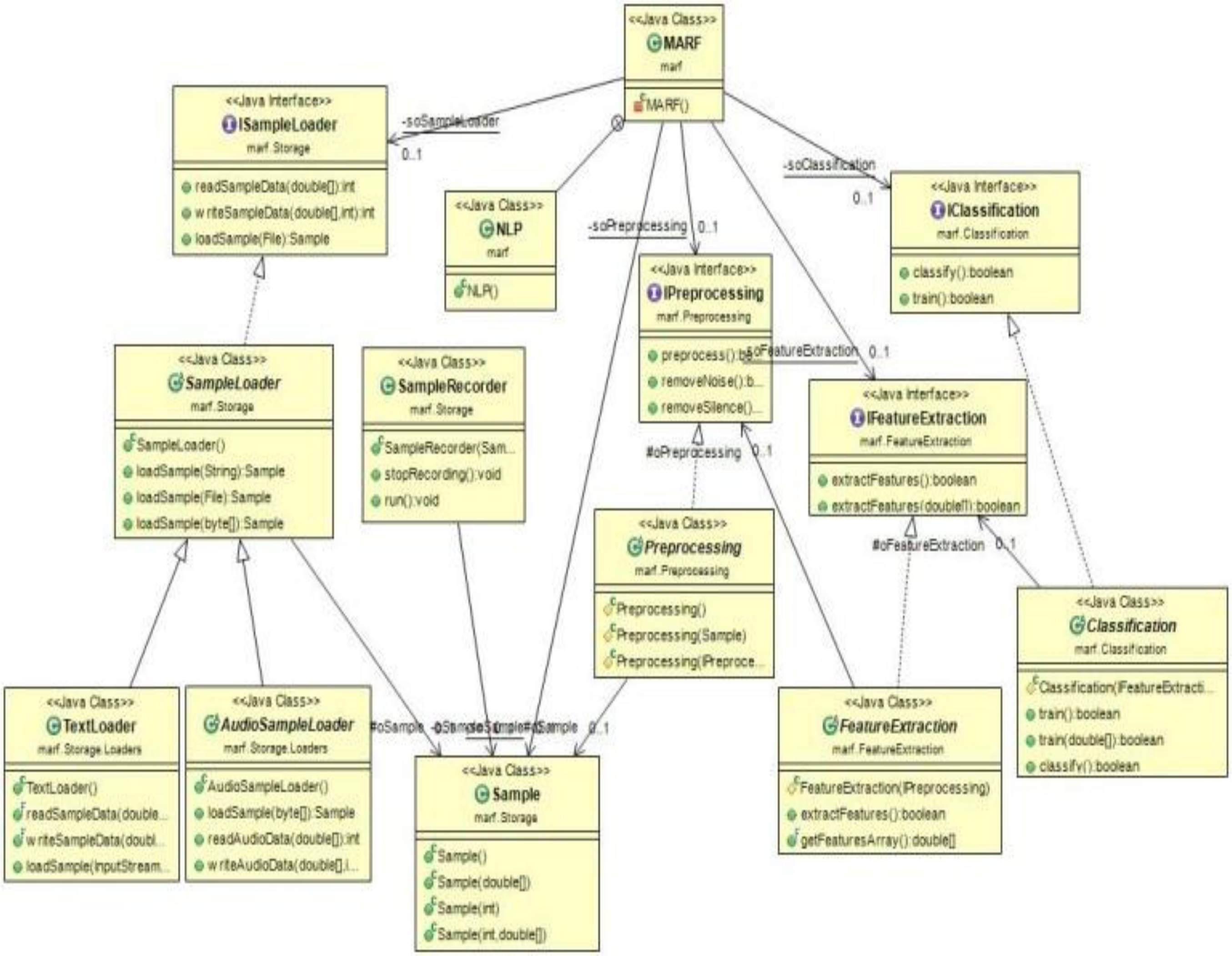

**Figure 9 - DMARF Class Diagram 1**

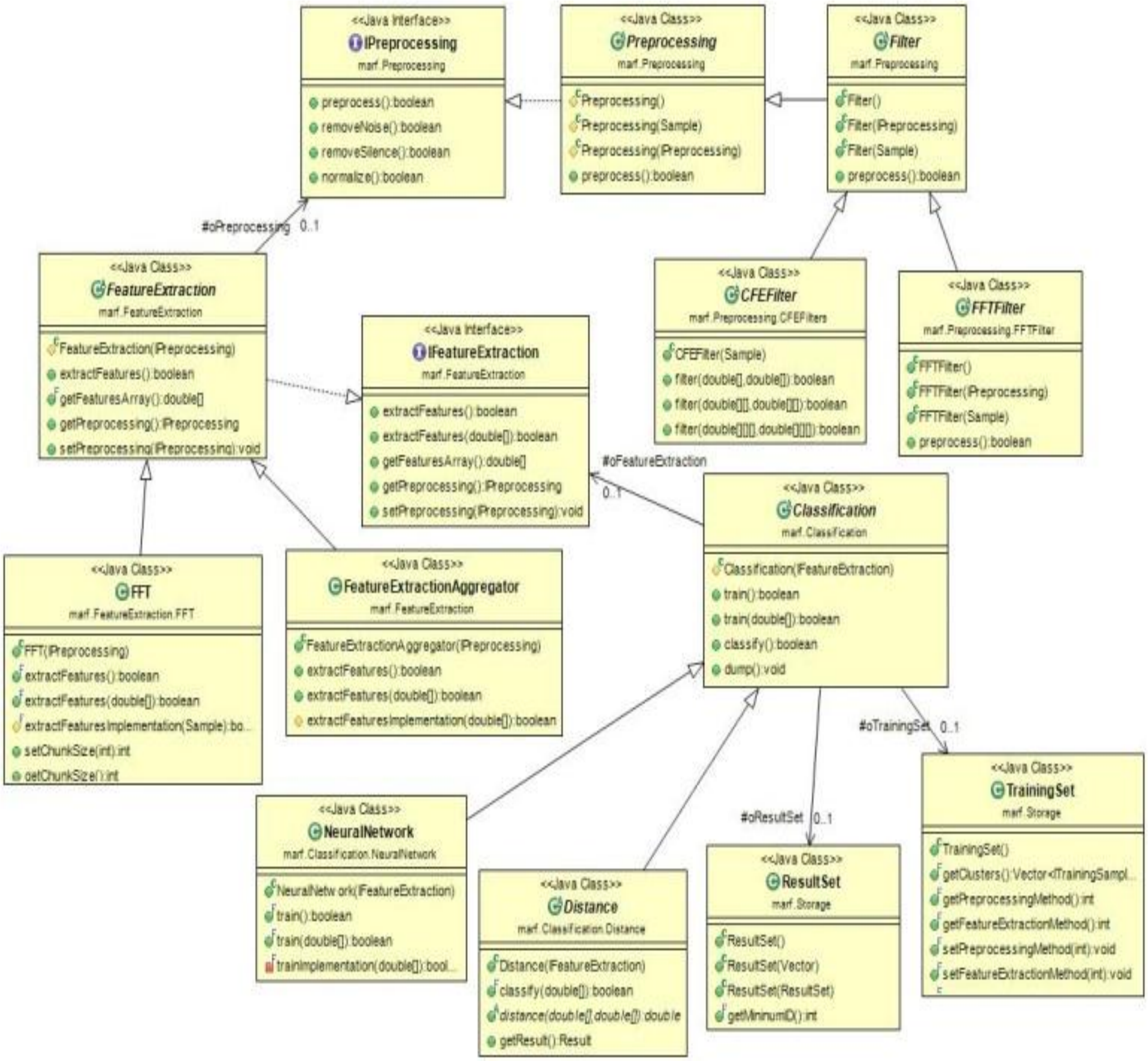

**Figure 10 - DMARF Class diagram 2**

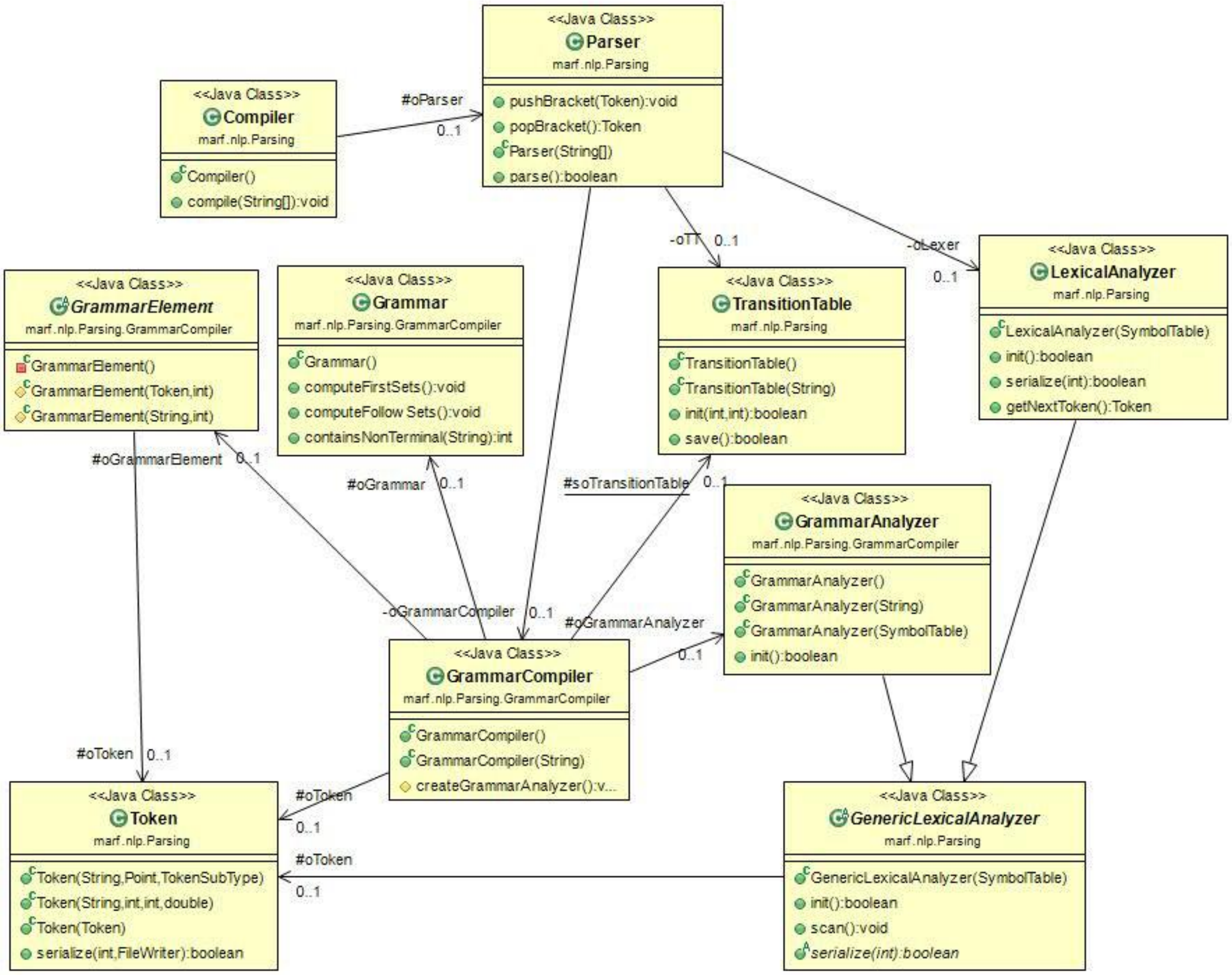

**Figure 11 - DMARF Class diagram 3**

| Conceptual Class | Actual Class | Description |
|---|---|---|
| Evidence sample | Sample.java | Evidence sample concept is specific to the application and takes as input text or audio just as sample.java does. |
| Sample Loader | SampleLoader.java | There are different types of sample loaders to load the samples. |
| Preprocessor | Preprocessing.java | Preprocessor does the initial filtering and noise removal. |
| Feature Extractor | FeatureExtraction.java | Specific features are extracted using filters |
| Classifier | Classification.java | Classifier is where the feature extracted samples are classified |

**Figure 12 - Conceptual classes vs Actual classes**

The Adaptor translator and forensic lucid expression concepts are not present in the actual system architecture because they are application specific concepts. Sample loader, preprocessor, Feature extractor and classifier concepts correspond to the core pipeline of the DMARF.

The concept called "DMARF system" is the composition of sample loaders, preprocessors, feature extractors and classifiers. Since the concept "DMARF system" represents the whole DMARF pipeline, there is no particular class in the actual architecture that maps to it. The mapping between conceptual and actual classes on comparing the concepts and the actual classes, we find no discrepancies except for the Adaptor translator concept and forensic lucid expression concept which are application specific. Adaptor translator is a program that encodes evidential samples into the Forensic lucid expression.

## 2) *GIPSY*

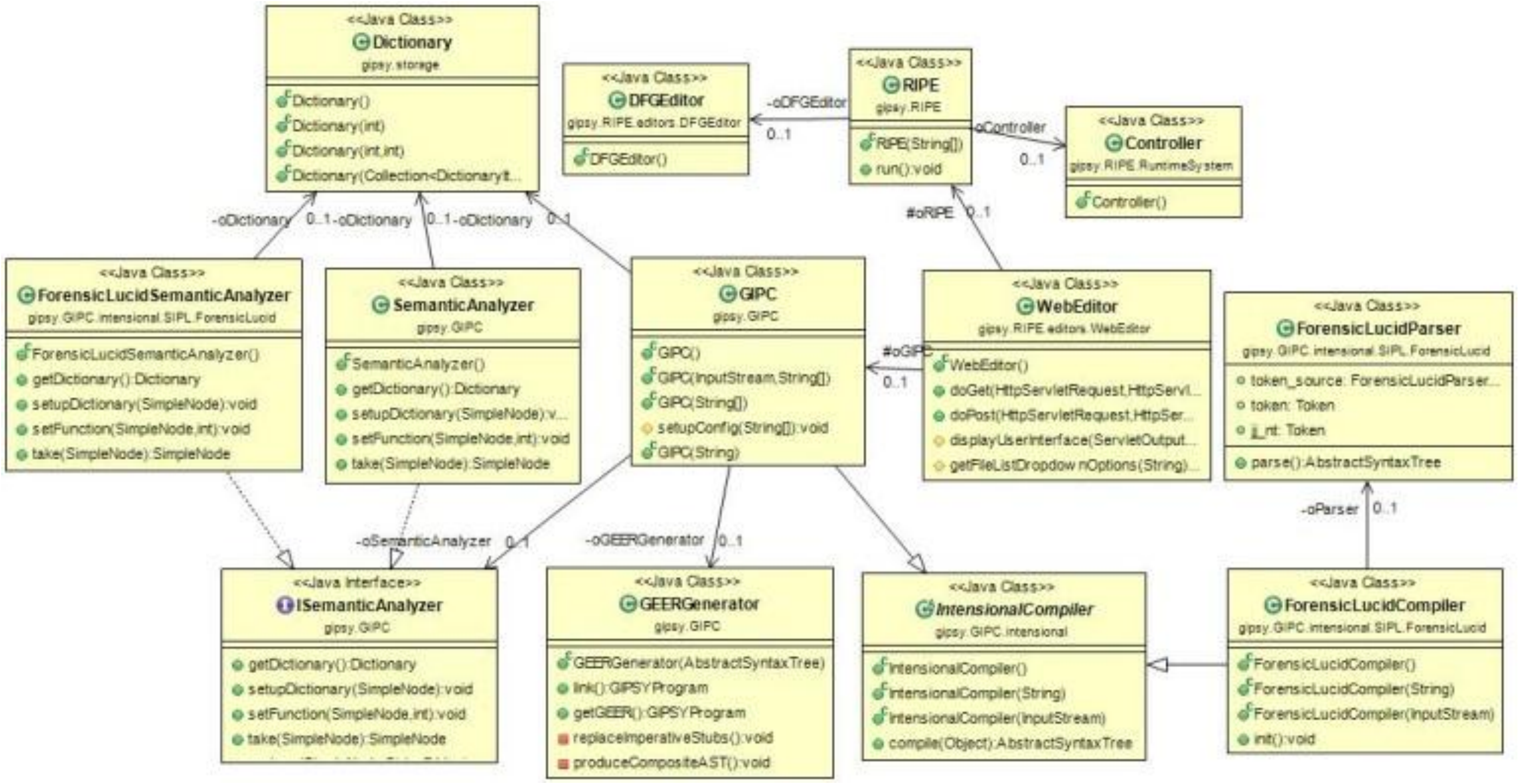

**Figure 14 - GIPSY Class diagram 1**

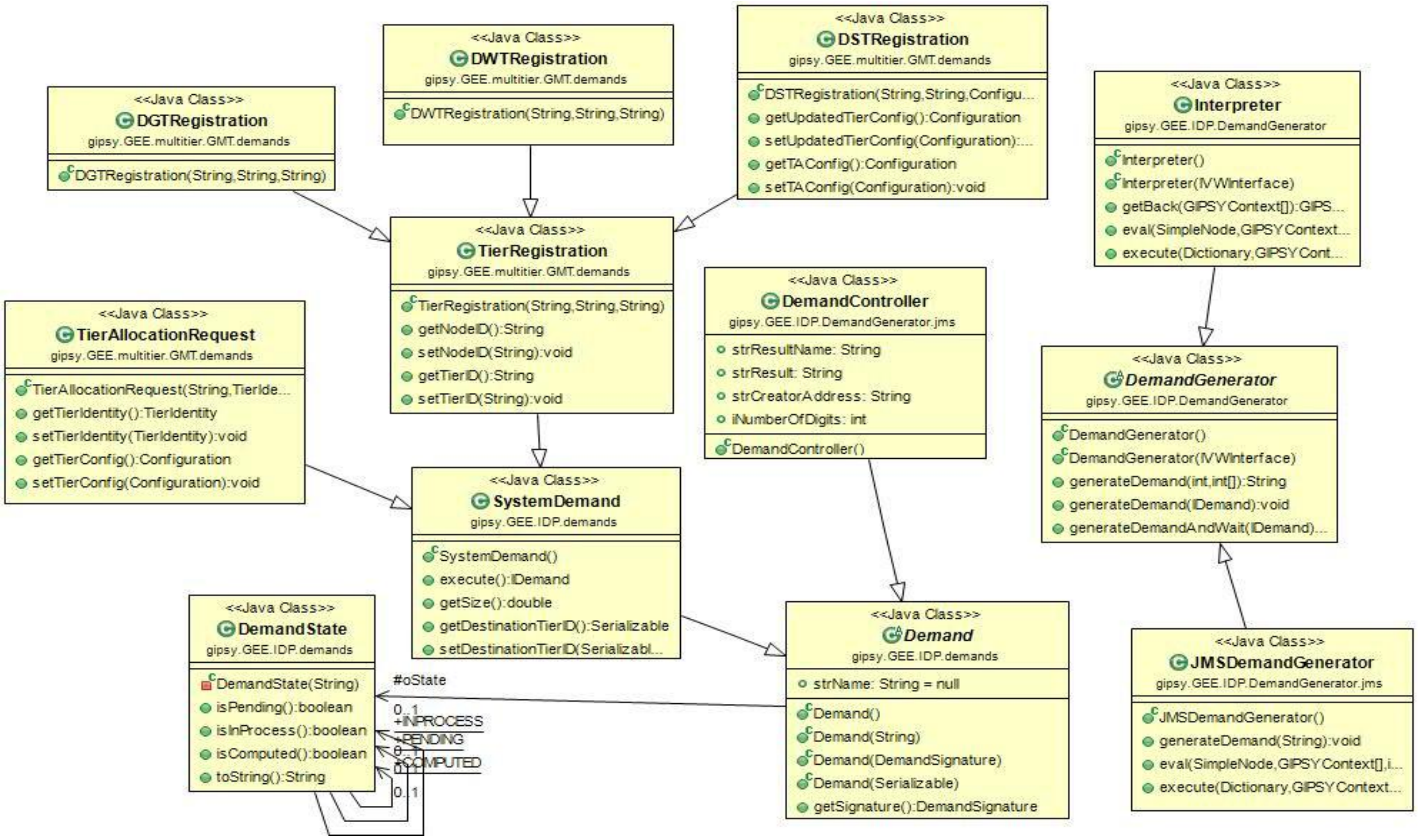

**Figure 13 - GIPSY Class Diagram 2**

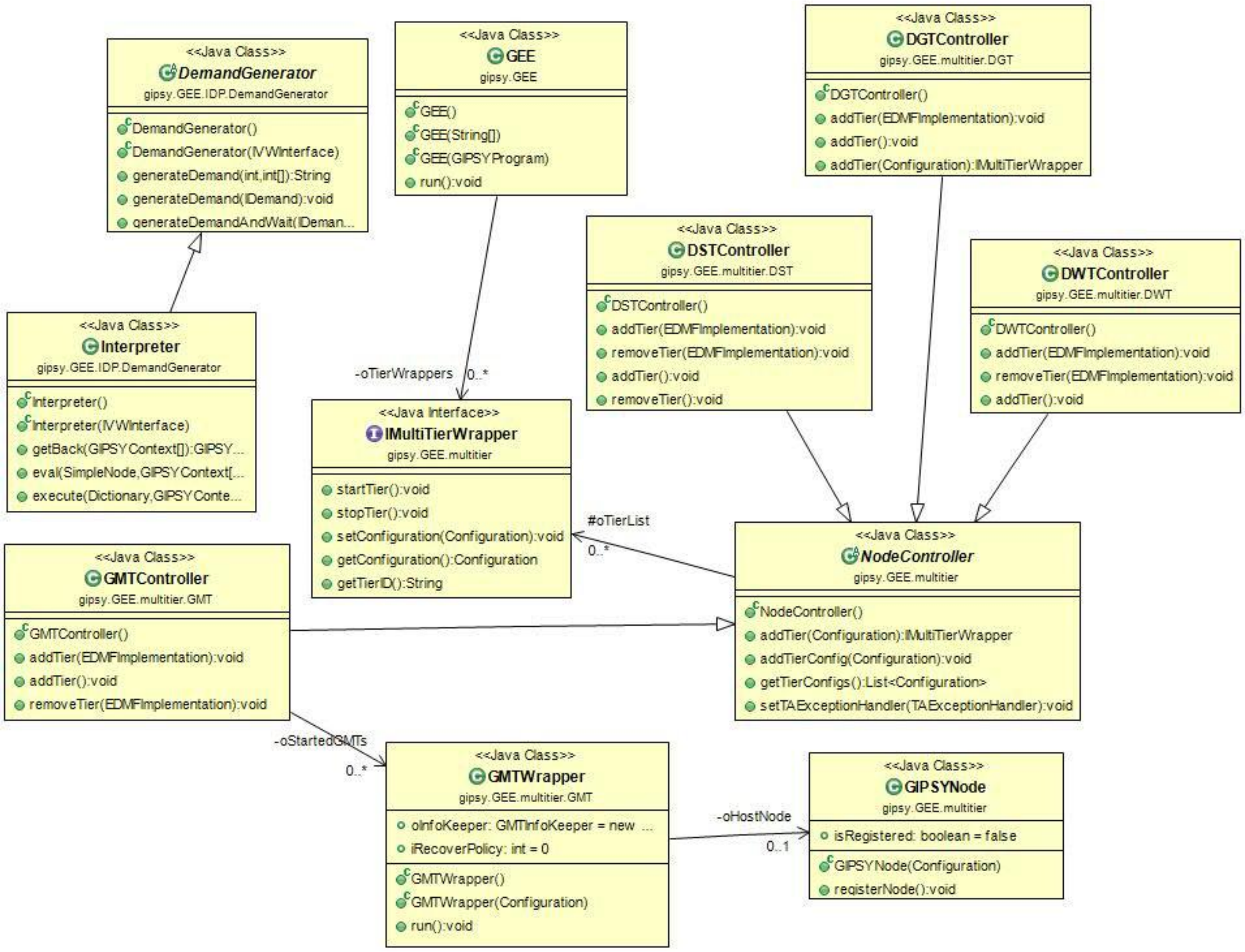

**Figure 15 - GIPSY Class Diagram 3**

| Conceptual Class | Actual Class | Description |
|---|---|---|
| Investigator Front End | WebEditor | Web Editor can serve as a front end for the police investigator. |
| RIPE | RIPE<br>WebEditor | Those actual classes together correspond to the domain concept called RIPE |
| GIPC | ForensicLucidCompiler<br>ForensicLucidParser<br>SemanticAnalyzer<br>Preprocessor<br>PreprocessorParser<br>ICompiler<br>GEERGenerator | Those actual classes together correspond to the domain concept named GIPC |
| GEE | IDemand<br>Demand<br>IntensionalDemand<br>DemandState<br>SystemDemand<br>GIPSYNode<br>NodeController<br>TierFactory<br>NodeController | These and other GEE classes shown in the UML Design Class diagram together correspond to the concept called GEE subsystem in the Domain Model. |

**Figure 16 Conceptual class and Actual Class**

We have a discrepancy between conceptual classes and design classes in the following points: 1. our conceptual class included whole the GEE subsystem of GIPSY as a single conceptual class because our Domain Model was more abstract than the actual architecture. The actual architecture of GEE subsystem is very big. Its key components are the tiers (DGT, DWT, GMT, DST) and demands, whose corresponding classes are shown on GIPSY design diagram. 2. GIPC and RIPE conceptual classes also have discrepancies with the actual architecture of the GIPC and RIPE components in the actual GIPSY system because our GIPC and RIPE domain concept were more abstract then the actual architecture.

Our GIPSY Conceptual classes differ from the design class in the following points. 1. Data Flow Graph Editor Conceptual class doesn't have the corresponding Design class. We have found a corresponding Java class DFGEditor.java in GIPSY package gipsy.RIPE.editors. DFGEditor, however that class is empty, which is why we didn't include it in the Design model. Apart from that, we have some conceptual classes that do not directly map to actual classes because they are specific to the application that our use case scenario describes. These include the following conceptual classes:

- Forensic Lucid Specification
- Encoder
- Encoded Evidence
- Transition Functions
- Crime Scene Report
- Investigator Front-end
- Text-based output
- GIPSY system

*Reverse engineering tool: ObjectAid UML Explorer*

We used ObjectAid UML Explorer for creating the class diagrams for DMARF and GIPSY case studies.

The ObjectAid UML Explorer is an agile and lightweight code visualization tool for the Eclipse IDE. It shows your Java source code and libraries in live UML class and sequence diagrams that automatically update as your code changes.

The class diagrams of the ObjectAid UML Explorer are based on the OMG's UML 2.0 specification. They can contain existing Java classes, interfaces, enumerations, annotations (collectively called classifiers henceforth in accordance with UML 2.0) as well as packages and package roots (i.e. JARs and source folders). Class diagrams only reflect the existing source code, which cannot be manipulated through the diagram. They are stored as XML files with the extension '.ucls'. The ObjectAid UML Explorer allows software developers to document and explore Java source code and libraries within the Eclipse IDE. It supports an agile approach to software development with seamless integration into the Eclipse IDE.

*Code snippet from two classes called 'Demand' and 'Demand Signature' from GIPSY*

Demand Class

```
package gipsy.GEE.IDP.demands;

public abstract class Demand extends FreeVector<Object>
implements IDemand
{
protected DemandSignature oSignature = null;

public Demand()
{
super();
this.oSignature = new DemandSignature();
this.oType = DemandType.PROCEDURAL;
this.oState = DemandState.PENDING;
this.oTimeLine = new TimeLine();

this.lAccessCounter = 0;
}


public Demand(DemandSignature poSignature)
{
this();
this.oSignature = poSignature;
}

public DemandSignature getSignature()
{
return this.oSignature;
```

```
}

public void setSignature(DemandSignature poSignatureID)
{
this.oSignature = poSignatureID;
}
public DemandSignature storeResult(Serializable poResult)
{
this.oWorkResult = poResult;
DemandSignature oSignature = this.oSignature;

if(poResult instanceof IDemand)
{
oSignature = ((IDemand)this.oWorkResult).getSignature();
}

return oSignature;
}

public synchronized boolean equals(Object poDemand)
{
if(poDemand != null && poDemand instanceof IDemand)
{
IDemand oOtherDemand = (IDemand)poDemand;

if
(
OtherDemand.getSignature().equals(this.oSignature) &&
OtherDemand.getContext().equals(this.oContextId) &&
OtherDemand.getType().equals(this.oType) &&
OtherDemand.getState().equals(this.oState)
)
{
return true;
}
}

return false;
}
}

DemandSignature

package gipsy.GEE.IDP.demands;

public class DemandSignature extends GIPSYSignature
{
```

```

public DemandSignature()
{
super();
this.oSignature = hashCode();
}


public DemandSignature(Serializable poSignature)
{
super(poSignature);
}
}
```

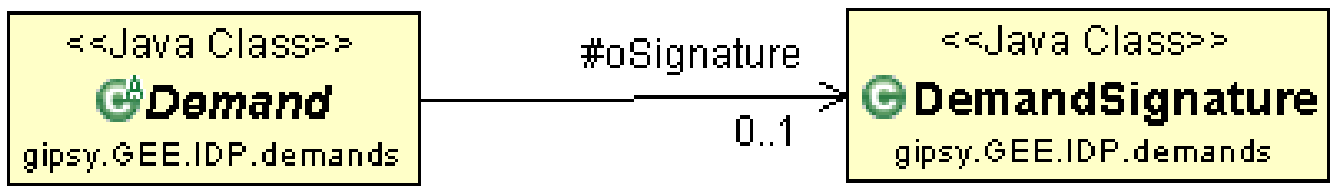

**Figure 17 - Class diagram for Demand and DemandSignature**

## V. METHODOLOGY

### *A. Refactoring*

#### *1) Identification of Code Smellsand System Level Refactorings*

*1.1 Overview*

We have employed various tools to sniff out codes smells in both DMARF and GIPSY, including JDeodorant [29] and Robusta [26] (as plug-ins to Eclipse IDE) and McCabe (as standalone tool) [28]. The following figure shows a summary of results obtained from these tools, intended to give the reader a general sense of the state of the systems (“Parser” classes excluded from analysis):

DMARF

| Code Smell | # of occurrences |
|---|---|
| God Class | 74 |
| Type Checking | 183 |
| Feature Envy | 23 |
| Long Method | 1190 |

GIPSY

| Code Smell | # of occurrences |
|---|---|
| God Class | 60 |
| Type Checking | 33 |
| Feature Envy | 218 |
| Long Method | 1085 |

JDeodorant

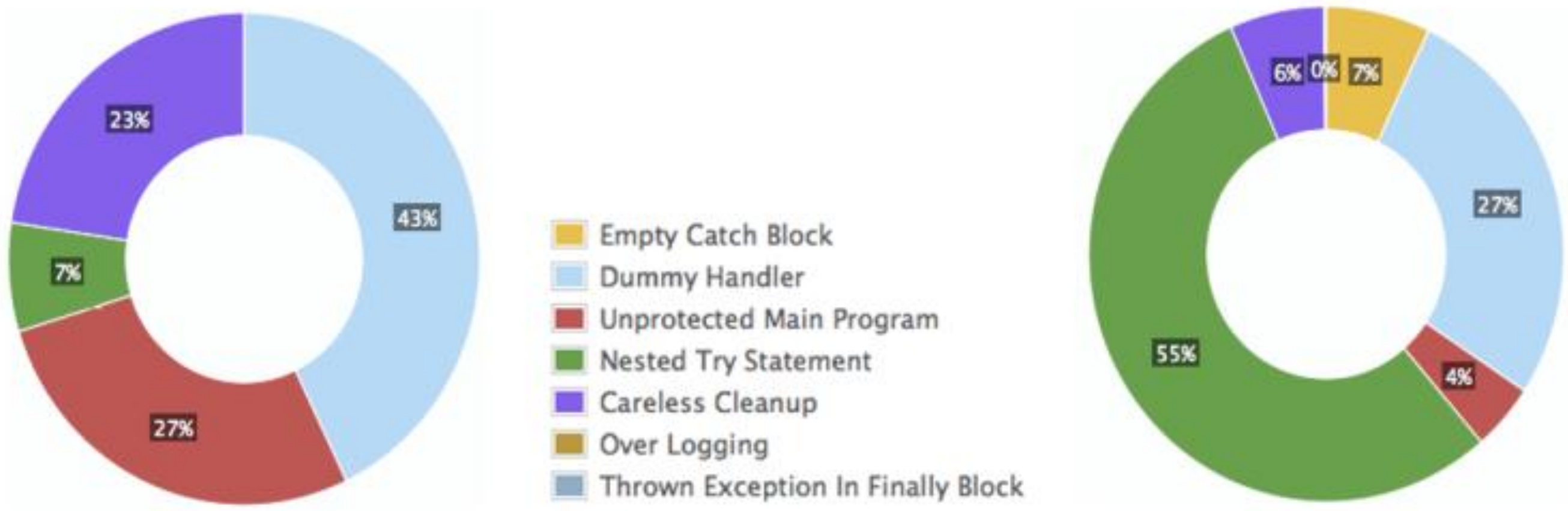

DMARF GIPSY

Robusta

| Code Smell | % Classes Exhibiting Code Smell | |
|---|---|---|
| | DMARF | GIPSY |
| Cyclomatic Complexity > 20 | 1.6% | 3.7% |
| Coupling Between Objects (CBO) > 2 | 5.0% | 0.0% |
| Weighted Methods per Class > 14 | 8.3% | 8.0% |
| Lack of Cohesion of Methods > 75% | 38% | 29.5% |

McCabe IQ

**Figure 18. Static analysis of DMARF and GIPSY. A high-level picture of the state of DMARF and GIPSY as hinted at by various refactoring and software measurement tools.**

In addition to the above code smells, we used logiscope[8] tool to find the classes with least maintainability in terms of number of out of bound metrics. The classes mentioned below exceeded the metrics' threshold values by a considerable level which has various impact on their quality characteristics.

Classes from DMARF with worst quality:

- **marf.Classification.NeuralNetwork.NeuralNetwork**
- **marf.Storage.StorageManager**

| Mnemonic | Metric Name | Min | Max | Class GIPSYGMTOperator | Class Demand |
|---|---|---|---|---|---|
| cl_comf | Class comment rate | 0.20 | +oo | 0.19 | 0.43 |
| cl_comm | Number of lines of comment | -oo | +oo | 160 | 179 |
| cl_data | Total number of attributes | 0 | 7 | 44 | 10 |
| cl_data_publ | Number of public attributes | 0 | 0 | 8 | 1 |
| cl_func | Total number of methods | 0 | 25 | 37 | 29 |
| cl_func_publ | Number of public methods | 0 | 15 | 19 | 29 |
| cl_line | Number of lines | -oo | +oo | 859 | 421 |
| cl_stat | Number of statements | 0 | 100 | 311 | 78 |
| cl_wmc | Weighted Methods per Class | 0 | 60 | 69 | 32 |
| cu_cdused | Number of direct used classes | 0 | 10 | 51 | 11 |
| cu_cdusers | Number of direct users classes | 0 | 5 | 2 | 13 |
| in_bases | Number of base classes | 0 | 3 | 6 | 5 |
| in_noc | Number of children | 0 | 3 | 0 | 7 |

**Figure 19 - Class level metrics for DMARF**

| Mnemonic | Metric Name | Min | Max | Class GIPSYGMTOperator | Class Demand |
|---|---|---|---|---|---|
| cl_comf | Class comment rate | 0.20 | +oo | 0.19 | 0.43 |
| cl_comm | Number of lines of comment | -oo | +oo | 160 | 179 |
| cl_data | Total number of attributes | 0 | 7 | 44 | 10 |
| cl_data_publ | Number of public attributes | 0 | 0 | 8 | 1 |
| cl_func | Total number of methods | 0 | 25 | 37 | 29 |
| cl_func_publ | Number of public methods | 0 | 15 | 19 | 29 |
| cl_line | Number of lines | -oo | +oo | 859 | 421 |
| cl_stat | Number of statements | 0 | 100 | 311 | 78 |
| cl_wmc | Weighted Methods per Class | 0 | 60 | 69 | 32 |
| cu_cdused | Number of direct used classes | 0 | 10 | 51 | 11 |
| cu_cdusers | Number of direct users classes | 0 | 5 | 2 | 13 |
| in_bases | Number of base classes | 0 | 3 | 6 | 5 |
| in_noc | Number of children | 0 | 3 | 0 | 7 |

**Figure 20 - Class level metrics for GIPSY**

High value of "Number of used classes" metric show that NeuralNetwork class more dependent on other classes which means the class is **highly coupled.** Higher value of "Number of Children" metric of StorageManager also implies that the class is **highly coupled.**
The "number of statements" metrics has gone out of bound in NeuralNetwork class which means that the **class does more work**. This makes it **complex and less maintainable**.

Classes from GIPSY with worst quality:

- **gipsy.RIPE.editors.RunTimeGraphEditor.ui.GIPSYGMTOperator**
- **gipsy.GEE.IDP.demands.Demand**

The table below shows the number of out of bound class level metrics (highlighted) for both the classes.

### *1.2 Definition of Code Smells*

***Type Checking***: this code smell typically arises when using conditional statements for to determine the run-time type of an object in an inheritance hierarchy (i.e. an object's run-time type maybe that of a child). However, such a conditional determination has been addressed by polymorphism, a fundamental aspect of OO programming paradigm. JDeodorant [8] is able to refactor this type of code smells using two approaches: 1. Replace Conditional with Polymorphism. 2. Replace Type Code with State/Strategy. The first approach is proposed when a piece of code identifies the type of subclass in conditional statement using ***instanceof*** operator and then casts the parent objects to one of its subclasses. The second approach is used when a conditional statement uses an attribute which represents a state to select desired subsequent execution flow.

***Long Method***: The longer the method the more likely it is to be doing more than one task, which could lead to violations of various OO principles including encapsulation, coherence, and low coupling. Furthermore, the longer the method the more difficult it becomes to understand what it does, or troubleshoot it. The refactoring suggested for this type of code smells is decomposition of the long method into smaller methods.

***Empty* Catch *block***: one of the main purposes of Java's enforcement of explicit exception handling is to provide useful and human-readable feedback that can help developers/maintainers identify the context and reasons behind an exception. Empty catch blocks effectively render the exception handling ineffective, since parties with no or little technical knowledge of the system (e.g. integration analysts) have no way of gaining hints into the causes of the exception.

***Feature Envy***: this code smell indicates that a method is more interested in another class rather than the class it is located in. JDeodorant [29] identifies Feature Envy by calculating the distances between methods and classes. These distances represent how many entities (methods and attributes) a method accesses in its versus those in other classes. A Feature Envy code smell is identified when a method accesses more entities of another class then entities of its own class. JDeodorant [29] proposes Move Method refactoring for such smell, resulting in the relocation of the method to the class it has more interest in.

### *1.3 Code Smells and Corresponding Refactorings in DMARF*

**Feature Envy**: using JDeodorant, we have identified 23 instances of this type of code smell in DMARF (see Figure 1 in the appendix). We present here a detailed example of the Feature Envy code smell where the affected classes are located in package marf.nlp.Parsing. The method addSymbol in class SymbolTable uses 3 times the method getLexeme and 2 times the method getPosition both from class Token, while using only 3 times one of the attributes (oSymTabEntries) from its own class. The code of this method is below.

```
/**
 * Adds a symbol token to the table. If the symbol is
 * already there, its additional location is recorded.
 * Else a new entry is created.
 * @param poToken symbol token to add
 * @return 0 on success
 */
public int addSymbol(Token poToken)
{
    if(this.oSymTabEntries.contains(poToken.getLexeme()))
    {
        SymTabEntry oSymTabEntry = (SymTabEntry)oSymTabEntries.get(poToken.getLexeme());
        oSymTabEntry.addLocation(poToken.getPosition());
    }
    else
    {
        SymTabEntry oSymTabEntry = new SymTabEntry(poToken);
        oSymTabEntry.addLocation(poToken.getPosition());
        this.oSymTabEntries.put(poToken.getLexeme(), oSymTabEntry);
    }

    return 0;
}
```

**Figure 21. Example of Feature Envy code smell before refactoring in SymbolTable class in DMARF.**

The following diagram depicts this Feature Envy code smell by showing which attributes and methods are accessed from the addSymbol method:

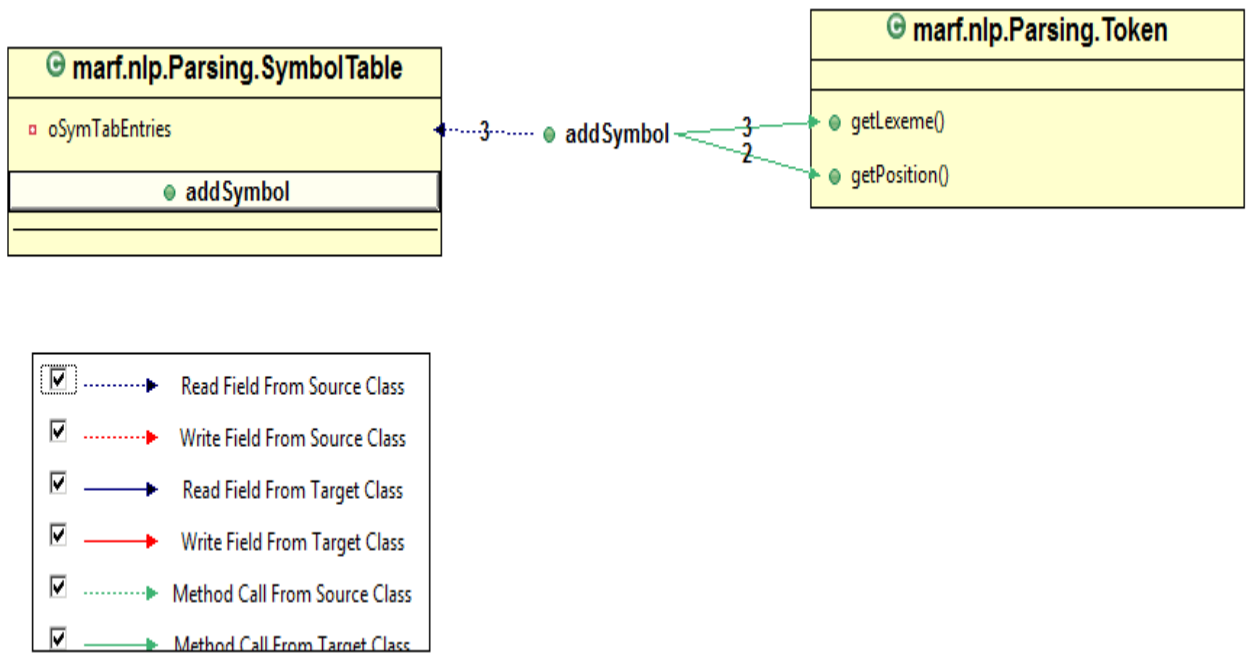

**Figure 22. Visualization of Feature Envy code smell in SymbolTable class in DMARF.**

Below UML Class diagram presents a subset of methods and attributes of the two mentioned classes before refactoring: SymbolTable and Token, both from marf.nlp.Parsing package. This UML diagram shows that the discussed classes have Dependency relation between themselves, and this is because the method addSymbol(Token) from SymbolTable class has parameter of type Token.

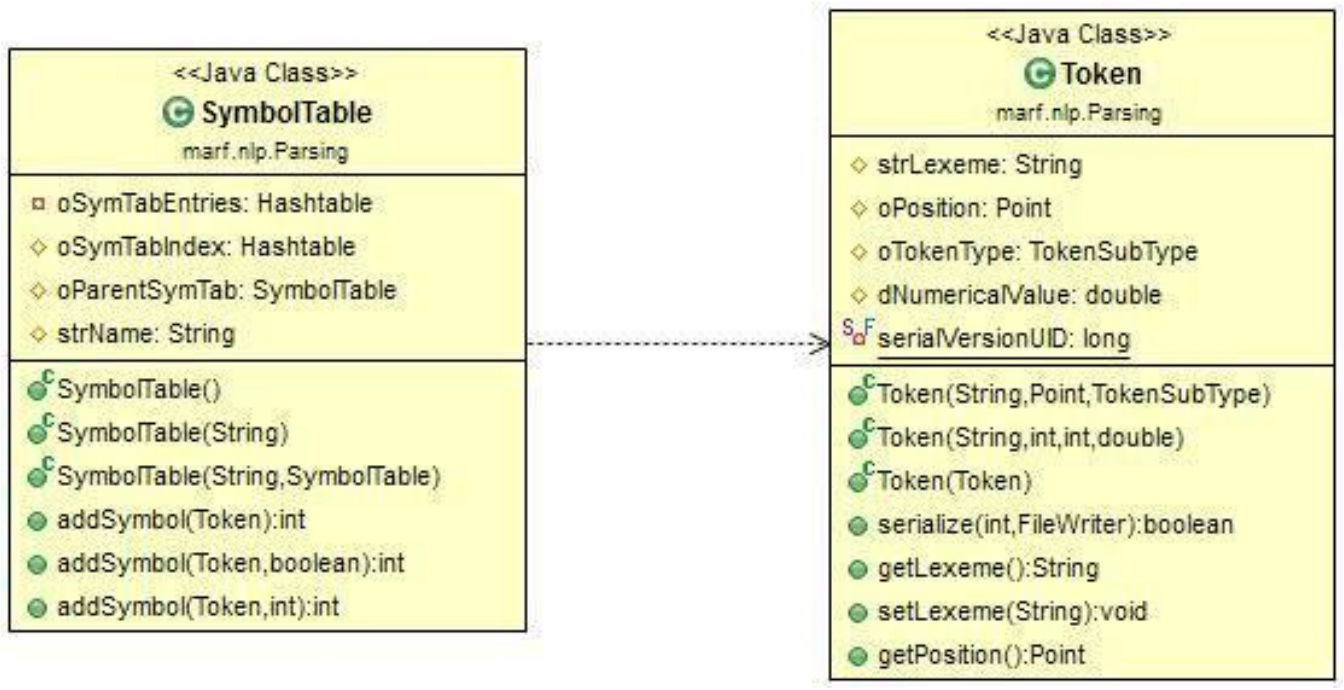

**Figure 23. UML Class Diagram of classes related to Feature Envy code smell in SymbolTable class in DMARF.**

To refactor this particular code smell, the method *int addSymbol(Token)* from SymbolTable class has to be moved to the Token class. This can be done as following: the method *addSymbol(Token)* in SymbolTable class has to become a delegate method for the moved method and the moved method has to be added to Token class. The moved method parameter will be of type Hashtable and the delegate method will be passing oSymTabEntries member from SymbolTable to the moved method. This change will eliminate the code smell because the moved method will only use the attributes and methods from Token class and will even not use any methods of its original SymbolTable class. Other code smells in DMARF of type Feature Envy can be fixed in similar fashion using Move Method Refactoring.

**Type Checking** code smells have been identified in DMARF using JDeodorant, which produced 183 such smells in total (see Figure 2 in the Appendix for a screenshot of example JDeodorant session). Here is a detailed example of Type Checking code smell in DMARF that can be fixed using

Replace Conditional with Polymorphism refactoring. The affected method is getComplexMatrix from the class complexMatrix located in marf.math package. Below is the code of this method.

```
public static ComplexMatrix getComplexMatrix(Matrix poMatrix)
{
    ComplexMatrix oMatrix;

    if(poMatrix instanceof ComplexMatrix)
    {
        oMatrix = (ComplexMatrix)poMatrix;
    }
    else
    {
        oMatrix = new ComplexMatrix(poMatrix);
    }

    return oMatrix;
}
```

**Figure 24. Example of method with Type Checking code smell in ComplexMatrix class in DMARF before refactoring.**

The below Class Diagram shows a subset of attributes and methods of the classes which are relevant to this code smell and to the proposed refactoring prior to refactoring. The method with code smell is getComplexMatrix from ComplexMatrix class. ComplexMatrix extends Matrix. Vector also extends Matrix.

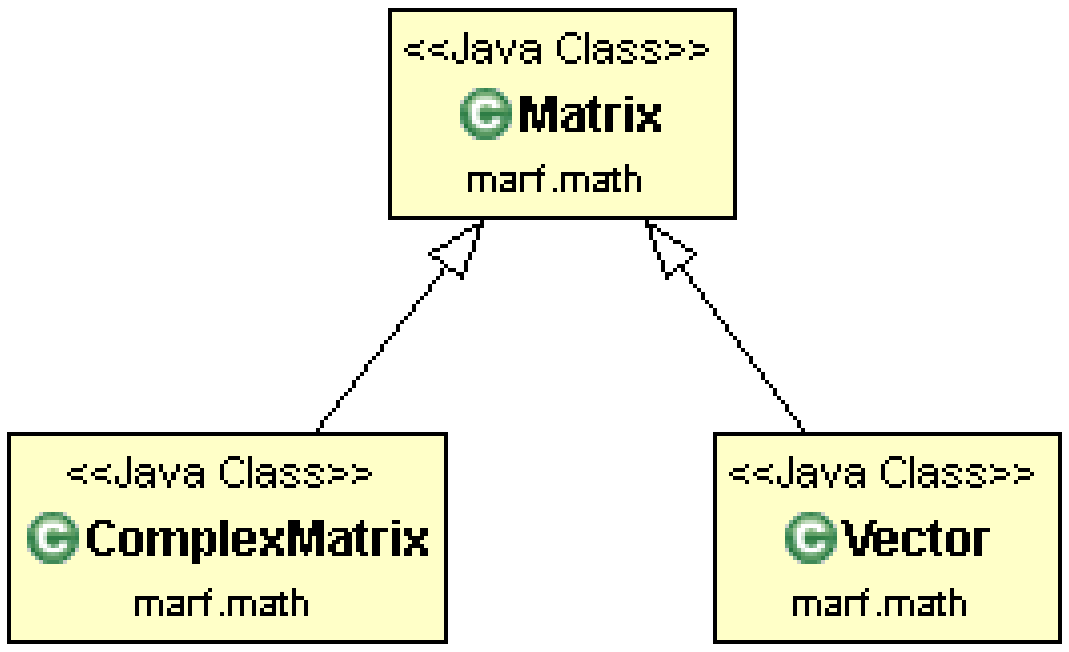

**Figure 25. UML Class Diagram of classes related to Type Checking code smell in ComplexMatrix class in DMARF.**

This code smell can be fixed in following steps: 1. Add method getComplexMatrix to Matrix class. The method should construct a new ComplexMatrix object initialized with the current Matrix object and then return the created ComplexMatrix object. 2. Override getComplexMatrix method in ComplexMatrix class which will return *this* (the calling object) as output. 3. Replace conditional statement in the static getComplexMatrix method in ComplexMatrix class by calling the new non-static getComplexMatrix method. This refactoring corrects the problem by introducing polymorphism instead of checking the type of subclass and emulating dynamic dispatch at runtime, which is in line with object oriented design principles. Other Type Checking smells in DMARF related to checking type of subclass at runtime can also be fixed using same approach.

**Long Method:** there are several instances of long method code smell in DMARF code. At the package level, we are going to analyze the code smell and describe a possible refactoring for marf.math package (see Figure 3 in the appendix for a screenshot of the Long Method identification in JDeodorant). JDeodorant tool finds a long method in algorithm.java file. The method is doFFT in the class FFT and the refactoring suggested is the Extract method. Below is the screenshot of the method which might be decomposed into smaller methods in order to resolve the Long Method code smell. The straight line indicates the place where the method might be decomposed into a sub method.

```
        padOutputReal[t] = padInputReal[i];
        padOutputImag[t] = padInputImag[i];
    }
_______________________________________________________

    // put it all back together (Danielson-Lanczos butterfly)
    int mmax = 2, istep, j, m;                              // counters
    double theta, wtemp, wpr, wr, wpi, wi, tempr, tempi;    // trigonometric recurrences

    n = iLength * 2;

    while(mmax < n)
    {
        istep = mmax * 2;
        theta = (piDirection * 2 * Math.PI) / mmax;
        wtemp = Math.sin(0.5 * theta);
        wpr   = -2.0 * wtemp * wtemp;
        wpi   = Math.sin(theta);
        wr    = 1.0;
        wi    = 0.0;
```

**Figure 26. Part of method of FFT class in DMARF which has Long Method code smell.**

The possible refactoring is decomposing the method into two so that the Danielson-Lanczos butterfly process would be done in a separate method (in the new method that will be created) and the return parameter would be changed accordingly. This refactroring would decompose the long method into two simpler methods which would improve comprehensibility and maintainability of the code and this is in line with object oriented design principles.

However, since this method is an implementation of a complex algorithm, we do not recommend implementing this refactoring, because if we split the method into two there would be additional memory operations associated with calling of another method with many parameters, and that would increase stack operations at runtime and hurt performance.

Below is the class diagram with a subset of attributes and methods of the FFT class and its close neighbors. The method doFFT is in the class FFT and its associations with other classes are shown by the class diagram below.

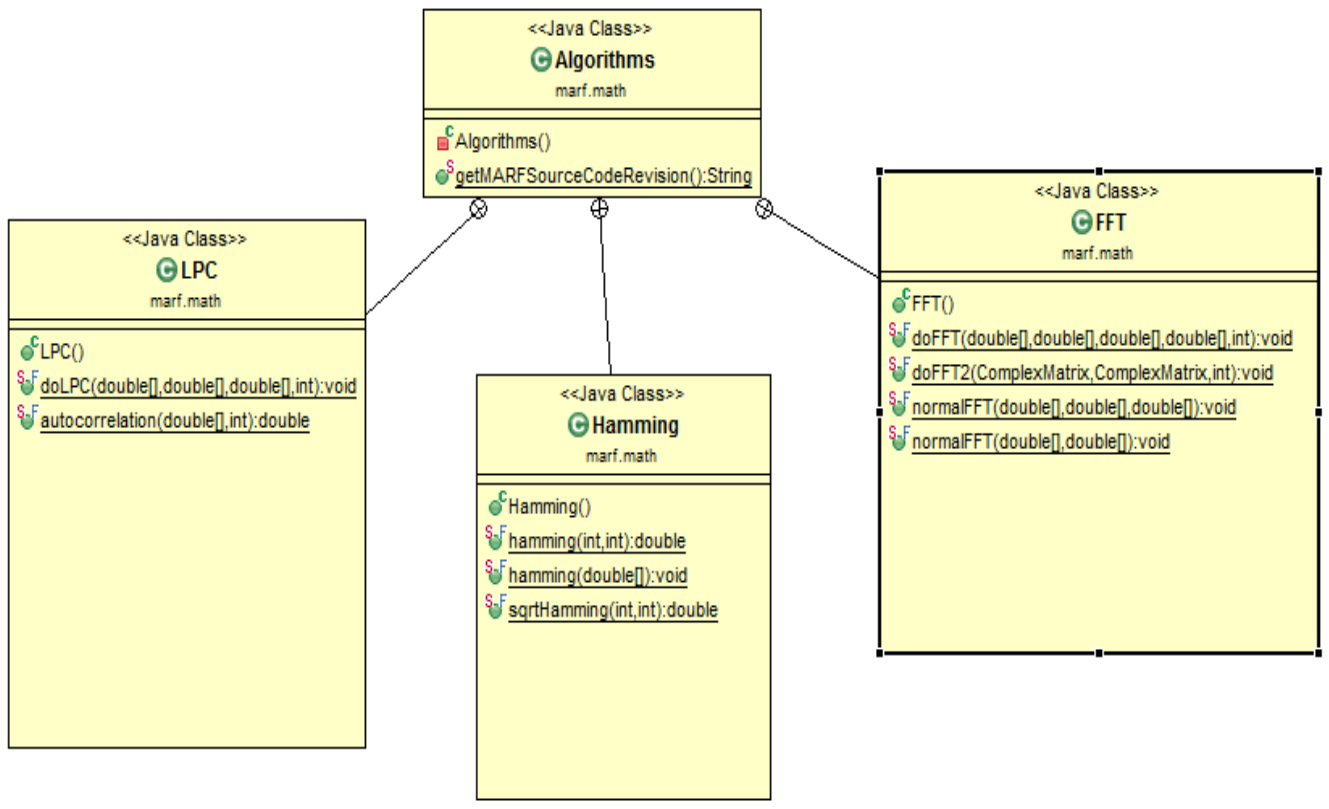

**Figure 27. UML Class Diagram of FFT class and its close neighbors in DMARF. FFT class's method doFFT has Long Method code smell.**

### *1.4 Code Smells and Corresponding Refactorings in GIPSY*

**1.4.1 Type Checking** code smell was identified in JINITA class located in package gipsy.GEE.IDP.DemandGenerator.jini.rmi. In the writeValue() method of JINITA class, the developer is using a conditional block to branch depending on what the run-time type of a certain IDemand interface object (passed as parameter to the writeValue()) happens to be. In JINITA class's writeValue() method, an IDemand object's run-time subtype needs to be determined as to which subtype it is: one of Demand (implementor of IDemand interface) subtypes: SystemDemand, ProceduralDemand, IntensionalDemand, or ResourceDemand. This is shown in the figure below:

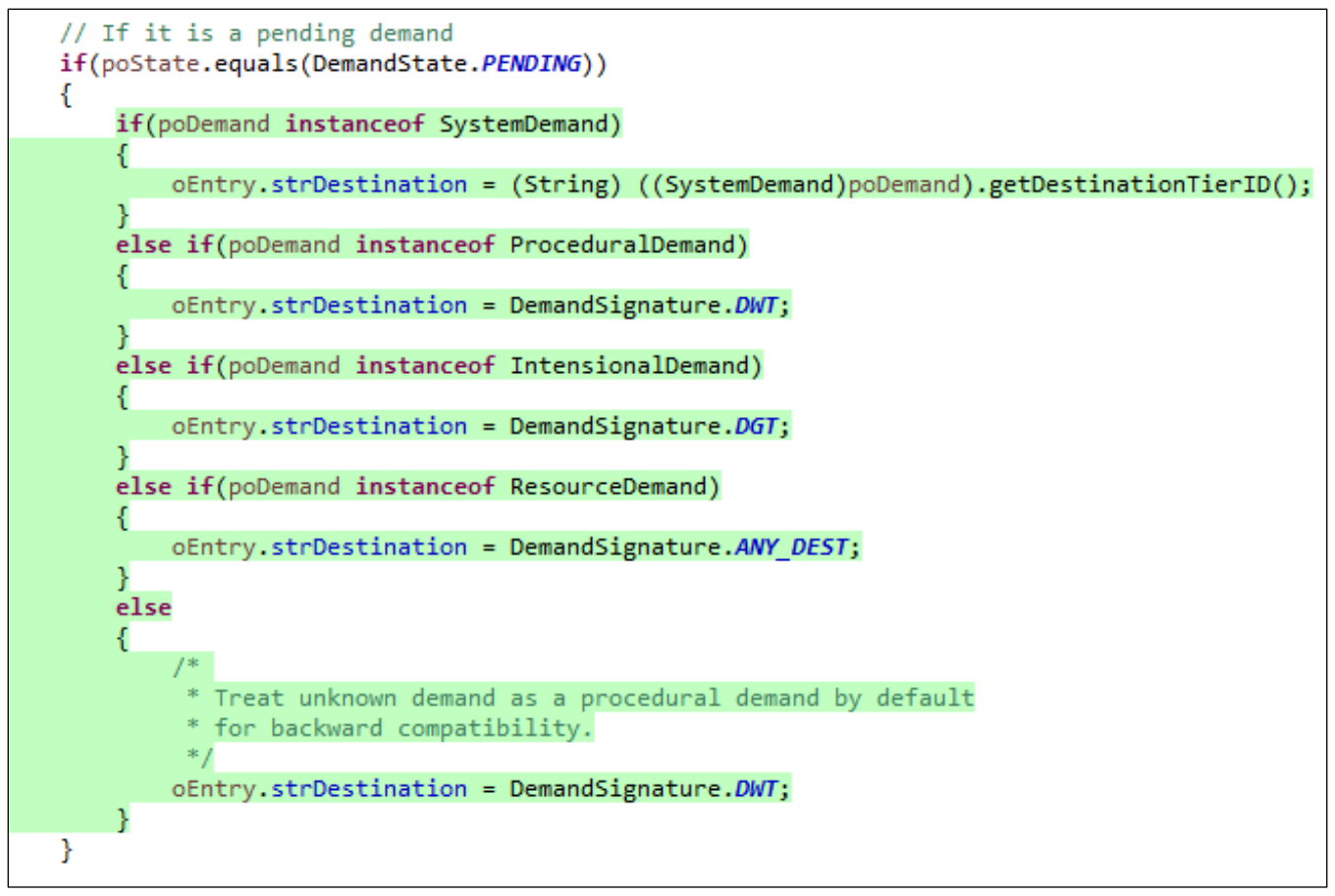

```
// If it is a pending demand
if(poState.equals(DemandState.PENDING))
{
    if(poDemand instanceof SystemDemand)
    {
        oEntry.strDestination = (String) ((SystemDemand)poDemand).getDestinationTierID();
    }
    else if(poDemand instanceof ProceduralDemand)
    {
        oEntry.strDestination = DemandSignature.DWT;
    }
    else if(poDemand instanceof IntensionalDemand)
    {
        oEntry.strDestination = DemandSignature.DGT;
    }
    else if(poDemand instanceof ResourceDemand)
    {
        oEntry.strDestination = DemandSignature.ANY_DEST;
    }
    else
    {
        /*
         * Treat unknown demand as a procedural demand by default
         * for backward compatibility.
         */
        oEntry.strDestination = DemandSignature.DWT;
    }
}
```

**Figure 28. A snapshot of implicated code in the identified Type checking code smell of class JINITA class. The type of a parameter object (poDemand) is being conditionally checked using instance of primitive operator to decide the appropriate next-step logic.**

The relationship between the involved classes is shown in the UML class diagram below, showing a *dependency* relationship betwen JINITA and IDemand (and its descendants):

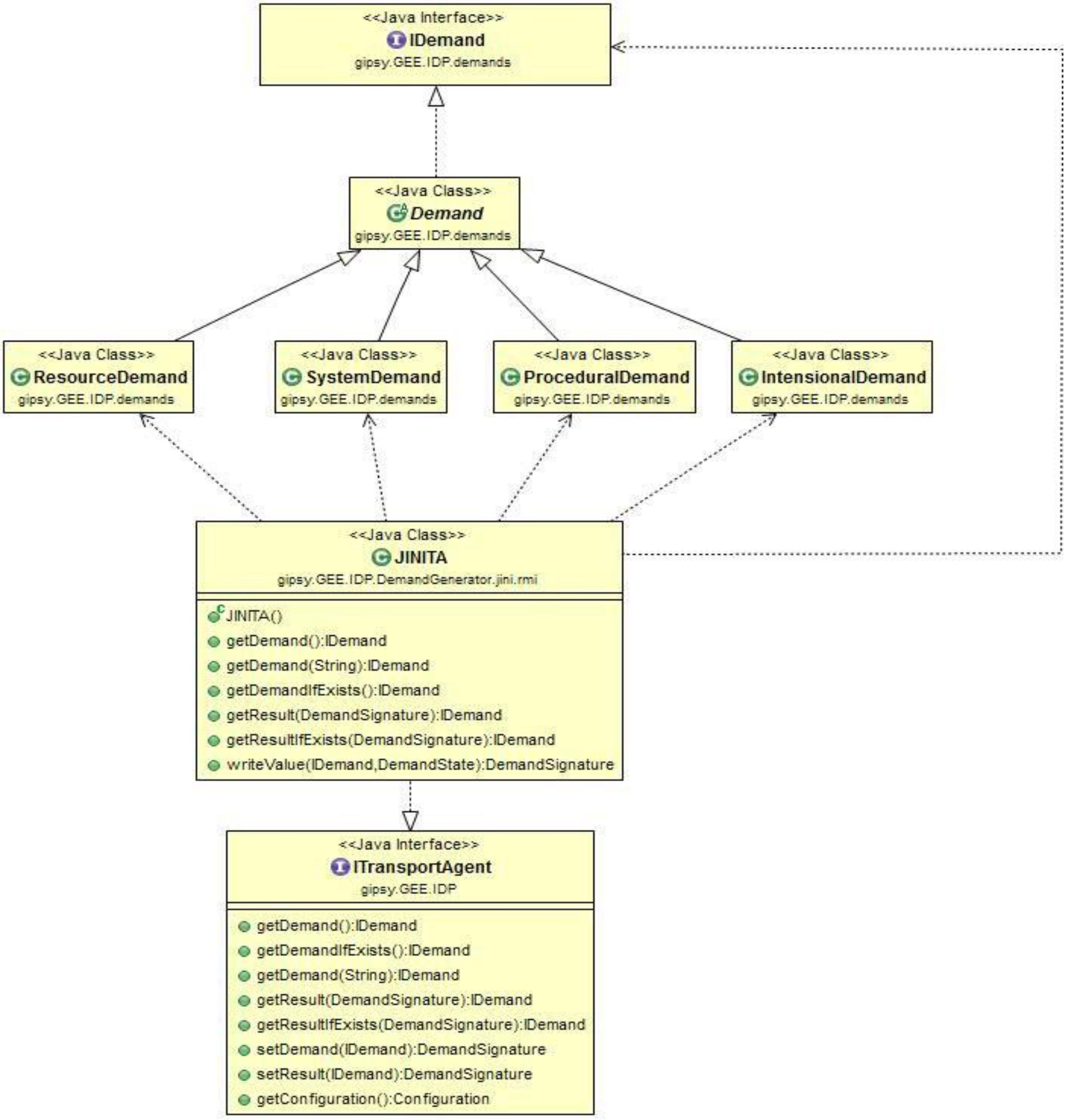

**Figure 29. Class diagram of relevant classes to the Type Checking code smell identified in JINITA class. The relationships depicted are before refactoring.**

Towards refactoring of this code smell, the conditional block is replaced with polymorphic method call, such that the conditional block is replaced with a call to IDemand's writeValue()method, as shown in the figure below:

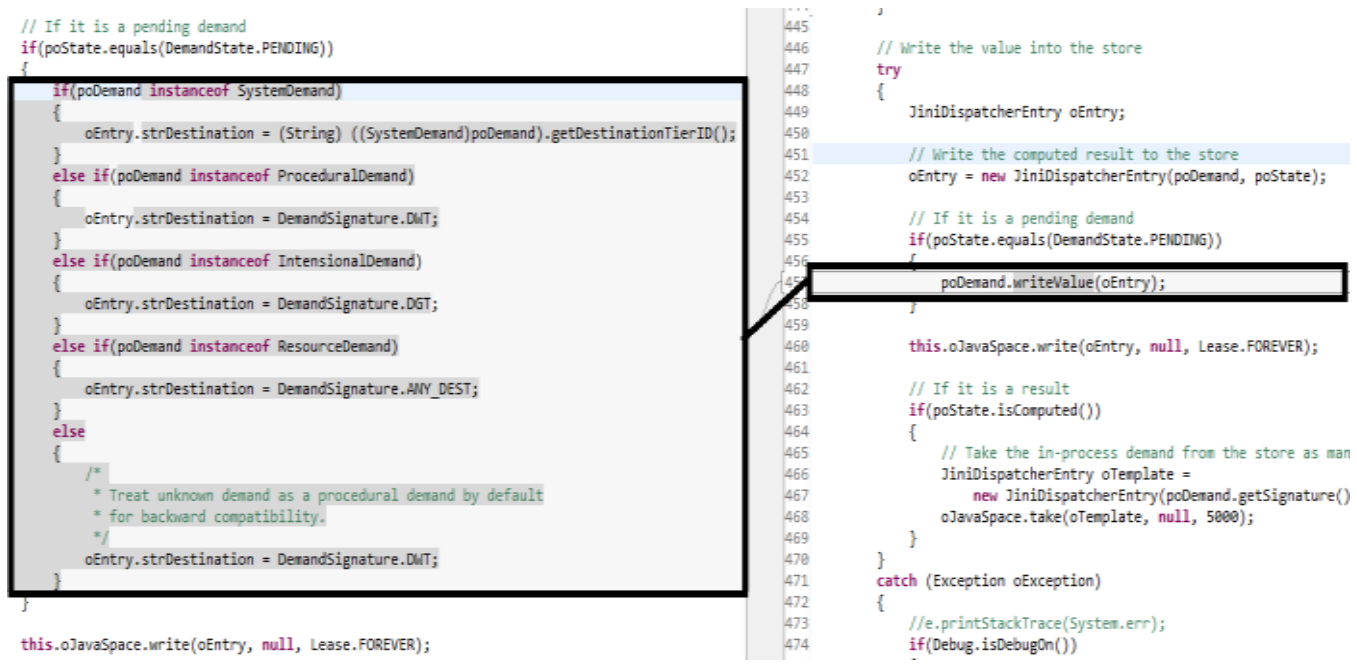

**Figure 30. : Refactoring of Type Checking code smell: conditional type checking code is replaced with a simple polymorphic call to a writeValue() method of IDemand interface (and its concrete implementations).**

Depending on the run-time subtype of the IDemand object, the correct method will polymorphically be called. Hence, the IDemand interface now declares a writeValue()abstract method, thereby requiring each of its implementer, Demand class, to implement it. Further, each of Demand 's children (SystemDemand,ProceduralDemand, IntensionalDemand, ResourceDemand) implement their own writeValue()method containing the logic that was originally in its corresponding conditional block before refactoring, as the figure below shows:

| Class | Refactoring |
|---|---|
| IDemand (interface) | `public abstract void writeValue(JiniDispatcherEntry oEntry);` |
| Demand (implements IDemand) | `public void writeValue(JiniDispatcherEntry oEntry) { oEntry.strDestination = (String) this.getDestinationTierID(); }` |
| ResourceDemand (extends Demand) | `public void writeValue(JiniDispatcherEntry oEntry) { oEntry.strDestination = DemandSignature.ANY_DEST; }` |
| IntensionalDemand (extends Demand) | `public void writeValue(JiniDispatcherEntry oEntry) { oEntry.strDestination = DemandSignature.DGT; }` |
| ProceduralDemand (extends Demand) | `public void writeValue(JiniDispatcherEntry oEntry) { oEntry.strDestination = DemandSignature.DWT; }` |
| SystemDemand (extends Demand) | `public void writeValue(JiniDispatcherEntry oEntry) { oEntry.strDestination = (String) this.getDestinationTierID(); }` |
| | |

**Figure 31 - Snapshots of consequential refactorings as a result of using polymorphism over conditional type checking. IDemand interface declares a writeValue() method, thus requiring all its descendants to provide concrete implementations. The appropriate method**

The following UML shows the relationship between classes involved **after** the refactoring of conditional block with polymorphism (notice that now JINITA has an established relationship with IDemand directly and its descendants indirectly):

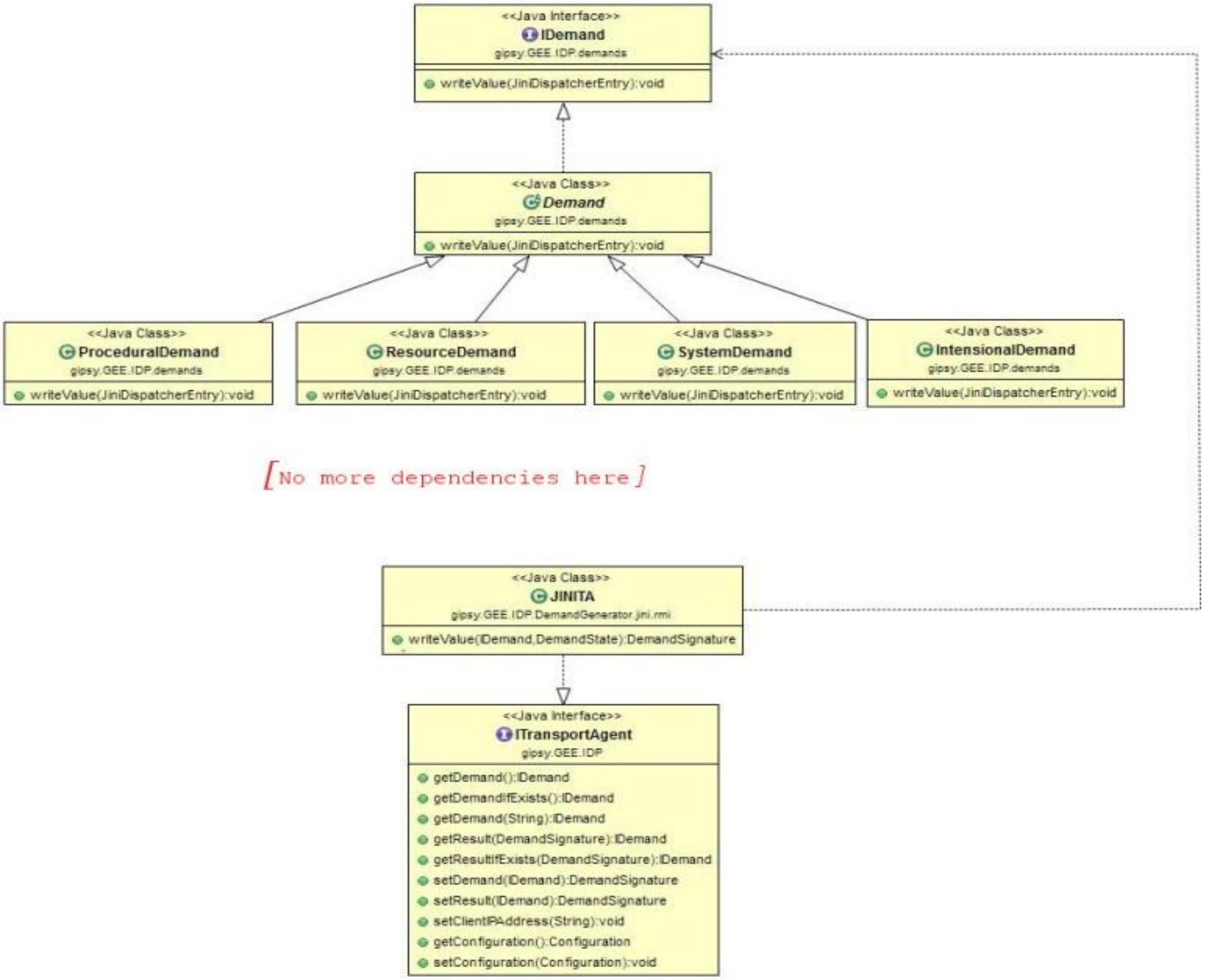

**Figure 32. Class diagram of relevant classes to the Type Checking code smell identified in JINITA class. The relationships depicted are after refactoring.**

### 1.4.2 Empty catch() block code smell: using Robusta plugin in Eclipse, we have identified 40 empty catch() blocks.

Ideally each catch() block must provide useful information about what the exception being caught. To refactor such an issue, we introduce a call to printStackTrace() method, which is part of Java's java.lang.Throwable class. The augmentation or replacement of System.err message with information relevant to the context of each class is beyond the scope of this project, as it requires detailed understanding of the purpose and logic of each of the 40 blocks.

### 1.4.3 Feature Envy Code smell in GIPSY

In this section we describe an example of Feature Envy code smell in GIPSY. This smell exists in LinkNode method located in LucidCodeGenerator class which is found in gipsy.GIPS.DFG.DFGAnalyzer package. This method uses more attributes and methods of LucidNodeItem class , more so than from its own class. The original code before refactoring of LinkNode is following:

```
public void linkNode(LucidNodeItem lcnode, LucidNodeItem lcparent, int nchild) {
  NodeDict.put(lcnode.ID, lcnode);
  lcnode.previous=lcparent;
  if(lcparent.ht.containsKey(itos(nchild))) {
    lcparent.ht.remove(itos(nchild));
  }
  lcparent.ht.put(itos(nchild), lcnode);
}
```

**Figure 33. Example of a method with Feature Envy code smell before refactoring which exists in LucidCodeGenerator class in GIPSY.**

This code smell is visualized below showing the methods and attributes of LucidCodeGenerator and LucidNodeItem classes that are accessed by LinkNode method. From this diagram we see that LinkNode method is more interested in LucidNodeItem class.

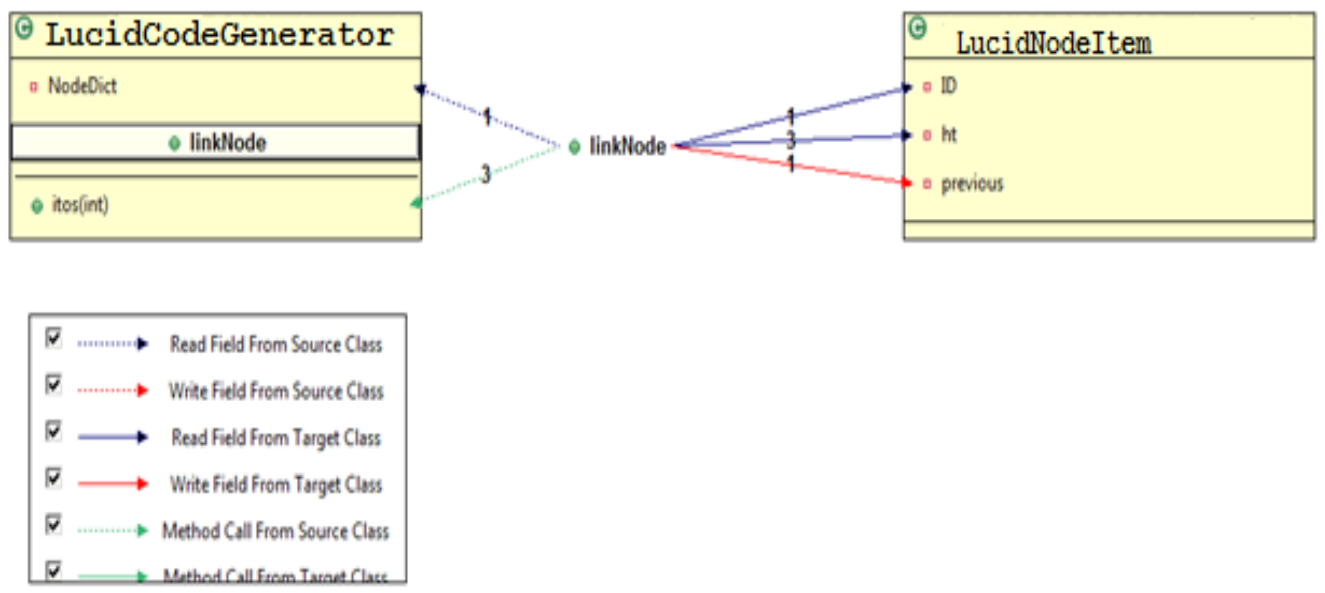

**Figure 34. Visualization of the sample Feature Envy code smell in GIPSY.**

This code smell can be refactored using Move Method refactoring. More concretely, the linkNode method should be moved to LinkNodeItem class and the LucidCodeGenerator class should be adjusted to use linkNode method from its new class, thereby resolving this code smell and lowering coupling and increasing cohesion. Below are more detailed steps for this refactoring:

1. Add linkNode method to LinkNode class:

```
public void linkNode(LucidNodeItem lcparent, int nchild, Hashtable NodeDict,LucidCodeGenerator lucidCodeGenerator)
{
		NodeDict.put(this.ID, this);
		this.previous = lcparent;
		if(lcparent.ht.containsKey(lucidCodeGenerator.itos(nchild))) {
		lcparent.ht.remove(lucidCodeGenerator.itos(nchild));
		}
		lcparent.ht.put(lucidCodeGenerator.itos(nchild), this);
}
```

2. Remove linkNode method from LucidCodeGenerator class.
3. linkNode method is used in several places in LucidCodeGenerator class. We need to replace the invocation of this method everywhere it is used.
   In method genAST()there is one invocation of linkNode method, in method genAST(LucidNodeItem) there are 6 invocations, in method getPoint(LucidNodeItem) there are 3 invocations and in method gendim(LucidNodeItem) there are also 3.

   Here is a sample replacement of the invocation of the moved method in genAST() method:

   Before refactoring:
   `linkNode(current2, LCnode, 1);`

   After refactoring:
   `current2.linkNode(LCnode, 1, NodeDict, this);`

The private method itos(int num) in LucidCodeGenerator class is used by linkNode method. Since linkNode method was moved to LucidNodeItem class, we need to change the access modifier of itos method to public to make it accessible from LucidNodeItem class.

## 2) *Specific Refactorings that You Will Implementin PM4*

### 2.1 *DMARF*

#### 2.1.1 *Type Checking code in getComplexMatrix method of complexMatrix class:*

This refactoring is done in marf.math package.
For DMARF, the first refactoring that we suggest to implement is the refactoring for Type Checking code smell found in the static getComplexMatrix method in complexMatrix class located in marf.math package. The exact steps of this refactoring are described in detail in previous section.
DMARF already has a Test app for marf.math package, its Eclipse project name is TestMath. All the tests of TestMath project are found in MathTestApp.java file and are written in the main method of MathTestApp class. However, the existing tests don’t cover directly the getComplexMatrix method and new tests have to be developed to test it. To unit test the getComplexMatrix method in isolation from other tests of the Math Test app, it is better to create a new junit package marf.junit.Math in DMARF project and to write a new java file with junit tests for the getComplexMatrix method.

The following test cases will be developed using junit for the static getComplexMatrix method from ComplexMatrix class:

Test Case 1:
1. Creation and filling of a Matrix object of a predefined size.
2. Acquisition of ComplexMatrix object using the static getComplexMatrix method of ComplexMatrix class executed against the Matrix object.
3. Verification that the structure and contents of the ComplexMatrix objects are in line with the source Matrix object.

Test Case 2:

1. Creation and filling of a Vector object of a predefined size.
2. Acquisition of ComplexMatrix object using the static getComplexMatrix method of ComplexMatrix class executed against the Vector object.
3. Verification that the structure and contents of the ComplexMatrix objects are in line with the source Vector object.

Test Case 3:
1.Creation and random filling of a ComplexMatrix object.
2. Acquisition of ComplexMatrix object using the static getComplexMatrix method of ComplexMatrix class executed against the ComplexMatrix object.
3. Verification that contents of the original ComplexMatrix and the new ComplexMatrix objects are same.

Below are only those parts from the source code (before refactoring) which will be involved in refactoring. The conditional statement which has the Type Checking smell is highlighted in yellow.

```
public class Matrix
implements Cloneable, Serializable
{…}

public class Vector
extends Matrix
{…}
public class ComplexMatrix
extends Matrix
{
        public static ComplexMatrix getComplexMatrix(Matrix poMatrix)
        {
                ComplexMatrix oMatrix;

                if(poMatrix instanceof ComplexMatrix)
                {
                        oMatrix = (ComplexMatrix)poMatrix;
                }
                else
                {
                        oMatrix = new ComplexMatrix(poMatrix);
                }

                return oMatrix;
        } …
}
```

*Feature Envy code smell refactoring:*

The second DMARF refactoring that we suggest implementing in PM4 is the Feature Envy refactoring which has been described in previous section, as part of which the method with signature int addSymbol(Token) from SymbolTable class should be moved to Token class.
We checked presence of test cases for the SymbolTable and Token classes from marf.nlp.Parsing package. There is a Test app ProbabalisticParsingApp, which uses some classes from marf.nlp.Parsing package where the classes that we are going to refactor are located, but that test app doesn't test directly our classes of interest. Therefore, we believe the best approach to unit test the addSymbol method is by developing and using new JUnit tests. We will create a new package in DMARF called marf.junit.nlp.Parsing and we will add a new java file which will implement the following test cases:

1. Addition of a new symbol *S1* to the Symbol Table using addSymbol method from SymbolTable class and verification of a successful return status of the method.
2. Retrieval of an entry from Symbol Table according to the lexeme of the token *S1* which had been inserted in step 1 and verification that the entry is not null.
3. Verification that the location of the retrieved entry corresponds to the location of the token which had been inserted into the Symbol Table in step 1.
4. Negative test - retrieval of an entry from Symbol Table according to a lexeme which hasn't been inserted into the Symbol Table and verification that the entry is null.
5. Addition of another symbol *S2* which has the same lexeme as the previously added symbol *S1*, but with a different location then *S1* to the Symbol Table using addSymbol method from SymbolTable class and verification of a successful return status of the method.
6. Retrieval of an entry from Symbol Table according to the lexeme of the token *S1 (same as the lexeme of S2)* and verification that the entry is not null.

Verification that the retrieved entry has both the locations of token *S1* and *S2* among the list of recorded locations.

*2.2 GIPSY*

*2.2.1 sType Checking code smell in* JINITA *class:*

The JINITA class is located in package gipsy.GEE.IDP.DemandGenerator.jini.rmi. In the writeValue() method of JINITA class, a conditional block is used to the run-time type of a an IDemand interface object, which is passed as parameter to the writeValue() method. We suggest replacing conditional type checking with polymorphism, and to that end we propose taking the following steps:

1. Declare a writeValue() method in the abstract class IDemand
2. Declare and implement a writeValue() method in class Demand (implementor of IDemand). The implementation code of this method contains the code inside the following JINITA's conditional block :

```
else
{
        oEntry.strDestination=DemandSignature.DWT;
}
```

3. Declare and implement a writeValue() method in class SystemDemand (child of Demand). The

implementation code of this method contains the code inside the following JINITA's conditional block :

```
else
{
oEntry.strDestination= (String) ((SystemDemand)poDemand).getDestinationTierID();
}
```

4. Declare and implement a writeValue() method in class ProceduralDemand (child of Demand). The implementation code of this method contains the code inside the following JINITA's conditional block :

```
else
{
else if(poDemand instanceof ProceduralDemand);
}
```

5. Declare and implement a writeValue() method in class IntensionalDemand (child of Demand). The implementation code of this method contains the code inside the following JINITA's conditional block :

```
else
{
else if(poDemand instanceof IntensionalDemand);
}
```

6. Declare and implement a writeValue() method in class ResourceDemand (child of Demand). The implementation code of this method contains the code inside the following JINITA's conditional block :

```
else
{
oEntry.strDestination=DemandSignature.ANY_DEST;
}
```

*2.2.2: Minimal Exception Feedback message in Empty* catch *blocks:*

In every identified empty catch block (as hinted by Eclipse's plugin JDeodorant [5]), a minimal code is introduced to provide feedback in the event of an exception being raised. A detailed context-conscious message of each exception is beyond the scope of this project as it requires knowledge of the details of each class in which such code smell occurs. However, as a minimal refactoring, we propose using Java's Throwable calss printStackTrace() method which prints the (throwable) object and its back-trace to the standard error stream. Hence, in each empty block we propose adding:

```
[throwable-object].printStackTrace();
```

### B. *Identification of Design Patterns*

#### 1) *DMARF*

##### 1.1 *Strategy Pattern:*

Strategy pattern is one of the Gang-of-Four's patterns [23]. The first example of Strategy pattern in DMARF is shown in the Class diagram below, which depicts a subset of attributes and methods of classes which contain an example of the Strategy pattern in DMARF. The class DMARF contains attribute soPreprocessing of type IPreprocessing, which is an Interface. There is one abstract class Preprocessing which implementats the IPreprocessing interface. Preprocessing class is subclassed by abstract class Filter class, and Filter is subclassed by abstract class CFEFilter. CFEFilter has 3 concrete subclasses: BandStopFilter, BandPassFilter and LowPassFilter. LowPassFilter has a subclass HighPassFilter.

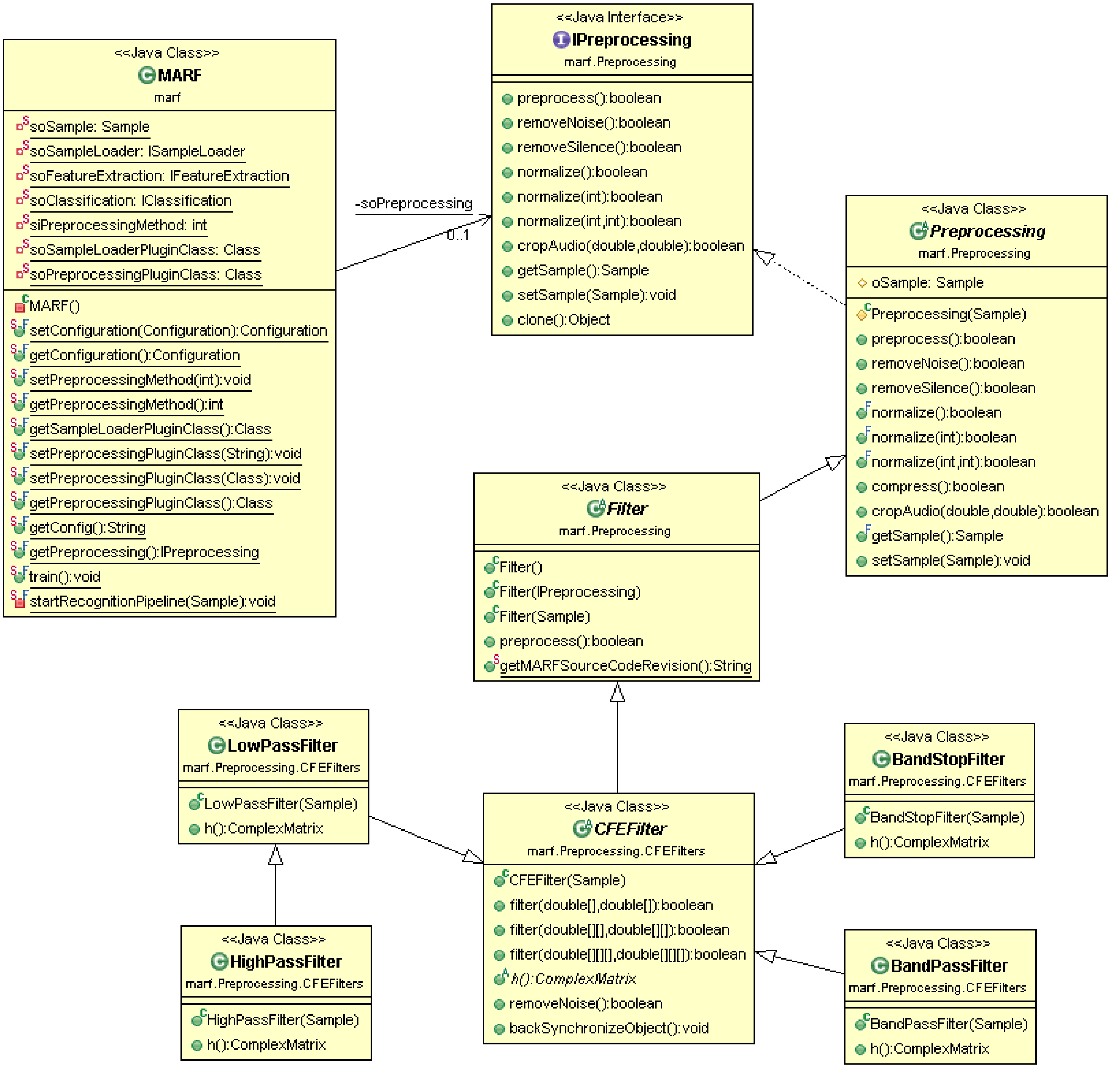

**Figure 35 - Class Diagram showing an example of Strategy pattern in DMARF where strategy's context class is MARF.**

This Strategy pattern is needed in order in order to select at runtime the type of preprocessing which will be used in the pipeline. soPreprocessing is a static member of the class of type the interfcace IPreprocessing and it contains the Preprocessing strategy. soPreprocessing is set dynamically in startRecognitionPipeline method according to the provided configuration. After setting the Preprocessor strategy, the startRecognitionPipeline method executes the preprocess method on the Preprocessor's object soPreprocessing.

This instance of Strategy design pattern was identified using Design Pattern Recognizer Eclipse plugin, which is a reverse engineering tool for identification of 5 types of design patterns: Composite, Façade, Observer, Singleton and State/Strategy.

Below are copy-pasted portions of DMARF code which are directly related to the described instance of Strategy design pattern:

```
public class MARF
{
   private static IPreprocessing soPreprocessing = null;

   private static final void startRecognitionPipeline(Sample
poSample)
   throws MARFException
```

```
    {   …
        soPreprocessing =
PreprocessingFactory.create(siPreprocessingMethod,
soSample); …
        soPreprocessing.preprocess();
    }
}
public interface IPreprocessing
extends Cloneable
{
    boolean preprocess()
            throws PreprocessingException;
}
public abstract class Preprocessing
extends StorageManager
implements IPreprocessing
{
    public boolean preprocess()
            throws PreprocessingException
    {…}
}
public abstract class Filter
extends Preprocessing
implements IFilter
{
    public boolean preprocess()
            throws PreprocessingException
    {…}
}
public abstract class CFEFilter
extends Filter
```

```
{…}
public class BandStopFilter
extends CFEFilter
{…}
public class BandPassFilter
extends CFEFilter
{…}
public class LowPassFilter
extends CFEFilter
{…}
public class HighPassFilter
extends LowPassFilter
{…}
```

The second example of Strategy pattern in DMARF: Below Class diagram shows a subset of attributes and methods of classes which form another example of the Strategy pattern in DMARF. The class MARFServant contains attribute oSampleLoaderRemote of type ISampleLoaderRMI, which is an Interface. There are two classes which implement ISampleLoaderRMI: ISampleLoaderWS_Stub and SampleLoaderServant.

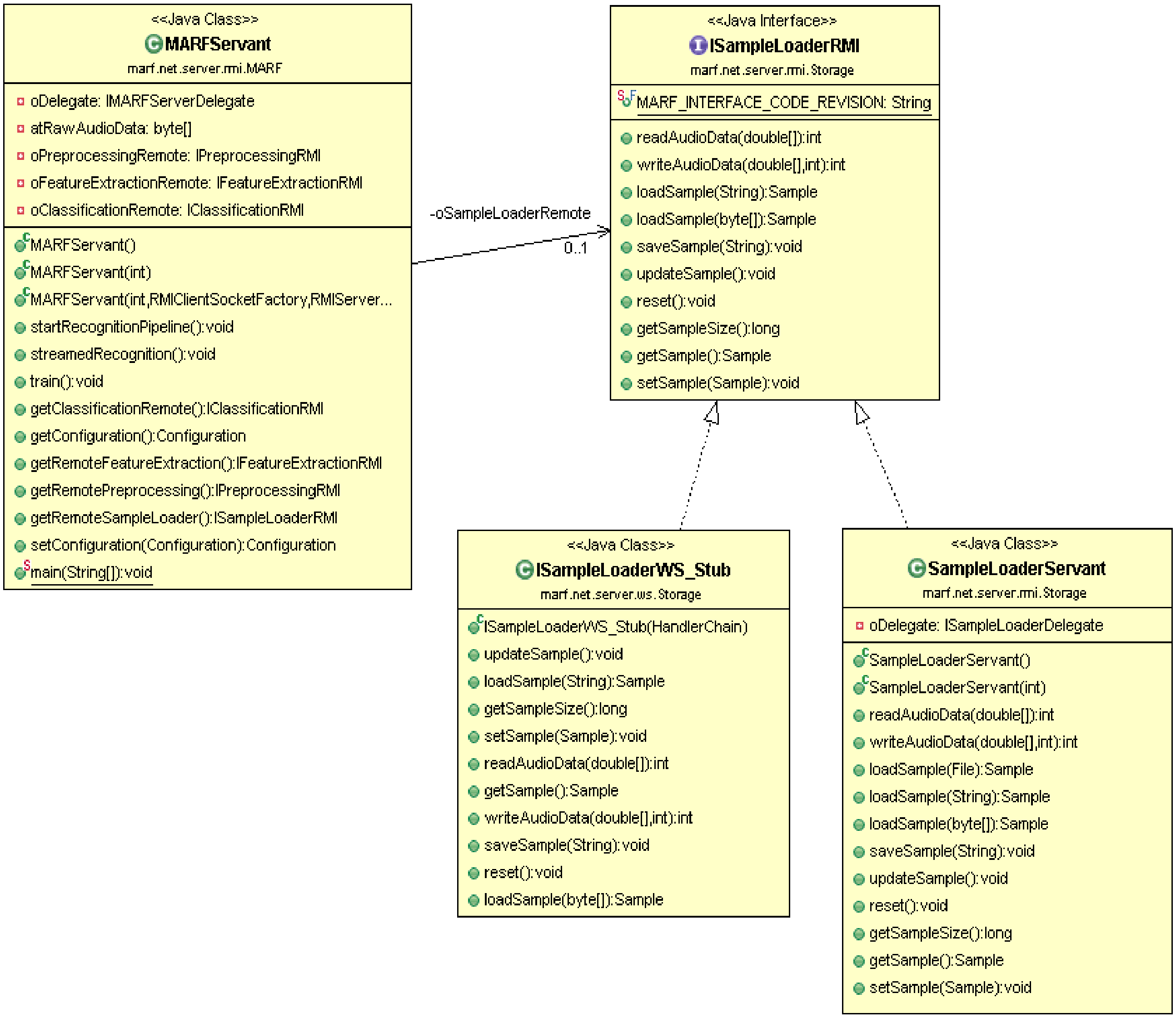

**Figure 36 -Class Diagram showing example of Strategy pattern in DMARF where strategy's context class is MARFServant.**

This Strategy pattern is needed in order to select at runtime one of the two implementations of sample loading RMI clients and then to load audio sample using the selected client.

This is done as following: the method startRecognitionPipeline in MARFServant class sets in runtime the oSampleLoaderRemote attribute with one of the two concrete implementations of ISampleLoaderRMI interface. The assignment of the oSampleLoaderRemote attribute is happening at runtime using method startRMISampleLoaderClient from RMIUtils class. Then startRecognitionPipeline method calls loadSample method on oSampleLoaderRemote attribute of the corresponding concrete implementation of ISampleLoaderRMI interface which has been instantiated. This instance of Strategy design pattern was also identified using Design Pattern Recognizer Eclipse plugin. Below are copy-pasted portions of DMARF code which are directly related to the described instance of Strategy design pattern.

```
public class MARFServant
    extends UnicastRemoteObject
    implements IMARFServerRMI, IRMIClient
{
    private ISampleLoaderRMI oSampleLoaderRemote = null;

    public void startRecognitionPipeline() throws
RemoteException, MARFException
    {…
        oSampleLoaderRemote =
RMIUtils.startRMISampleLoaderClient();
        Sample soSample =
oSampleLoaderRemote.loadSample(this.atRawAudioData);
    …}
}
```

```
public interface ISampleLoaderRMI
    extends IRMIServer, Cloneable
{
    Sample loadSample(final String pstrFilename)
        throws RemoteException, StorageException;
}

public class ISampleLoaderWS_Stub
    extends com.sun.xml.rpc.client.StubBase
    implements marf.net.server.ws.Storage.ISampleLoaderWS,
            marf.net.server.ws.IWSServer,
            marf.net.server.rmi.Storage.ISampleLoaderRMI
{
    public marf.Storage.Sample loadSample(byte[]
arrayOfbyte_1)
        throws marf.Storage.StorageException,
java.rmi.RemoteException {…}
}

public class SampleLoaderServant
    extends UnicastRemoteObject
    implements ISampleLoaderRMI
{
    public Sample loadSample(byte[] patFileAudioData)
        throws RemoteException, StorageException {…}
}
```

*1.2 Adapter Pattern*

The design pattern **Adapter Pattern** identified in DMARF is one of Gang-of-Four's Structural patterns. Adapter pattern converts the interface of a class into another interface. Adapter pattern enables those classes to work together, which otherwise were unable to because of incompatible interfaces. Adapter pattern works as a bridge between two incompatible interfaces. An adapter helps two incompatible interfaces to work together.

There are two types of Adapter patterns – Object Adapter Pattern and Class Adapter Pattern. The Adapter pattern identified in the following segment is an example of a Class Adapter Pattern. In this case, the adapter uses multiple polymorphic interfaces to achieve its goal. The adapter is created by implementing or inheriting both the interface that is expected and the interface that is pre-existing. It is typical for the expected interface to be created as a pure interface class, especially in languages such as Java that do not support multiple inheritance.

The reason for using the adapter pattern, in the example below, is to help the feature extraction interface compatible with preprocessing interface. The adapter is **FeatureExtraction,** the adaptee is **IPreprocessing.** The class **FeatureExtraction** class contains the ExtractFeatures() method and an internal reference is made to the preprocessing. Thus the feature is extracted.

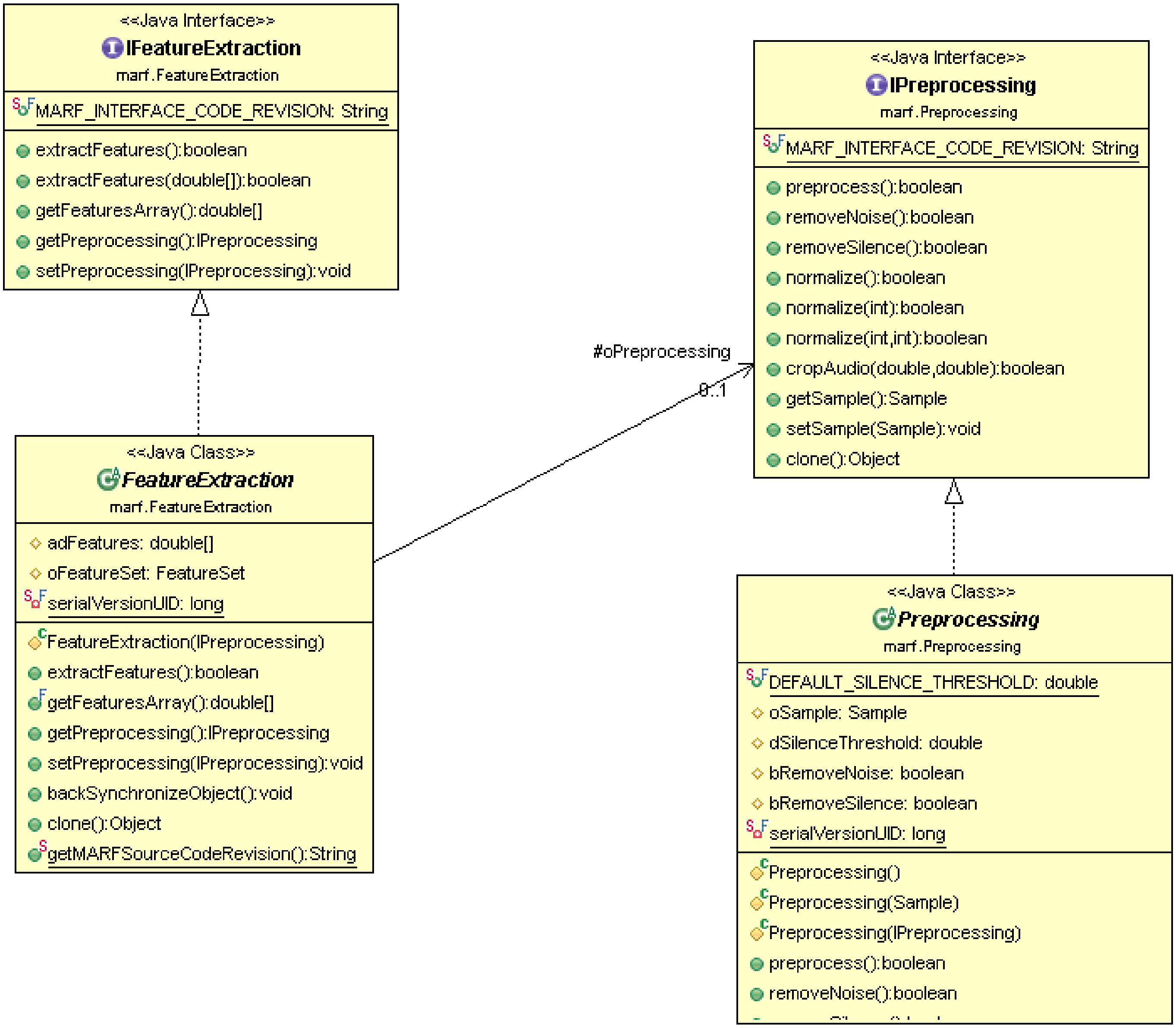

**Figure 37 - Adapter pattern class diagram**

The reason for using the adapter pattern here is to help the feature extraction interface compatible with preprocessing interface. The ExtractFeatures() method in FeatureExtraction class, an internal reference is made to the preprocessing and thus the feature is extracted.
Following is the code snippet,

```
public abstract class FeatureExtraction
extends StorageManager
implements IFeatureExtraction
{
----- Other methods----
protected IPreprocessing oPreprocessing = null;
```

```
public boolean extractFeatures()
        throws FeatureExtractionException
        {
                return
this.extractFeatures(this.oPreprocessing.getSample().getSamp
leArray());
        }
---Other methods ------
{

public interface IPreprocessing
extends Cloneable
{
--- Other methods ---
boolean preprocess();
Sample getSample();
```

```
-- Other methods -------
}
```

### *1.3 Decorator Pattern:*

The decorator pattern is a structural design pattern which enables us to add new or additional behavior to an object during runtime, depending on the situation. Decorators provide a flexible alternative to sub classing for extending functionality. The goal is to make it so that the extended functions can be applied to one specific instance, and, at the same time, still be able to create an original instance that doesn't have the new functions. We need the decorator pattern when we want to add responsibilities to individual objects, not to an entire class[32].

The following UML class diagram describes the Decorator design Pattern in MARF. The component here is the interface IFeatureExtractionCORBAOperations and the Decorator is the IFeatureExtractionCORBAPOATie.

The component pointer is the "private marf.net.server.corba.FeatureExtraction.IFeatureExtractionCORBAOperations _impl".

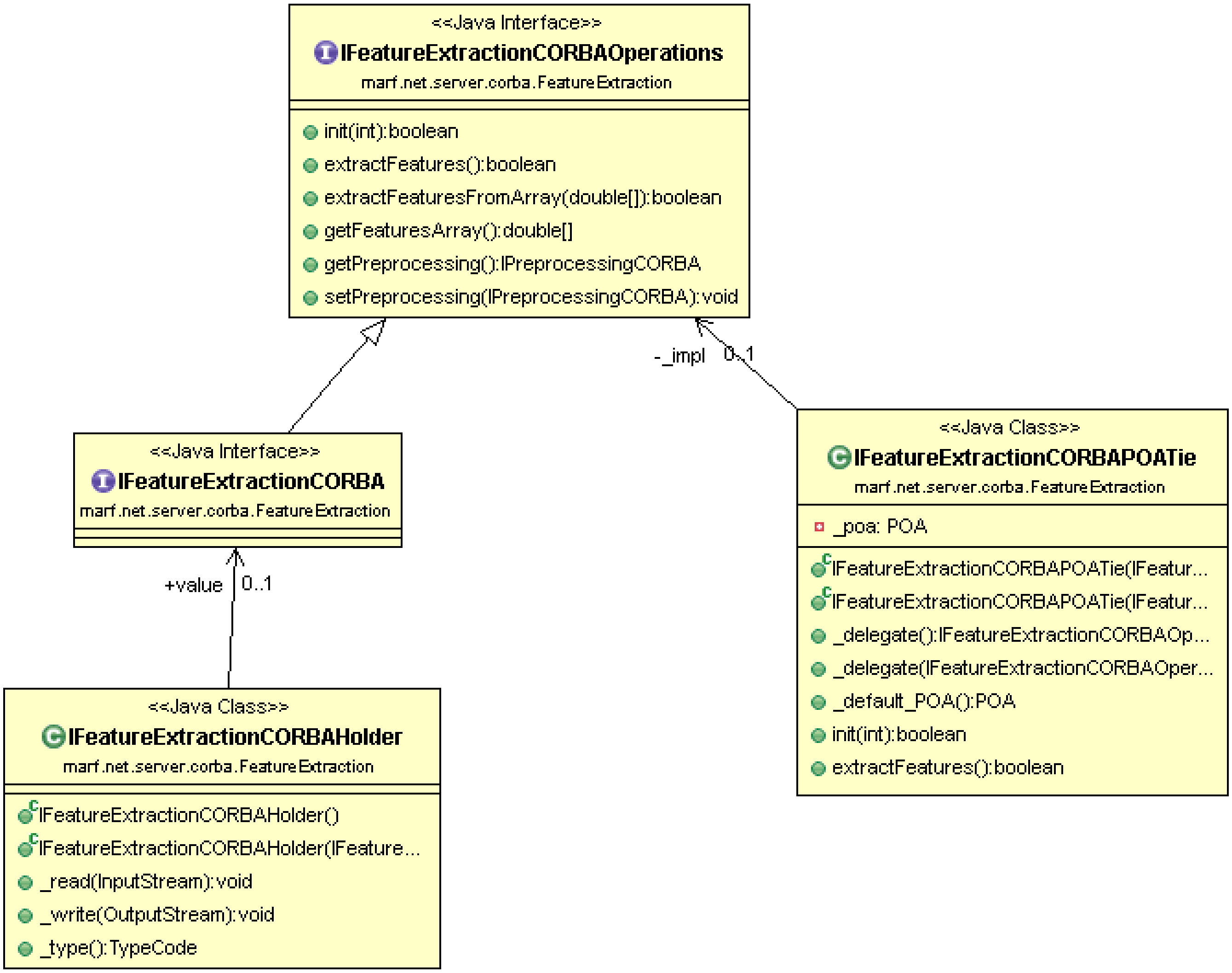

**Figure 38 - Decorator pattern Class diagram**

The reason for using the decorator pattern here is to eliminate adding the behavior of classes at run time which is done when there are sub classes. This pattern provides new behavior to classes at run time. The Decorator here wraps the component. The method getpreprocessing() in the Decorator class retrieves the inner preprocessing reference and the setpreprocessing() method allows setting the source preprocessing module.

The corresponding code snippet is as follows:

```
public class IFeatureExtractionCORBAPOATie extends
IFeatureExtractionCORBAPOA
{
public IFeatureExtractionCORBAPOATie (
marf.net.server.corba.FeatureExtraction.IFeatureExtractionC
ORBAOperations delegate ) {
      this._impl = delegate;
  }
private
marf.net.server.corba.FeatureExtraction.IFeatureExtractionC
ORBAOperations _impl;

public
marf.net.server.corba.Preprocessing.IPreprocessingCORBA
getPreprocessing ()
  {
    return _impl.getPreprocessing();
  }
public void setPreprocessing
(marf.net.server.corba.Preprocessing.IPreprocessingCORBA
poPreprocessing)
  {
    _impl.setPreprocessing(poPreprocessing);
  }
public boolean extractFeaturesFromArray (double[]
padData) throws
marf.net.server.corba.CORBACommunicationException
  {
    return _impl.extractFeaturesFromArray(padData);
  }
}

            public interface
IFeatureExtractionCORBAOperations
{
marf.net.server.corba.Preprocessing.IPreprocessingCORBA
getPreprocessing ();

void setPreprocessing
(marf.net.server.corba.Preprocessing.IPreprocessingCORBA
poPreprocessing);
boolean extractFeatures () throws
marf.net.server.corba.CORBACommunicationException;

}
```

### *1.4 Singleton Pattern:*

Instantiation of a class to one object is restricted by a design pattern called the singleton pattern. When one object tries to coordinate the actions across the system this kind of pattern is useful. In many cases the singleton pattern is criticized to be anti-pattern. Many patterns like abstract factory, builder and prototype singleton pattern is used in their implementations. Singleton is considered as global variables. This is effectively accomplished concealing the constructor and giving a different system that calls the constructor just after acceptance that the case does not exist. Typically, the singleton example is utilized as a placeholder for setups, or static information, where the exertion of getting the information is significant (the trade-off is the memory utilization, since the occurrence stays in the memory for the application lifecycle, once instantiated). The example of the class is recovered by a static technique. Sample of how a class is characterized as a singleton is given below

```
public class MySingleton{
MySingleton instance=null;
private MySingleton(){ ….}
public static synchronized getInstance()
{
            if(instance == null)
{
                        /* Invoke the constructor only if the
instance is null
                        instance = new MySingleton();
            }
            return instance;
}
}
```

Example of a singleton pattern in DMARF is class marf.util.OptionFileLoader

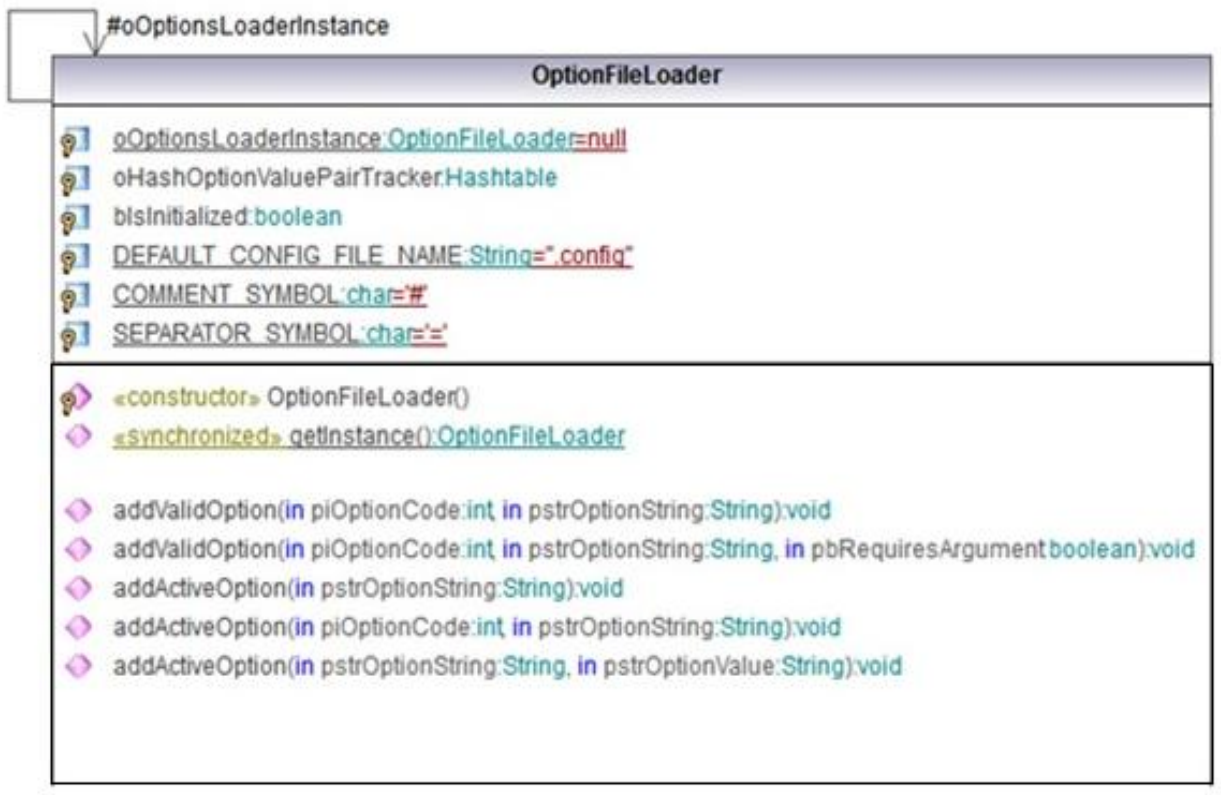

**Figure 39 - Singleton Pattern Class diagram**

```
package marf.util;
public class OptionFileLoader implements IOptionProvider
{
   protected static OptionFileLoader oOptionsLoaderInstance
= null;
            ……………….
   protected OptionFileLoader() {
```

```
/**/
 }
   public static synchronized OptionFileLoader getInstance()
   {
            if(oOptionsLoaderInstance == null)
            {
    oOptionsLoaderInstance = new OptionFileLoader();
            }

            return oOptionsLoaderInstance;
   }
/**/
}
```

*2) GIPSY*

*2.1 Factory Pattern*

The design patter **Factory Method**, one of the Gang-of-Four's patterns [23] has been identified in GIPSY. The essence of Factory Method pattern is to create an interface for creating an object, but leaving the decision as to which concrete class this objects belongs to is left for subclasses to decide. The following UML class diagram depicts the general structure of the Factory Method pattern [24]

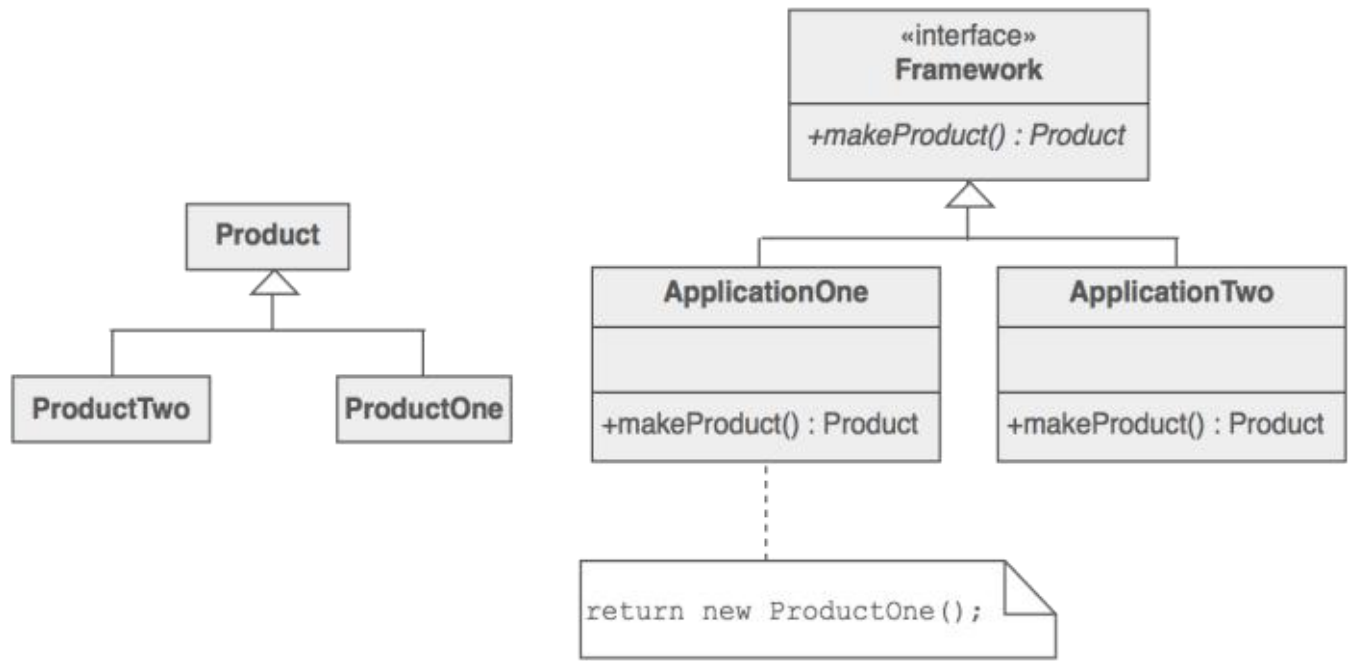

**Figure 40 - A factory interface masks the concrete classes creating objects, here referred to as 'Applications'. Each application (a concrete factory) is responsible for creating a 'product' (an object)[24].**

In GIPSY's gipsy.lang package, the IArithmeticOperatorsProvider interface's add() and substract () are the 'factory' methods for the 'prodcut' GIPSYType object. Which concrete GIPSYType object is 'manufactured' is left for IArithmeticOperatorsProvider's subclasses (representing the 'application' classes): GenericArithmeticOperatorsDelegate and GIPSYInteger to decide. Below is a UML class diagram depicting the relationships between the relevant classes

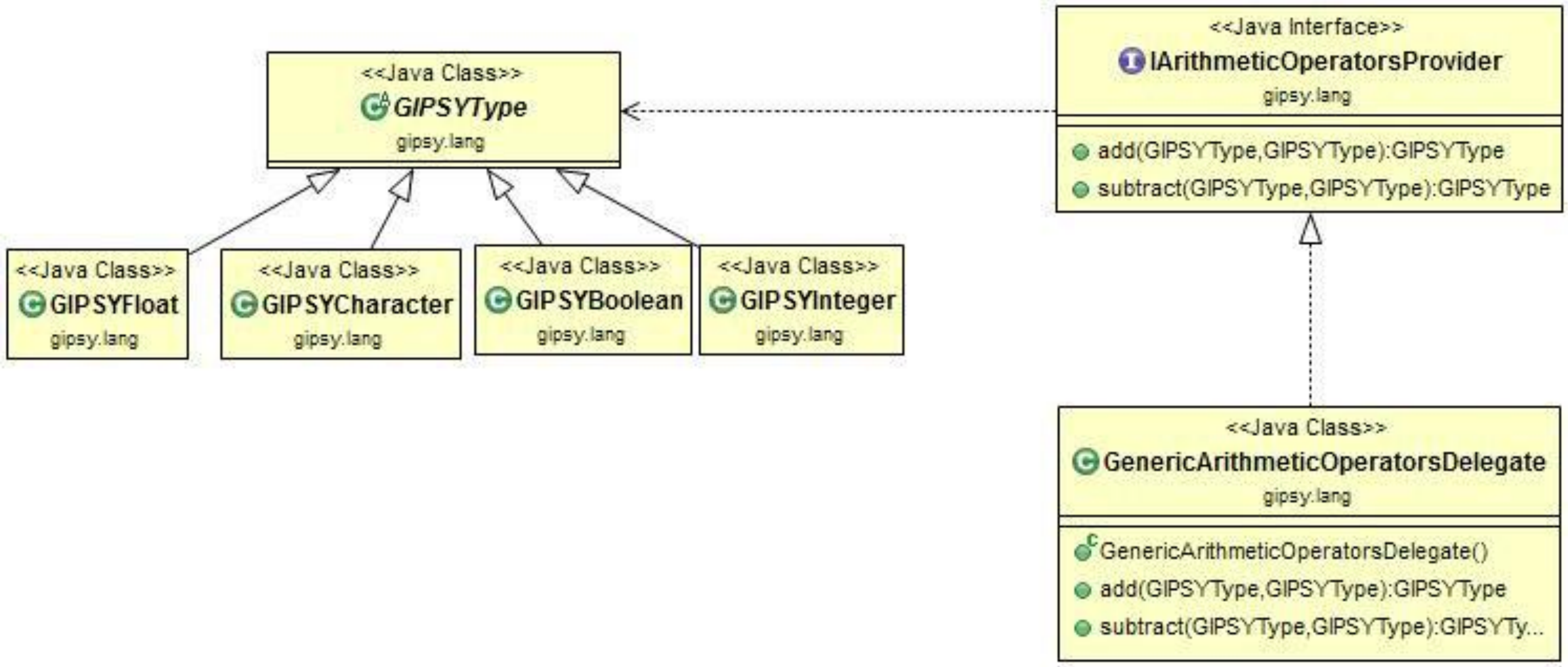

**Figure 41 - A class diagram of classes involved in the identified Factory Method in GIPSY. The IArithmeticOperationsProvider 's add() and substract () are the factory methods, manufacturing objects ('products') of GIPSY type (and its descendants).**

The justification for this pattern is due to the fact that there are many GIPSYTypes and for an application class to define one concretely is to expose the instantiation logic. Furthermore, if GIPSYType's were created using new, then the declaring class is tied up with a reference to that class. Below are (psudo) code snippets relevant to this instance of Factory Method design pattern (modified for concision and clarity)

```
public interface IArithmeticOperatorsProvider{
        GIPSYType add(GIPSYType poLHS, GIPSYType
poRHS);
        GIPSYType subtract(GIPSYType poLHS,
```

```
GIPSYType poRHS);
}


public class GenericArithmeticOperatorsDelegate
implements IArithmeticOperatorsProvider {

public GIPSYType add(GIPSYType poLHS, GIPSYType
poRHS) {
        //Concrete implementation

        //conditional on the left operand
        if(poLHS instanceof [GIPSYInteger, GIPSYFloat,
        GIPSYCharacter,GIPSYBolean]){
        //conditional on the right operand
        if(poRHS instanceof GIPSYInteger){
                return [GIPSYInteger,
        GIPSYFloat,GIPSYCharacter,GIPSYBolean])
        }
        .
        .
}

public GIPSYType subtract(GIPSYType poLHS, GIPSYType
poRHS){
                //Concrete implementation
                .
                .
                // Same as add() above.
                .
                .

        }
}
```

*Abstract Factory:*

The Abstract Factory design pattern is intended to provide an interface or an abstract class for "creating families of related or dependent objects, without specifying their concrete classes" [25]. In an abstract sense, this design pattern is concerned with creating many "platforms", to facilitate the creation of many "products", as shown in Figure 6.2.2.1 below

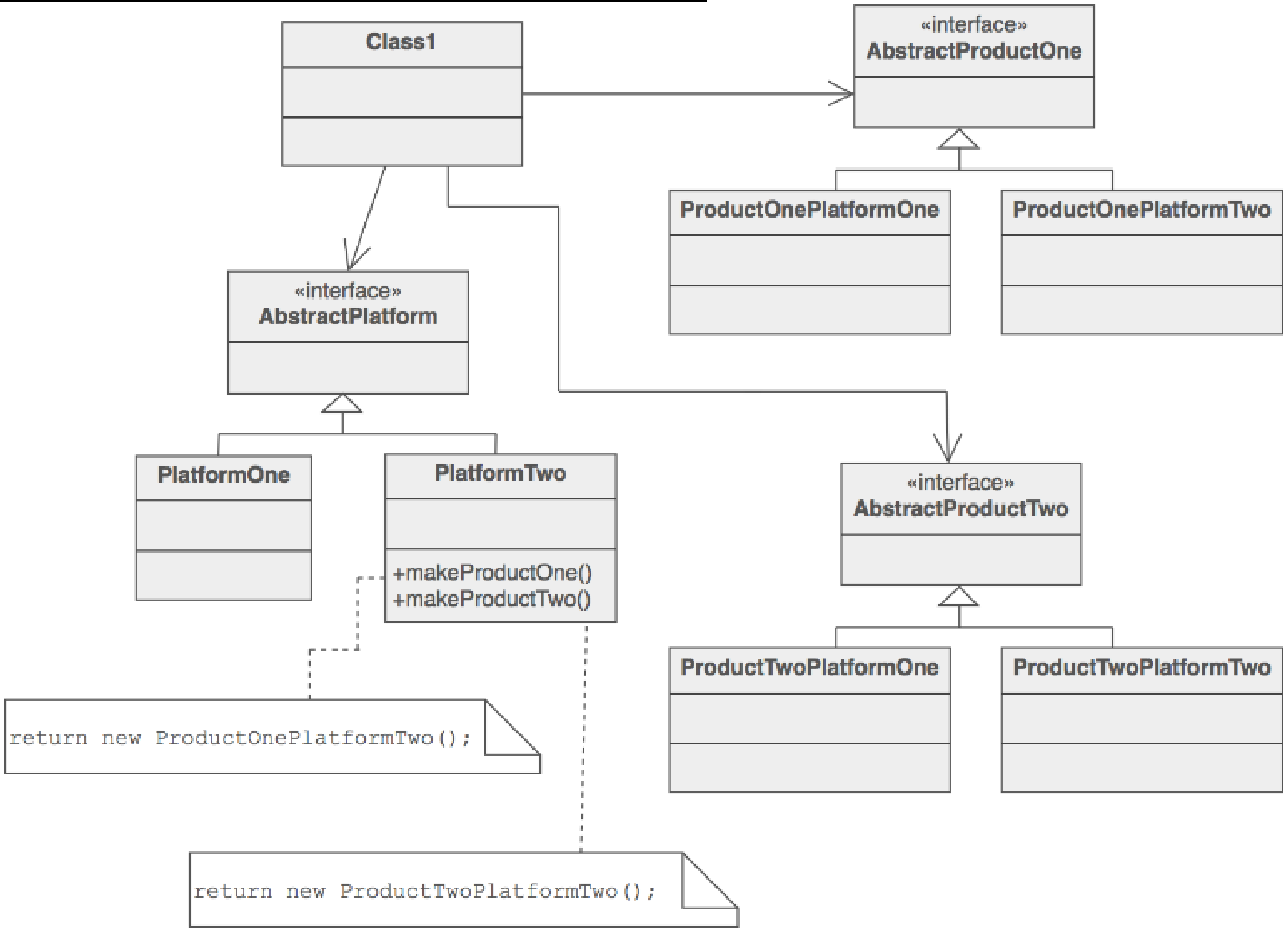

**Figure 42 - an abstract view of the structure and components involved in Abstract Factory design pattern [25]. The aim is to create an interface masking a collection of "platforms", and an interface masking a collection of "products". Subsequently, a concrete prod**

In GIPSY, an Abstract Factory design pattern has been identified in gipsy.GEE.multitier package, whereby TierFactory abstract class and IMultiTierWrapper interface serve as the "platform" and "product" interfaces, respectively. As can be seen in Figure 6.2.2.1, there are a number of platforms, which corresponds to the subtypes of TierFactory:

- The Demand Worker Tier Factory: DWTFactory
- The Demand Generator Tier Factory: DGTFactory
- The Demand Store Tier Factory: DSTFactory
- Tier Wrapper Factory: TierWrapperFactory, complements the Tier Factory, masking the differences in instatiation of different types of tiers.

Furthermore, there are a number of products, namely the subtypes of IMultiTierWrapper:

- Generic Tier Wrapper: GenericTierWrapper, implements IMultiTierWrapper's operation common to all wrappers
- The Demand Store Tier (DST) DSTWrapper, stores and migrates demands between tiers
- The Demand Generator Tier DGTWrapper, generates intensional demands.
- The Demand Worker Tier DWTWrapper, processes procedural demands.

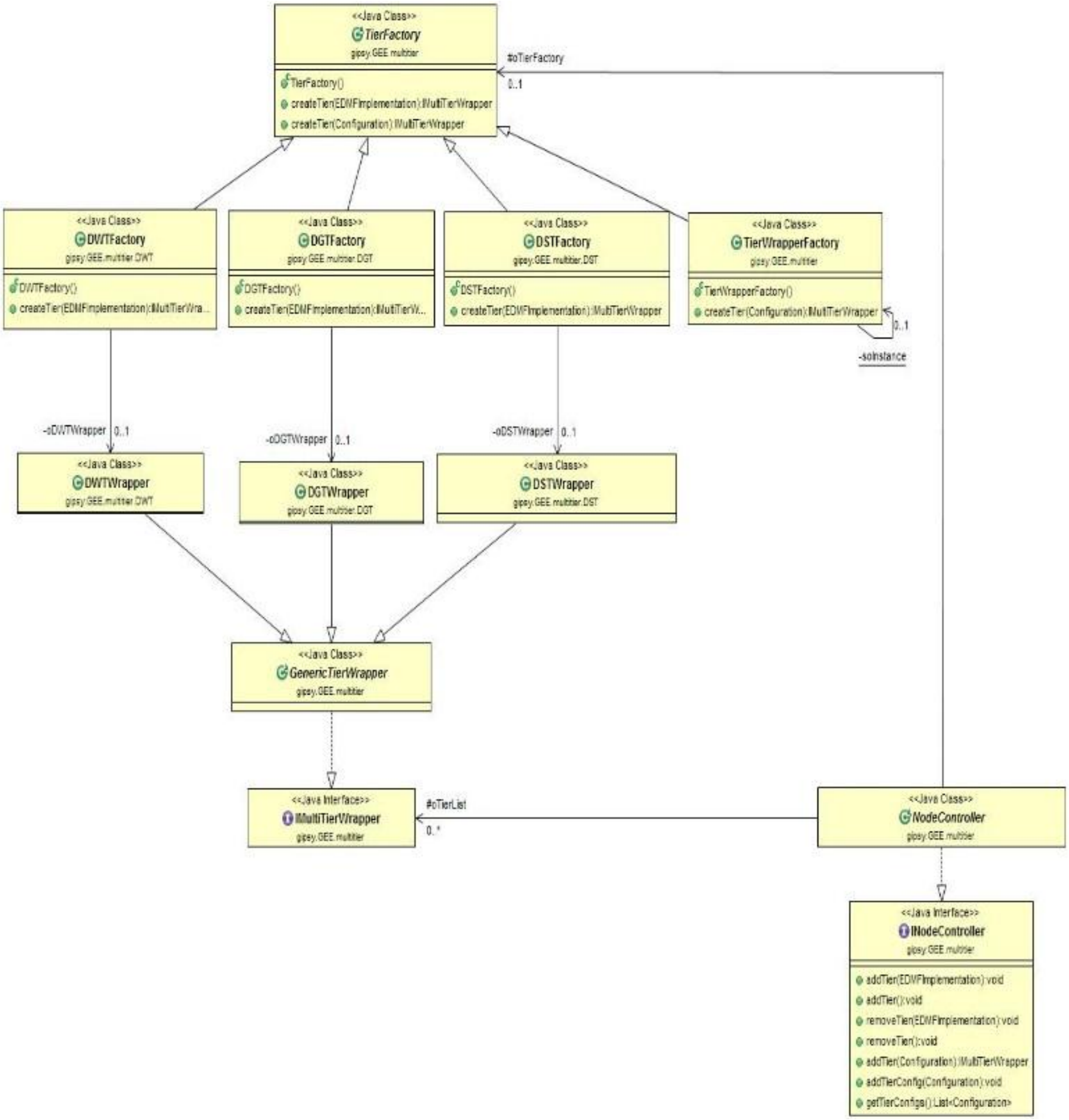

**Figure 43 - Identified Abstract Factory design pattern in GIPSY's gipsy.GEE.multitier package. The TierFactory and IMultiTierWrapper serve as the platform and product interfaces, respectively, and the subtypes of each correspond to the concrete platforms (contexts)**

The (pseudo) code snippets below shows a slice of Abstract Factory-related classes, modified for concision and clarity

```java
public abstract class TierFactory{
	public IMultiTierWrapper createTier(EDMFImplementation poDMFImp){}
	.
	.
	//other operations
}
```

**Abstract Factory (platform's interface)**

```java
public class DSTFactory extends TierFactory
{
	 private DSTWrapper oDSTWrapper = null;
	 public IMultiTierWrapper createTier(EDMFImplementation poDMFImp){
		.
		.
		//instanstiation logic that eventually leads to:
		this.oDSTWrapper = new DSTWrapper();
		.
		.
		//finally, a product is created:
		return this.oDSTWrapper;
	 }
```

**Concrete Factory (the "Demand Store Platform")**

```java
public class DSTWrapper extends GenericTierWrapper {
	 public DSTWrapper()
	 {
		 //pump up to super-constructor in GenericTierWrapper
		 //(debugging logic only)
	 }

}
```

**Concrete Product**

```java
public abstract class GenericTierWrapper implements IMultiTierWrapper{
	 public GenericTierWrapper() {...}
	 public GenericTierWrapper(...){...}
}
```

**Generic Product**

```java
public interface IMultiTierWrapper extends Runnable
{
		//declaration of operations common to all products
```

**Abstract Product (product's**

**Figure 44 -(pseudo) code snippets showing relevant code segments in various classes involved in the identified Abstract Factory design pattern. The Concrete Factory/Product used here are DSTFactory and DSTWrapper, but similar parallel code snippets are also implemented in the rest of concrete factories ("platforms")/products, namely: DWTFactory/DGTFactory and DWTWrapper/ DGTWrapper**

### *2.2 Observer Pattern*

Observer Pattern define one to many dependency between objects so that when one object changes the state, all its dependents are notified and updated automatically.

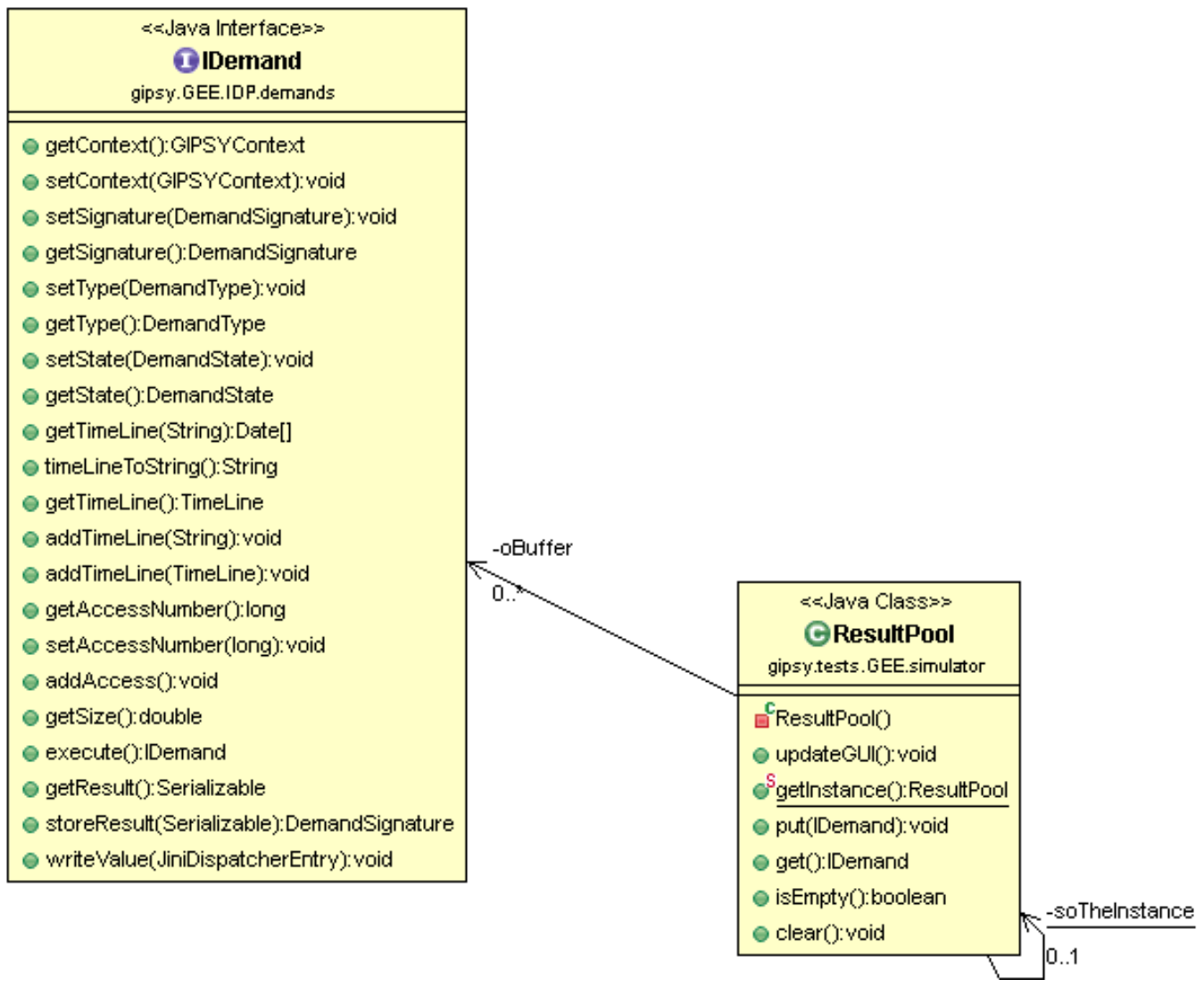

**Figure 45 - Observer pattern Class diagram**

The above diagram shows the class diagram of observer pattern in GIPSY. Here in this diagram we have gipsy.GEE.IDP.demands.IDemand class as an observer clas and gipsy.tests.GEE.simulator.ResultPool class as a subject to notify through gipsy.tests.GEE.simulator.ResultPool::updateGUI():void .

IDemand get a pending demand to be picked up by any tier from the store and set it to the destination. The result of the demand is then specified with the specified signature from the store

Below is the source code showing the Observer pattern in GIPSY.

```
public interface IDemand
extends ISequentialThread, Cloneable
{

public GIPSYContext getContext();
public void setContext(GIPSYContext poContext);
void setSignature(DemandSignature poSignatureID);
DemandSignature getSignature();
void setType(DemandType poType);
DemandType getType();
void setState(DemandState poState);
DemandState getState();
TimeLine getTimeLine();
/*
*Accumulates timeline points per tier ID for this demand
*/
void addTimeLine(String pstrTierID);
/*
* Reference counting
*/
void addTimeLine(TimeLine poTimeLine);
long getAccessNumber();
void setAccessNumber(long plAccessNumber);
void addAccess();
double getSize();
/**
* Evaluate a sequential thread.

*/
IDemand execute();
Serializable getResult();
DemandSignature storeResult(Serializable poResult);
public abstract void writeValue(JiniDispatcherEntry oEntry);
}
```

### *2.3 Prototype Pattern*

**Prototype** pattern is a creational pattern that creates a new object by cloning an existing object. It allows an object to create customized objects without knowing their class or any details of how to create them. We use Prototype pattern when a system should be independent of how it's products are created, composed, and represented, and:[31]

- Classes to be instantiated are specified at run-time
- Avoiding the creation of a factory hierarchy is needed
- It is more convenient to copy an existing instance than to create a new one.

In GIPSY system the prototype pattern is found to be used in the class Configuration and GIPSYNode.

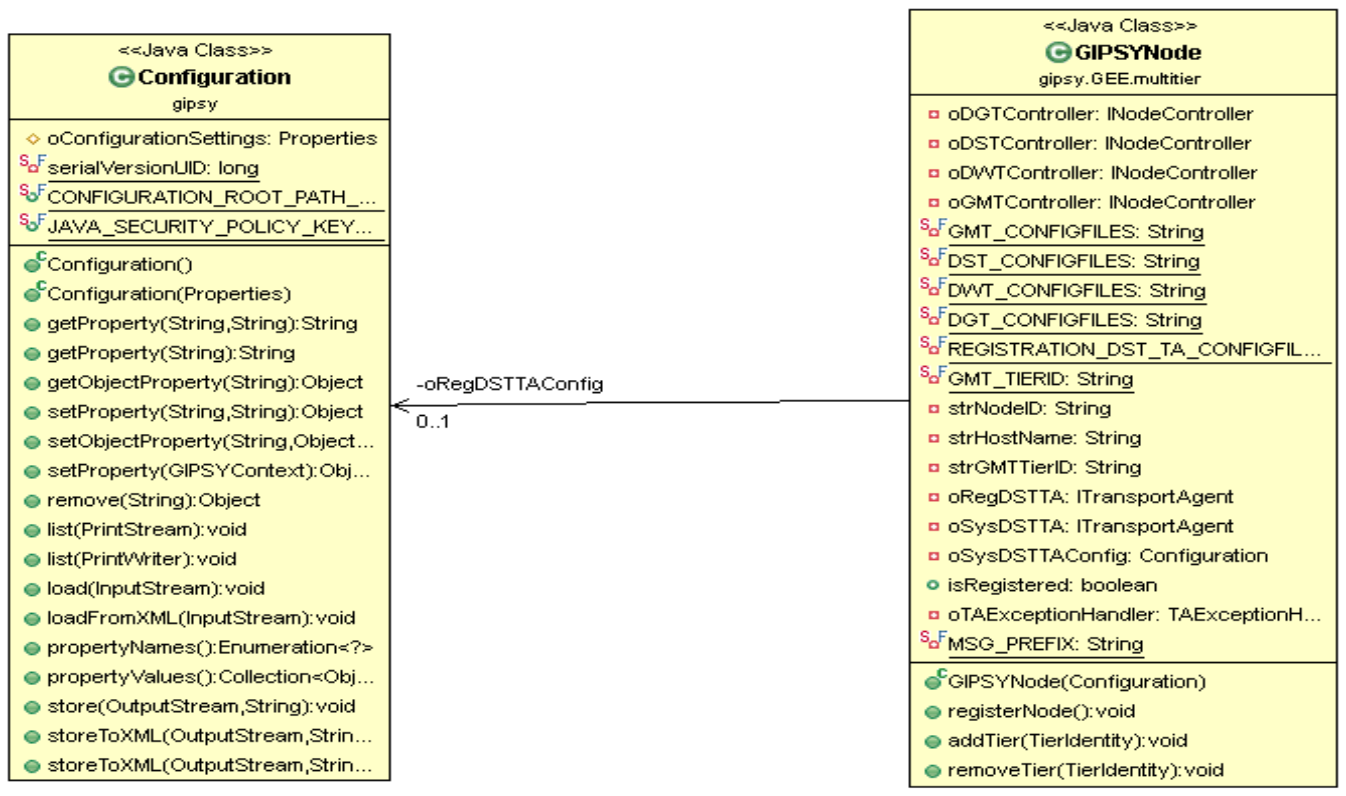

**Figure 46 - Prototype pattern Class diagram**

The reason for using the Prototype pattern here is to hide the complexity of creating new instance from the client. In the above Class diagram, The client is the GIPSYNode class, Prototype is the configuration class where the configuration object is cloned in the GIPSYNode.

In the class *Configuration,*

The following code represents the clone object created for the configuration class.

```
public class Configuration
implements Serializable
{
public synchronized Object clone()
        {
                Configuration oNewConfig = new
Configuration();

        oNewConfig.setConfigurationSettings((Properties)
this.oConfigurationSettings.clone());
                return oNewConfig;
        }
-------Other methods-----
}
```

The following code represents the clone object being created in the *GIPSYNode* class

```
public class GIPSYNode
extends Thread
{
public void run()
        {
------ Couple of methods --------
oTierConfig = (Configuration)
oRequest.getTierConfig().clone();

switch(oTierIdentity)
{
--Case statements for DST, DWT and DGT tiers------
}
}
```

*2.4 Proxy Pattern*

The access to an object has to be controlled sometimes. Sometimes the accessing the objects entirely will cost us. So untill that point we need some light weight objects to access those attributes. Those light weight objects are called proxies.

The proxy design pattern implementation[33] can be as follows:

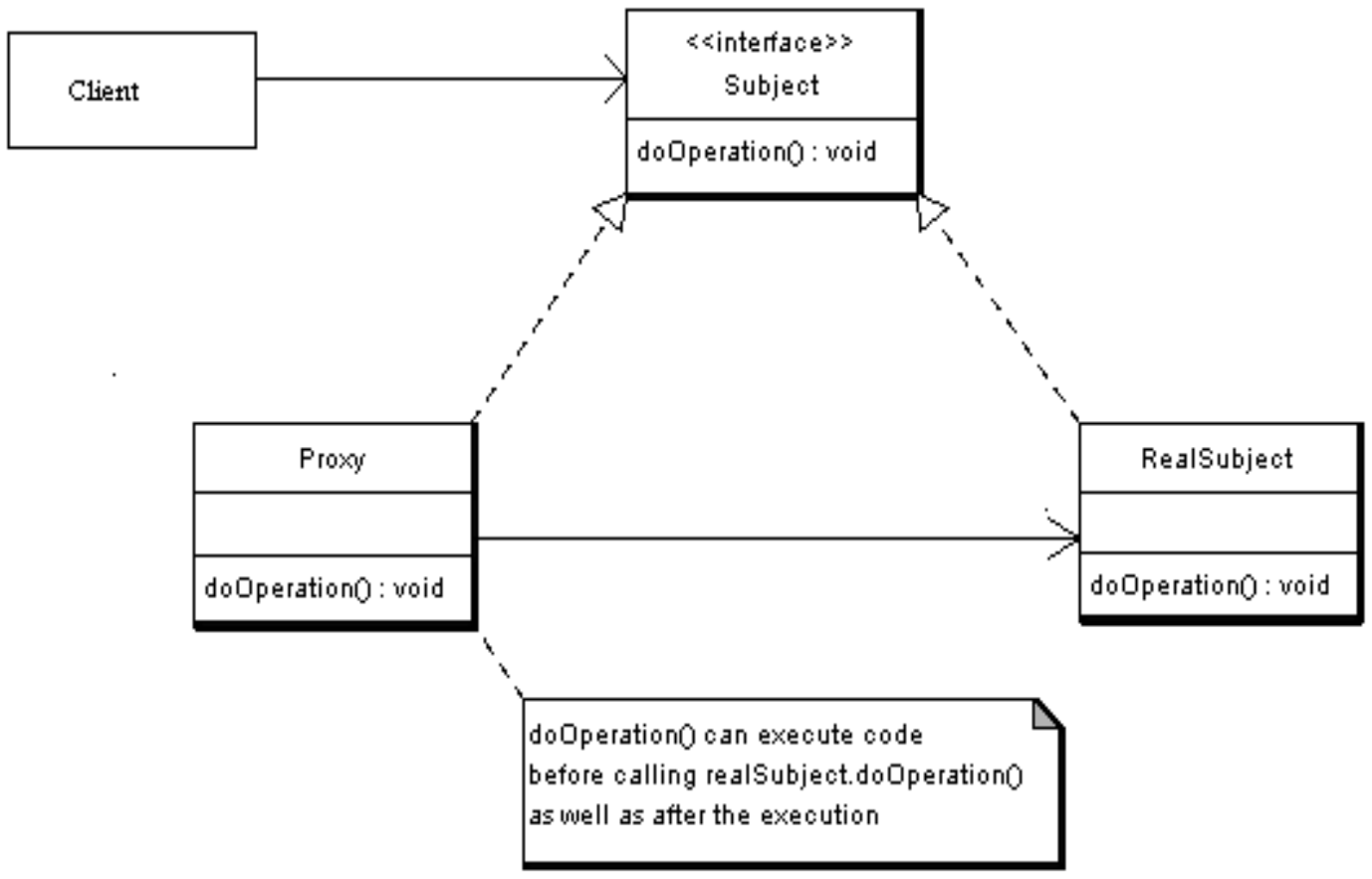

**Figure 47 - Proxy design pattern**

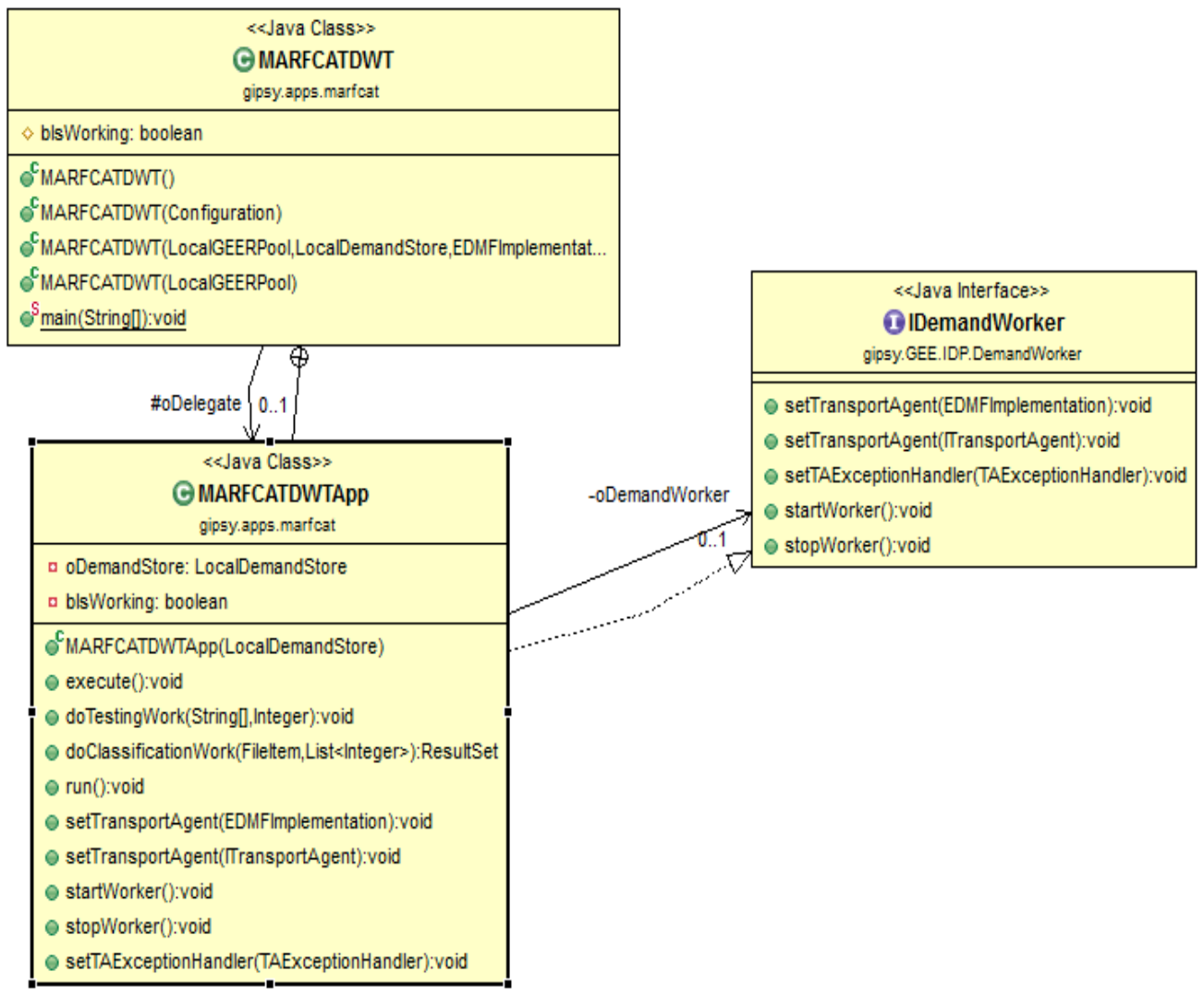

**Figure 48 - Proxy class diagram**

In the GIPSY system there exists a proxy pattern at MARFCATDWTApp, MARFCATDWT and IDemandWorker packages. The methods in IDemandWorker and MARFCTDWT is accessed by MARFCATDWTApp package, but initially the startWorker() method is accessed by created an object to that class but whereas the stopWorker() methos is accessed without some lightweight object the same with the metods in MARFCATDWT package also. This kind access to these methods gives proxy patterns in the code. The code structure for the above mentioned code pattern is given below.

```
public interface IDemandWorker extends Runnable
{
void setTransportAgent(EDMFImplementation
poDMFImp);
void setTransportAgent(ITransportAgent poTA);
void setTAExceptionHandler(TAExceptionHandler
poTAExceptionHandler);
```

```
void startWorker();
void stopWorker();
}
public class MARFCATDWT extends DWTWrapper
{

@Override
public void setTransportAgent(EDMFImplementation
poDMFImp)
{
this.oDemandWorker.setTransportAgent(poDMFImp);
}
@Override
public void setTransportAgent(ITransportAgent poTA)
{
this.oDemandWorker.setTransportAgent(poTA);
}
@Override
public void startWorker()
{
this.oDemandWorker.startWorker();
this.bIsWorking = true;
}
@Override
public void stopWorker()
{
this.oDemandWorker.stopWorker();
this.bIsWorking = false;
}
@Override
public void
setTAExceptionHandler(TAExceptionHandler
poTAExceptionHandler)
{
this.oDemandWorker.setTAExceptionHandler(poTAExc
eptionHandler);
}
}
```

## VI. Implementation

### A. *Refactoring Changesets and Diffs*

#### 1) *DMARF*

**Feature Envy code smell refactoring in DMARF**

**Change 0/3: Refactor SymbolTable class to resolve Feature Envy code smell in addSymbol method.**

The method addSymbol in class SymbolTable from package marf.nlp.Parsing is more interested in class Token from the same package rather than in its own class SymbolTable, because it uses more facilities from class Token then from its own class. The code smell is refactored using Move Method refactoring by moving the method addSymbol to Token class.
Relation of the changes (patches) in the patch set: Change #1 adds method addSymbol to Token class. This serves as the base for change #2 in which the original addSymbol method in SymbolTable class is changed to delegate its task to the new addSymbol method in Token class. Change #3 contains unit tests implemented using junit to test the addSymbol method of SymbolTable class before and after the refactoring in order to verify that the external behavior didn't change during refactoring.

**Change 1/3: Add method addSymbol to Token class.**
Reason for change: addSymbol method from SymbolTable class from marf.nlp.Parsing package should be moved to Token class located in the same package in order to resolve Feature Envy code smell which arises from addSymbol method. As the first step in the refactoring, addSymbol method is added to Token class, in order to delegate the work of the original addSymbol method from SymbolTable class to this new method in the subsequent steps.
Impact on the system: addSymbol method is added to Token class. The method performs equivalent logic as the original addSymbol method in SymbolTable class. It accepts a Hashtable which maps Tokens' lexemes to SymTabEntry objects as input parameter and then checks if the lexeme of *this* Token is already present in the Hashtable, it records its additional location in the mapped SymTabEntry object, else a new entry is created in the Hashtable.

Diff of the change:

```
Index: src/marf/nlp/Parsing/Token.java
===================================================
==================
RCS file:
/groups/m/me_soen6471_1/cvs_repository/DMARF/src/marf/nlp/Par
sing/Token.java,v
retrieving revision 1.3
diff -u -r1.3 Token.java
--- src/marf/nlp/Parsing/Token.java	21 Aug 2014 01:08:34 -0000
	1.3
+++ src/marf/nlp/Parsing/Token.java	21 Aug 2014 01:18:21 -0000
@@ -9,6 +9,7 @@
 import java.io.FileWriter;
 import java.io.IOException;
 import java.io.Serializable;
+import java.util.Hashtable;


 /**
@@ -238,6 +239,24 @@
	{
		return "$Revision: 1.20.4.2 $";
	}
+
+	/**
+	 * Adds a symbol token to the table. If the symbol is
already there, its additional location is recorded. Else a new entry is
created.
```

```
+		* @param oSymTabEntries
+		* @return  0 on success
+		*/
+		public int addSymbol(Hashtable oSymTabEntries) {
+			if (oSymTabEntries.contains(getLexeme())) {
+				SymTabEntry oSymTabEntry =
(SymTabEntry) oSymTabEntries
+					.get(getLexeme());
+
		oSymTabEntry.addLocation(getPosition());
+			} else {
+				SymTabEntry oSymTabEntry = new
SymTabEntry(this);
+
		oSymTabEntry.addLocation(getPosition());
+				oSymTabEntries.put(getLexeme(),
oSymTabEntry);
+			}
+			return 0;
+		}
 }

 // EOF
```

**Change 2/3: Replace addSymbol method's body in SymbolTable class with the delegation to the new addSymbol Method in Token class.**

Reason for change: to finalize the refactoring of Feature Envy code smell related to addSymbol method in SymbolTable class, the body of this method is replaced with the delegation to the new addSymbol method in Token class.

Impact on the system: now additions of new symbols to Symbol Table will be delegated and handled by Token class instead of SymbolTable class. This way the Feature Envy code smell will be removed, because the addSymbol method is more interested in Token class then in SymbolTable class.

Diff of the change:

```
Index: src/marf/nlp/Parsing/SymbolTable.java
================================================
==================
RCS file:
/groups/m/me_soen6471_1/cvs_repository/DMARF/src/marf/nlp/Par
sing/SymbolTable.java,v
retrieving revision 1.3
diff -u -r1.3 SymbolTable.java
--- src/marf/nlp/Parsing/SymbolTable.java	21 Aug 2014
01:08:34 -0000	1.3
+++ src/marf/nlp/Parsing/SymbolTable.java	21 Aug 2014
01:56:34 -0000
@@ -83,19 +83,7 @@
		*/
		public int addSymbol(Token poToken)
		{
-
		if(this.oSymTabEntries.contains(poToken.getLexeme()))
-			{
-				SymTabEntry oSymTabEntry =
(SymTabEntry)oSymTabEntries.get(poToken.getLexeme());
-
		oSymTabEntry.addLocation(poToken.getPosition());
-			}
-			else
-			{
-				SymTabEntry oSymTabEntry = new
SymTabEntry(poToken);
-
		oSymTabEntry.addLocation(poToken.getPosition());
-
		this.oSymTabEntries.put(poToken.getLexeme(),
oSymTabEntry);
-			}
-
-			return 0;
+			return poToken.addSymbol(oSymTabEntries);
		}

		/**
```

**Change 3/3: Add tests for addSymbol method from SymbolTable class.**

Reason for change: to test the code which was involved in refactoring to make sure that its external behavior stayed the same as before refactoring.

Impact on the system: created new package marf.junit.nlp.Parsing, added new class SymbolTableTest and added junit tests to test addSymbol method in SymbolTable class.

During unit testing we found that the original method addSymbol in SymbolTable class has a code defect (this wasn't changed by refactoring because refactoring is not supposed to change external behavior): the method tries to check if the given lexeme (lexeme is a Key in Hashtable) is mapped to some value (SymTabEntry) in the Hashtable. If it does, the location of the new Token is supposed to be added to the list of locations of the corresponding SymTabEntry to which the lexeme maps in the Hashtable. If it doesn't, a new entry is created in the Hashtable. The problem is in the following line: if(this.oSymTabEntries.contains(poToken.getLexeme())). The result of this conditional will always be false because the contains() method of Hashtable class checks if a given value, not a key, is present in the Hashtable. As a result, when adding a token whose lexeme is already present in the symbol table, its lexeme will always be remapped to a new SymTabEntry instead of adding its location to the list of locations of existing SymTabEntry. As a fix it is possible to replace contains() with containsKey() method.

Because of this defect in the original addSymbol method, there was no possibility to implement all the test cases proposed for this refactoring, because when adding into SymbolTable 2 tokens with different locations, but with the same lexeme, only the last location of the common lexeme will be present in the SymbolTable.

```
Index: src/marf/junit/nlp/Parsing/SymbolTableTest.java
================================================
```

```
==================
RCS file: src/marf/junit/nlp/Parsing/SymbolTableTest.java
diff -N src/marf/junit/nlp/Parsing/SymbolTableTest.java
--- /dev/null	1 Jan 1970 00:00:00 -0000
+++ src/marf/junit/nlp/Parsing/SymbolTableTest.java	1 Jan 1970 00:00:00 -0000
@@ -0,0 +1,53 @@
+package marf.junit.nlp.Parsing;
+
+import java.awt.Point;
+import java.util.Vector;
+
+import marf.nlp.Parsing.SymTabEntry;
+import marf.nlp.Parsing.SymbolTable;
+import marf.nlp.Parsing.Token;
+import marf.nlp.Parsing.TokenSubType;
+import junit.framework.TestCase;
+
+public class SymbolTableTest extends TestCase
+{
+	public static void main(String[] args)
+	{
+		junit.textui.TestRunner.run(SymbolTableTest.class);
+	}
+
+	public final void testAddSymbol()
+	{
+		String scopeName = "scope1";
+		SymbolTable symTab = new SymbolTable(scopeName);
+
+		String lexeme1 = "Hello World!";
+		String lexeme2 = "Hello World Again!";
+
+		Point point1 = new Point(10,20);
+		Point point2 = new Point(3,4);
+
+		Token token1 = new Token(lexeme1, point1, new TokenSubType());
+		Token token2 = new Token(lexeme1, point2, new TokenSubType());
+
+		// Check that addSymbol method returns 0, which is expected on success
+		assertTrue(symTab.addSymbol(token1) == 0);
+
+		// Check that token which was added is returned successfully from Symbol Table
+		SymTabEntry symTabEntry = symTab.getSymTabEntry(lexeme1);
+		assertTrue(symTabEntry != null);
+
+		// Negative test - Trying to get token for a wrong lexeme, expecting null
+		symTabEntry = symTab.getSymTabEntry("not present lexeme");
+		assertTrue(symTabEntry == null);
+
+		// Add token2 to Symbol Table
+		assertTrue(symTab.addSymbol(token2) == 0);
+
+		// Verify that SymTabEntry contains point2 of token2 among the list of locations
+		symTabEntry = symTab.getSymTabEntry(lexeme1);
+		Vector vec = symTabEntry.getLocationsList();
+		assertTrue(vec.contains(point2));
+	}
+
+}
```

**Type Checking code smell refactoring in DMARF**

**Change 0/4: Refactor ComplexMatrix class to resolve Type Checking code smell in getComplexMatrix method.**

The static method getComplexMatrix in class ComplexMatrix from package marf.math has a Type Checking code smell, because it checks the type of subclass using *instanceof* operator in a conditional expression and then according to the subclass type branches the program flow. The code smell is refactored using *Replace Conditional with Polymorphism* refactoring method.

Relation of the changes (patches) in the patch set: Change #1 adds new method getComplexMatrix to parent class (Matrix). Change #2 overrides the new method in subclass ComplexMatrix with ComplexMatrix-specific behavior. Change #3 fixes the code smell by replacing conditional with polymorphism using the new methods which were added in previous changes #1 and #2. Change #4 adds unit tests for the static getComplexMatrix method. This test will be used before and after refactoring to make sure that the behavior of the method doesn't change after refactoring.

**Change 1/4: Add getComplexMatrix method to Matrix class.**

Reason for change: To add new method getComplexMatrix to Matrix class. The new method will create and return a new ComplexMatrix object instantiated from *this* Matrix object. It will also serve as the base method for polymorphic calls. This change is needed for Type Checking code smell refactoring in static getComplexMatrix method of ComplexMatrix class.

Impact on the system: extended functionality of Matrix class with the ability to produce a ComplexMatrix from the current Matrix object. It will serve as the base for the next steps in the process of refactoring the Type Checking code smell.

Diff of the change:

```
Index: src/marf/math/Matrix.java
===================================================================
RCS file: /groups/m/me_soen6471_1/cvs_repository/DMARF/src/marf/math/Matrix.java,v
retrieving revision 1.5
diff -u -r1.5 Matrix.java
--- src/marf/math/Matrix.java 23 Aug 2014 02:26:25 -0000 1.5
+++ src/marf/math/Matrix.java	23 Aug 2014 02:28:13 -0000
@@ -2304,6 +2304,12 @@
				this.iDirection = piDirection;
			}
		}
+
```

```
+		public ComplexMatrix getComplexMatrix()
+		{
+			ComplexMatrix oMatrix = new
ComplexMatrix(this);
+			return oMatrix;
+		}
 }

 // EOF
```

**Change 2/4: Add implementation of getComplexMatrix method to ComplexMatrix class.**

Reason for change: To add new method getComplexMatrix to ComplexMatrix class which overrides the same method of its parent class Matrix with ComplexMatrix-specific behavior. This change is needed for Type Checking code smell refactoring in static getComplexMatrix method of ComplexMatrix class.

Impact on the system: created a new method getComplexMatrix in ComplexMatrix class. The method returns the calling instance of ComplexMatrix itself as the output.

```
Diff of the change:
Index: src/marf/math/ComplexMatrix.java
==================================================
==================
RCS file:
/groups/m/me_soen6471_1/cvs_repository/DMARF/src/marf/math/C
omplexMatrix.java,v
retrieving revision 1.1.1.1
diff -u -r1.1.1.1 ComplexMatrix.java
--- src/marf/math/ComplexMatrix.java	30 Jul 2014 13:44:49 -0000
	1.1.1.1
+++ src/marf/math/ComplexMatrix.java	22 Aug 2014
03:14:22 -0000
@@ -1303,6 +1303,10 @@
			// TODO Auto-generated method stub
			return super.transpose();
		}
+
+		public ComplexMatrix getComplexMatrix() {

+			return this;
+		}
 }

 // EOF
```

**Change 3/4: Replace conditional with polymorphism to fix Type Checking code smell.**

Reason for change: To fix the Type Checking code smell in the static getComplexMatrix method of ComplexMatrix class by replacing the type checking with polymorphism using the new getComplexMatrix method which was added to Matrix class and its subclasses.

Impact on the system: replaced type checking condition in the body of the static getComplexMatrix method of ComplexMatrix class and removed the code smell from the method.

```
Diff of the change:
Index: src/marf/math/ComplexMatrix.java
==================================================
==================
RCS file:
/groups/m/me_soen6471_1/cvs_repository/DMARF/src/marf/math/C
omplexMatrix.java,v
retrieving revision 1.2
diff -u -r1.2 ComplexMatrix.java
--- src/marf/math/ComplexMatrix.java	22 Aug 2014 03:15:47 -0000
	1.2
+++ src/marf/math/ComplexMatrix.java	22 Aug 2014
03:19:08 -0000
@@ -489,17 +489,7 @@
	 */
	public static ComplexMatrix getComplexMatrix(Matrix
poMatrix)
	{
-		ComplexMatrix oMatrix;
-
-		if(poMatrix instanceof ComplexMatrix)
-		{
-			oMatrix = (ComplexMatrix)poMatrix;
-		}
-		else
-		{
-			oMatrix = new
ComplexMatrix(poMatrix);
-		}
-
+		ComplexMatrix oMatrix =
poMatrix.getComplexMatrix();
		return oMatrix;
	}
```

**Change 4/4: Add junit tests for the static getComplexMatrix method from ComplexMatrix class.**

Reason for change: to test the static getComplexMatrix method from ComplexMatrix class to make sure that its external behavior before refactoring of Type Checking code smell is equal to its behavior after the refactoring.

Impact on the system: created new package marf.junit.math, added new class ComplexMatrixTest and added junit tests to test the static getComplexMatrix method from ComplexMatrix class. The tests cover the following: 1. Getting ComplexMatrix from Matrix class 2. Getting ComplexMatrix from Vector class. 3. Getting ComplexMatrix from ComplexMatrix class.

```
Diff of the change:
Index: src/marf/junit/math/ComplexMatrixTest.java
==================================================
==================
RCS file: src/marf/junit/math/ComplexMatrixTest.java
diff -N src/marf/junit/math/ComplexMatrixTest.java
--- /dev/null	1 Jan 1970 00:00:00 -0000
+++ src/marf/junit/math/ComplexMatrixTest.java1 Jan 1970
```

```
00:00:00 -0000
@@ -0,0 +1,92 @@
+package marf.junit.math;
+
+import junit.framework.TestCase;
+import marf.math.ComplexMatrix;
+import marf.math.Matrix;
+import marf.math.Vector;
+
+public class ComplexMatrixTest extends TestCase
+{
+	public static void main(String[] args)
+	{
+
	junit.textui.TestRunner.run(ComplexMatrixTest.class);
+	}
+
+	public final void testGetComplexMatrixFromMatrix()
+	{
+		//-----------------------------------------------------------
------
+		// Test for getting ComplexMatrix from Matrix
+		//-----------------------------------------------------------
------
+
+		final int MATRIX_ROWS = 13;
+		final int MATRIX_COLS = 17;
+
+		// Constructing and randomly filling a Matrix
+		Matrix matrix = new Matrix(MATRIX_ROWS,
MATRIX_COLS);
+		matrix.setAllRandom();
+
+		// Getting ComplexMatrix from the Matrix
+		ComplexMatrix complexMatrix =
ComplexMatrix.getComplexMatrix(matrix);
+
+		// Checking that ComplexMatrix's number of
rows is equal to source Matrix's number of rows
+		assertTrue(complexMatrix.getRows() ==
MATRIX_ROWS);
+
+		// Checking that ComplexMatrix's number of
columns is equal to source Matrix's number of columns
+		assertTrue(complexMatrix.getCols() ==
MATRIX_COLS);
+
+		// Verifying that ComplexMatrix's elements are
equal to the source Matrix's elements
+		for (int i=0; i<MATRIX_ROWS; i++)
+			for (int j=0; j<MATRIX_COLS; j++)
+
	assertTrue(complexMatrix.getElement(i, j) ==
matrix.getElement(i,j));
+	}
+
+	public final void testGetComplexMatrixFromVector()
+	{
+		//-----------------------------------------------------------
------
+		// Test for getting ComplexMatrix from Vector
+		//-----------------------------------------------------------
------
+
+		// Constructing and filling a Vector
+		final int VECTOR_LENGTH = 25;

+		Vector vec = new Vector(VECTOR_LENGTH);
+
+		for (int i=0; i<VECTOR_LENGTH; i++)
+		{
+			vec.setElement(i, i * 1.234);
+		}
+
+		// Getting ComplexMatrix from the Vector
+		ComplexMatrix complexMatrix =
ComplexMatrix.getComplexMatrix(vec);
+
+		// Checking that ComplexMatrix's number of
rows is equal to Vector's length
+		assertTrue(complexMatrix.getRows() ==
VECTOR_LENGTH);
+
+		// Checking that ComplexMatrix's number of
columns is equal to 1
+		assertTrue(complexMatrix.getCols() == 1);
+
+		// Verifying that Vector's elements are equal to
the ComplexMatrix's elements
+		for (int i=0; i<VECTOR_LENGTH; i++)
+		{
+
	assertTrue(complexMatrix.getElement(i, 0) ==
vec.getElement(i));
+		}
+	}
+
+	public final void
testGetComplexMatrixFromComplexMatrix()
+	{
+		//-----------------------------------------------------------
------
+		// Test for getting ComplexMatrix from
ComplexMatrix
+		//-----------------------------------------------------------
------
+
+		// Creating new ComplexMatrix
+		ComplexMatrix complexMatrix1 = new
ComplexMatrix();
+
+		// Randomly filling the ComplexMatrix
+		complexMatrix1.setAllRandom();
+
+		// Calling getComplexMatrix method on the
ComplexMatrix
+		ComplexMatrix complexMatrix2 =
ComplexMatrix.getComplexMatrix(complexMatrix1);
+
+		// Verifying that complexMatrix1 and
complexMatrix2 are equal
+
	assertTrue(complexMatrix1.equals(complexMatrix2));
+	}
+}
```

*2) GIPSY*

**Replacing Conditional Type Checking with Polymorphic Method Calls:**

**Change 0/3: Refactoring JINITA class to employ polymorphic method call in place of conditional type checking**:

We implement here a refactoring of the Type Checking code smell found in GIPSY's JINITA class, as described in Planned Refactorings section above. In the writeValue() of JINITA class located in package gipsy.GEE.IDP.DemandGenerator.jini.rmi, the developer is using a conditional block to branch depending on what the run-time type of a certain IDemand interface object (passed as parameter to the writeValue()) happens to be. In JINITA class's writeValue() method, an IDemand object's run-time subtype needs to be determined as to which subtype it is: one of Demand (implementor of IDemand interface) subtypes: SystemDemand, ProceduralDemand, IntensionalDemand, or ResourceDemand.

Such scenario brings to mind polymorphism, a fundamental concept of OO paradigm, the motivation of which is to address a solution to this scenario: give the same name different operations, call the appropriate operation depending on the calling object's type at run-time. Below we describe the actual steps taken to replace conditional type checking with polymorphic method calls.

**Change 1/3: Replacing conditional block with polymorphic method call to writeValue()**:

Reason for change: in order to replace the conditional-based type checking of object type (as a prequel to subsequent execution branch at run-time) with polymorphic method calls. This adheres to a fundamental OO design principle and improves reusability (more concrete implementation methods can be added in the future without requiring changes the calling class(s), in this case JINITA).

Impact on the system: the type of Demand object in JINITA class is determined dynamically at run-time and the corresponding implementation of writeValue() method is polymorphically called.

Diff of the changes:

```
### Eclipse Workspace Patch 1.0
#P GIPSY
Index: src/gipsy/GEE/IDP/DemandGenerator/jini/rmi/JINITA.ja
================================================
================
RCS file:
/groups/m/me_soen6471_1/cvs_repository/GIPSY/src/gipsy/GE
DemandGenerator/jini/rmi/JINITA.java,v
retrieving revision 1.1.1.1
diff -u -r1.1.1.1 JINITA.java
--- src/gipsy/GEE/IDP/DemandGenerator/jini/rmi/JINITA.java
        5 Aug 2014 18:02:53 -0000        1.1.1.1
+++ src/gipsy/GEE/IDP/DemandGenerator/jini/rmi/JINITA.java
        23 Aug 2014 18:04:05 -0000
@@ -38,7 +38,7 @@
  * @author Serguei Mokhov
  * @author Yi Ji
  *
- * @version $Id: JINITA.java,v 1.51 2012/06/17 16:58:55 mok
$
+ * @version $Id$
  */
 public class JINITA
 implements Serializable, ITransportAgent
@@ -454,30 +454,7 @@
                        // If it is a pending demand
                        if(poState.equals(DemandState.PENI
                        {
-                               if(poDemand instanceof
SystemDemand)
-                               {
-                                       oEntry.strDestinati
(String) ((SystemDemand)poDemand).getDestinationTierID();
-                               }
-                               else if(poDemand instanceo
ProceduralDemand)
-                               {
-                                       oEntry.strDestinati
DemandSignature.DWT;
-                               }
-                               else if(poDemand instanceo
IntensionalDemand)
-                               {
-                                       oEntry.strDestinati
DemandSignature.DGT;
-                               }
-                               else if(poDemand instanceo
ResourceDemand)
-                               {
-                                       oEntry.strDestinati
DemandSignature.ANY_DEST;
-                               }
-                               else
-                               {
-                                       /*
-                                        * Treat unknown d
as a procedural demand by default
                                         * for backward
-
compatibility.
-                                        */
-                                       oEntry.strDestinati
DemandSignature.DWT;
-                               }
+                               poDemand.writeValue(oEnt
                        }

                        this.oJavaSpace.write(oEntry, null,
```

```
Lease.FOREVER);
```

**Change 2/3: Declaring an abstract writeValue () method in IDemand interface:**

Reason for change: to force each subclass of IDemand to provide a concrete implementation of the writeValue() logic.

Impact on the system: ensures no run-type exception is thrown as a result of missing concrete implementation of writeValue() method, when called polymorphically from JINITA class.

Diff of the change:

```
### Eclipse Workspace Patch 1.0
#P GIPSY
Index: src/gipsy/GEE/IDP/demands/IDemand.java
=============================================
===============
RCS file:
/groups/m/me_soen6471_1/cvs_repository/GIPSY/src/gipsy/GE
mands/IDemand.java,v
retrieving revision 1.1.1.1
diff -u -r1.1.1.1 IDemand.java
--- src/gipsy/GEE/IDP/demands/IDemand.java	5 Aug 201
18:02:54 -0000	1.1.1.1
+++ src/gipsy/GEE/IDP/demands/IDemand.java	23 Aug 20
18:09:24 -0000
@@ -1,8 +1,8 @@
 package gipsy.GEE.IDP.demands;

+import gipsy.GEE.IDP.DemandGenerator.jini.rmi.JiniDispatch
 import gipsy.interfaces.ISequentialThread;
 import gipsy.lang.GIPSYContext;
-
 import java.io.Serializable;
 import java.util.Date;

@@ -13,7 +13,7 @@
  * @author Emil Vassev
  * @author Serguei Mokhov
  * @since 1.0.0
- * @version 2.0.0, $Id: IDemand.java,v 1.18 2010/09/09 18:21:
mokhov Exp $
+ * @version 2.0.0, $Id$
  */
 public interface IDemand
 extends ISequentialThread, Cloneable
@@ -194,6 +194,8 @@
	//void storeResult(String pstrDirName);
	//DemandSignature storeResult(Object poResult);
	DemandSignature storeResult(Serializable poResult);
+
+	public abstract void writeValue(JiniDispatcherEntry oE
 }

 // EOF
```

**Change 3/3: Impementing writeValue () in IDemand's subtypes :**

Reason for change: to implement the concrete logic writeValue() in each of Demand subtype. These are the methods that are called polymorphically from JINITA class. The content of a writeValue() method in a given class corresponds to the logic originally in the corresponding conditional block before refactoring.

Impact on the system: whatever the run-time type of a Demand object happens to be in JINITA's writeValue() method, the corresponding polymorphically-called writeValue in the corresponding Demand object will be available. The system risks stumbling upon a run-time error if this implementation was missing.

Demand's subtypes (SystemDemand, ProceduralDemand, IntensionalDemand, or ResourceDemand) were similarly altered, whereby each subtype's writeValue() method contains the logic from the corresponding conditional block before refactoring.

Diff of the changes:

```
### Eclipse Workspace Patch 1.0
#P GIPSY
Index: src/gipsy/GEE/IDP/demands/Demand.java
==================================================
==================
RCS file:
/groups/m/me_soen6471_1/cvs_repository/GIPSY/src/gipsy/GEE/ID
P/demands/Demand.java,v
retrieving revision 1.1.1.1
diff -u -r1.1.1.1 Demand.java
--- src/gipsy/GEE/IDP/demands/Demand.java	5 Aug 2014
18:02:54 -0000	1.1.1.1
+++ src/gipsy/GEE/IDP/demands/Demand.java	23 Aug 2014
18:10:19 -0000
@@ -1,11 +1,10 @@
 package gipsy.GEE.IDP.demands;

+import
gipsy.GEE.IDP.DemandGenerator.jini.rmi.JiniDispatcherEntry;
 import gipsy.lang.GIPSYContext;
-
 import java.io.Serializable;
 import java.lang.reflect.Method;
 import java.util.Date;
-
 import marf.util.FreeVector;


@@ -435,6 +434,10 @@

			return strDemandData + "\n" + super.toString();
		}
+
```

```
+		public void writeValue(JiniDispatcherEntry oEntry) {
+			oEntry.strDestination = DemandSignature.DWT;
+		}
 }

 // EOF
Index: src/gipsy/GEE/IDP/demands/IntensionalDemand.java
===================================================================
RCS file:
/groups/m/me_soen6471_1/cvs_repository/GIPSY/src/gipsy/GEE/IDP/demands/IntensionalDemand.java,v
retrieving revision 1.1.1.1
diff -u -r1.1.1.1 IntensionalDemand.java
--- src/gipsy/GEE/IDP/demands/IntensionalDemand.java	5 Aug 2014 18:02:54 -0000	1.1.1.1
+++ src/gipsy/GEE/IDP/demands/IntensionalDemand.java	23 Aug 2014 18:10:19 -0000
@@ -1,5 +1,6 @@
 package gipsy.GEE.IDP.demands;

+import gipsy.GEE.IDP.DemandGenerator.jini.rmi.JiniDispatcherEntry;
 import gipsy.lang.GIPSYContext;
 import gipsy.lang.GIPSYIdentifier;

@@ -12,7 +13,7 @@
  *
  * @author Bin Han
  * @author Serguei Mokhov
- * @version $Id: IntensionalDemand.java,v 1.12 2010/12/06 13:38:39 mokhov Exp $
+ * @version $Id$
  * @since
  */
 public class IntensionalDemand
@@ -88,6 +89,10 @@
			// TODO Auto-generated method stub
			return 0;
		}
+
+		public void writeValue(JiniDispatcherEntry oEntry) {
+			oEntry.strDestination = DemandSignature.DGT;
+		}
 }

 // EOF
Index: src/gipsy/GEE/IDP/demands/ProceduralDemand.java
===================================================================
RCS file:
/groups/m/me_soen6471_1/cvs_repository/GIPSY/src/gipsy/GEE/IDP/demands/ProceduralDemand.java,v
retrieving revision 1.1.1.1
diff -u -r1.1.1.1 ProceduralDemand.java
--- src/gipsy/GEE/IDP/demands/ProceduralDemand.java	5 Aug 2014 18:02:54 -0000	1.1.1.1
+++ src/gipsy/GEE/IDP/demands/ProceduralDemand.java	23 Aug 2014 18:10:19 -0000
@@ -1,9 +1,9 @@
 package gipsy.GEE.IDP.demands;


+import gipsy.GEE.IDP.DemandGenerator.jini.rmi.JiniDispatcherEntry;
 import gipsy.lang.GIPSYIdentifier;
 import gipsy.lang.GIPSYType;
-
 import java.io.Serializable;


@@ -17,7 +17,7 @@
  *
  * @author Bin Han
  * @author Serguei Mokhov
- * @version $Id: ProceduralDemand.java,v 1.16 2012/06/16 03:10:37 mokhov Exp $
+ * @version $Id$
  * @since
  */
 public class ProceduralDemand
@@ -126,6 +126,10 @@

			return strDemandData + "\n" + super.toString();
		}
+
+		public void writeValue(JiniDispatcherEntry oEntry) {
+			oEntry.strDestination = DemandSignature.DWT;
+		}
 }

 // EOF
Index: src/gipsy/GEE/IDP/demands/ResourceDemand.java
===================================================================
RCS file:
/groups/m/me_soen6471_1/cvs_repository/GIPSY/src/gipsy/GEE/IDP/demands/ResourceDemand.java,v
retrieving revision 1.1.1.1
diff -u -r1.1.1.1 ResourceDemand.java
--- src/gipsy/GEE/IDP/demands/ResourceDemand.java	5 Aug 2014 18:02:54 -0000	1.1.1.1
+++ src/gipsy/GEE/IDP/demands/ResourceDemand.java	23 Aug 2014 18:10:19 -0000
@@ -1,5 +1,6 @@
 package gipsy.GEE.IDP.demands;

+import gipsy.GEE.IDP.DemandGenerator.jini.rmi.JiniDispatcherEntry;
 import gipsy.interfaces.GEERSignature;
 import gipsy.interfaces.ResourceSignature;

@@ -9,7 +10,7 @@
  *
  * @author Bin Han
  * @author Serguei Mokhov
- * @version $Id: ResourceDemand.java,v 1.11 2013/01/07 19:16:07 mokhov Exp $
+ * @version $Id$
  */
 public class ResourceDemand
 extends Demand
@@ -132,6 +133,10 @@
			// TODO Auto-generated method stub
			return super.toString();
		}
+
+		public void writeValue(JiniDispatcherEntry oEntry) {
```

```
+                       oEntry.strDestination =
DemandSignature.ANY_DEST;
+               }
 }

 // EOF
Index: src/gipsy/GEE/IDP/demands/SystemDemand.java
===================================================================
RCS file:
/groups/m/me_soen6471_1/cvs_repository/GIPSY/src/gipsy/GEE/ID
P/demands/SystemDemand.java,v
retrieving revision 1.1.1.1
diff -u -r1.1.1.1 SystemDemand.java
--- src/gipsy/GEE/IDP/demands/SystemDemand.java         5 Aug
2014 18:02:54 -0000             1.1.1.1
+++ src/gipsy/GEE/IDP/demands/SystemDemand.java        23 Aug
2014 18:10:19 -0000
@@ -1,7 +1,7 @@
 package gipsy.GEE.IDP.demands;

+import
gipsy.GEE.IDP.DemandGenerator.jini.rmi.JiniDispatcherEntry;
 import gipsy.util.NotImplementedException;
-
 import java.io.Serializable;


@@ -10,7 +10,7 @@
  *
  * @author Bin Han
  * @author Serguei Mokhov
- * @version $Id: SystemDemand.java,v 1.10 2013/01/07 19:16:08
mokhov Exp $
+ * @version $Id$
  * @since
  */
 public class SystemDemand
@@ -158,6 +158,10 @@
                {
                        this.iPriority = piPriority;
                }
+
+               public void writeValue(JiniDispatcherEntry oEntry) {
+                       oEntry.strDestination = (String)
this.getDestinationTierID();
+               }
 }

 // EOF
```

***Populating Empty* catch () *Blocks with Minimal Feedback Messages:***

**Change 0/2: The 'Empty catch () block' code smell in GIPSY**:

The principal of mandatory enforcement of exception handling adopted in Java's original implementation is to provide useful and human-readable feedback that improves testing and maintainability efforts. Exception handling helps developers/maintainers under the context and instigating factors leading to an exception. Furthermore, stakeholders with limited knowledge of the low-level implementation details (e.g. integration analysts and developers) necessarily need useful exception feedback in order to begin to address the issue.

**Change 1/2: The Identification of Empty catch () blocks in GIPSY**:

Using Robusta plugin in Eclipse [26], 40 empty catch blocks have been identified. Ideally each catch() block provides information relevant and specific to the context in which the exception was raised. To refactor such an issue, we introduce a call to printStackTrace() method, which is part of Java's java.lang.Throwable class. The augmentation or replacement of System.err message with information relevant to the context of each class is beyond the scope of this project, as it requires detailed understanding of the purpose and logic of each of the 40 blocks.

In every identified empty catch block (as hinted by Eclipse's plugin JDeodorant [5]), a minimal code is introduced to provide feedback in the event of an exception being raised. A detailed context-conscious message of each exception is beyond the scope of this project as it requires knowledge of the details of each class in which such code smell occurs. However, as a minimal refactoring, we propose using Java's Throwable calss printStackTrace() method which prints the (throwable) object and its back-trace to the standard error stream. Hence, in each empty block we propose adding:

[*throwable-object*].printStackTrace();

where [throwable-object] is a the Exception object being passed to the catch block in a given context.

**Change 2/2: supplementing a feedback message in each empty catch block**:

Reason for change: to provide a minimal feedback message in the event of an exception being caught. This meets the minimum requirement of satisfying the basic principle of exception handling, thus enhancing reliability and maintainability of the system as a whole. A more context-specific exception handling of the exception would ideal, providing class-, module-, and system-level explanation of the context and the instigations leading to the exception. However, such an ambitious refactoring is beyond the scope of this project as it requires in-depth and in-breadth knowledge of GIPSY.

Impact on the system: this refactoring should improve the reliability of the system by meeting minimal expectations of exception handling principle.

A total of 40 refactored empty catch () blocks have been refactored by embedding a method call to Java's Exception (or any of its descendants) printStackTrace() method by the Exception object thrown into the catch () block. It should be noted however, that there are two classes in particular, JINITA.java and JMSTRansportAgent, in which GIPSY's developers have made explicit documentation in the form of comments advising not to handle the exception, and therefore we have obliged and left them unchanged.

Diff of the change: a snippet is shown here, the full patch can be found in the Appendix, as well as under the 'patchset' folder submitted along with the project report and repo's tar.

```
### Eclipse Workspace Patch 1.0
#P GIPSY
Index:
src/gipsy/GEE/IDP/DemandGenerator/jini/rmi/JINITransportAgent.j
ava
===================================================
==================
RCS file:
/groups/m/me_soen6471_1/cvs_repository/GIPSY/src/gipsy/GEE/ID
P/DemandGenerator/jini/rmi/JINITransportAgent.java,v
retrieving revision 1.2
diff -u -r1.2 JINITransportAgent.java
---
src/gipsy/GEE/IDP/DemandGenerator/jini/rmi/JINITransportAgent.j
ava        22 Aug 2014 19:57:46 -0000  1.2
+++
src/gipsy/GEE/IDP/DemandGenerator/jini/rmi/JINITransportAgent.j
ava        24 Aug 2014 04:05:46 -0000
@@ -159,6 +159,7 @@
                                }
                                catch (InterruptedException
ex)
                                {
+
                ex.printStackTrace();
                                }
                        }
                }
@@ -750,6 +751,7 @@
        }
                        catch(InterruptedException ex)
                        {
+                               ex.printStackTrace();
        }
    }
  }
Index: src/gipsy/GIPC/DFG/DFGAnalyzer/LucidCodeGenerator.java
===================================================
==================
RCS file:
/groups/m/me_soen6471_1/cvs_repository/GIPSY/src/gipsy/GIPC/D
FG/DFGAnalyzer/LucidCodeGenerator.java,v
retrieving revision 1.2
diff -u -r1.2 LucidCodeGenerator.java
--- src/gipsy/GIPC/DFG/DFGAnalyzer/LucidCodeGenerator.java
        22 Aug 2014 19:57:46 -0000  1.2
+++ src/gipsy/GIPC/DFG/DFGAnalyzer/LucidCodeGenerator.java
        24 Aug 2014 04:05:46 -0000
@@ -396,7 +396,7 @@
      try {
        rf1.writeBytes(LCnode.Scope_label.substring(9)+": ");
      }
-     catch(Exception e) {}
+      catch(Exception e) {e.printStackTrace();}
   }


@@ -427,7 +427,7 @@
       try {
         rf1.writeBytes(prtspace2+"(");
       }
-      catch(Exception e) {}
+       catch(Exception e) {e.printStackTrace();}

      }

@@ -448,7 +448,7 @@
         try {
           rf1.writeBytes(prtspace2+ifstr);
         }
-        catch(Exception e) {}
+         catch(Exception e) {e.printStackTrace();}


        }
@@ -470,7 +470,7 @@
```

•

•

•

•

•

•

*[The rest of the diff can be found in the Appendix, as well as under the 'patchset' directory submitted with the final report and repo's tar]*

•

•

*[A total of 40 catch() blocks are patched, in 13 classes]*

## VII. Conclusion

This study has been a hands-on learning experience through which we got to apply software architecture principles on two large OSS systems: DMARF and GIPSY. It began with series of investigations into the background, motivations and high-level architecture of the two systems. After having done a thorough research literature review of the two systems, we were able to elucidate on various aspects of their respective requirements and design specifications, including the identification of actors and stakeholders, some use cases, and domain models. We have learned a great deal during this phase since we are working on living and breathing systems that have already been realized, and as such we got to place ourselves back in time in the shoes of the original developers. This phase further prepared us to move on to the next phase, in which we dug deeper in the implementation (the *gut*) of both systems. This latter phase was a challenge. It required the ability to link concrete implementation components, with previously learned design components, and –more importantly– succeed in identifying legitimate and potentially useful areas of improvements. To this end we employed a myriad of software dynamic and static code analysis tools, which in and of itself was a great learning experience. We do hope that the insights we provided into the two systems can be of value to the original developers of DMARF and GIPSY, and that the refactorings we suggested and implemented can be of use.

## <u>Group ID</u>: mes64711

| Student name | Paper | |
|---|---|---|
| | **DMARF** | **GIPSY** |
| Dhivyaa Nandakumar | Towards autonomic specification of Distributed MARF with ASSL: Self-healing | Towards autonomic GIPSY |
| Srikanth Suryadevara | Self-Forensics Through Case Studies of Small to Medium Software Systems | Advances in the Design and Implementation of Multi-tier Architecture in GIPSY Environment |
| Riya Ray | Self-Optimization Property in Autonomic Specification of Distributed MARF with ASSL | Towards a Self-Forensics Property in the ASSL Toolset |
| Ali Alshamrani | On Design and Implementation of Distributed Modular Audio Recognition Framework | The GIPSY Architecture |
| Dmitriy Fingerman | Managing Distributed MARF with SNMP | Unifying and refactoring DMF to support concurrent Jini and JMS DMS in GIPSY |
| Aaradhna Goyal | Autonomic specification of self-protection for Distributed MARF with ASSL. | Distributed eductive execution of hybrid intensional programs. |
| Dileep Vanga | Towards Security Hardening of Scientific Demand-Driven and Pipelined Distributed Computing Systems | Using the General Intensional Programming System (GIPSY) for Evaluation of Higher-Order Intensional Logic (HOIL) Expressions |
| Parul Gupta | Distributed Modular Audio Recognition Framework (DMARF) and its Applications Over Web Services | An Interactive Graph-Based Automation Assistant: A Case Study to Manage the GIPSY's Distributed Multi-tier Run-Time System |

## Individual contributions on Design Patterns:

**DMARF**

| S. No. | Design Pattern | Person Responsible |
|---|---|---|
| 1. | Strategy | Dmitriy Fingerman |
| 2. | Adapter | Riya Ray |
| 3. | Decorator | Aaradhna Goyal |
| 4. | Singleton | Dileep Vanga |

**GIPSY**

| S. No. | Design Pattern | Person Responsible |
|---|---|---|
| 1. | Factory, Abstract Factory | Ali Alshamrani |
| 2. | Observer | Parul Gupta |
| 3. | Prototype | Dhivyaa Nandakumar |
| 4. | Proxy | Srikanth Suryadevara |

# APPENDIX A

1. The Metrics (v. 1.3.6) plugin is basically a metrics calculation and dependency analyzer for the Eclipse platform. It measures various metrics with average and standard deviation and also detects cycles in packages and type dependencies and graphs them. [17]

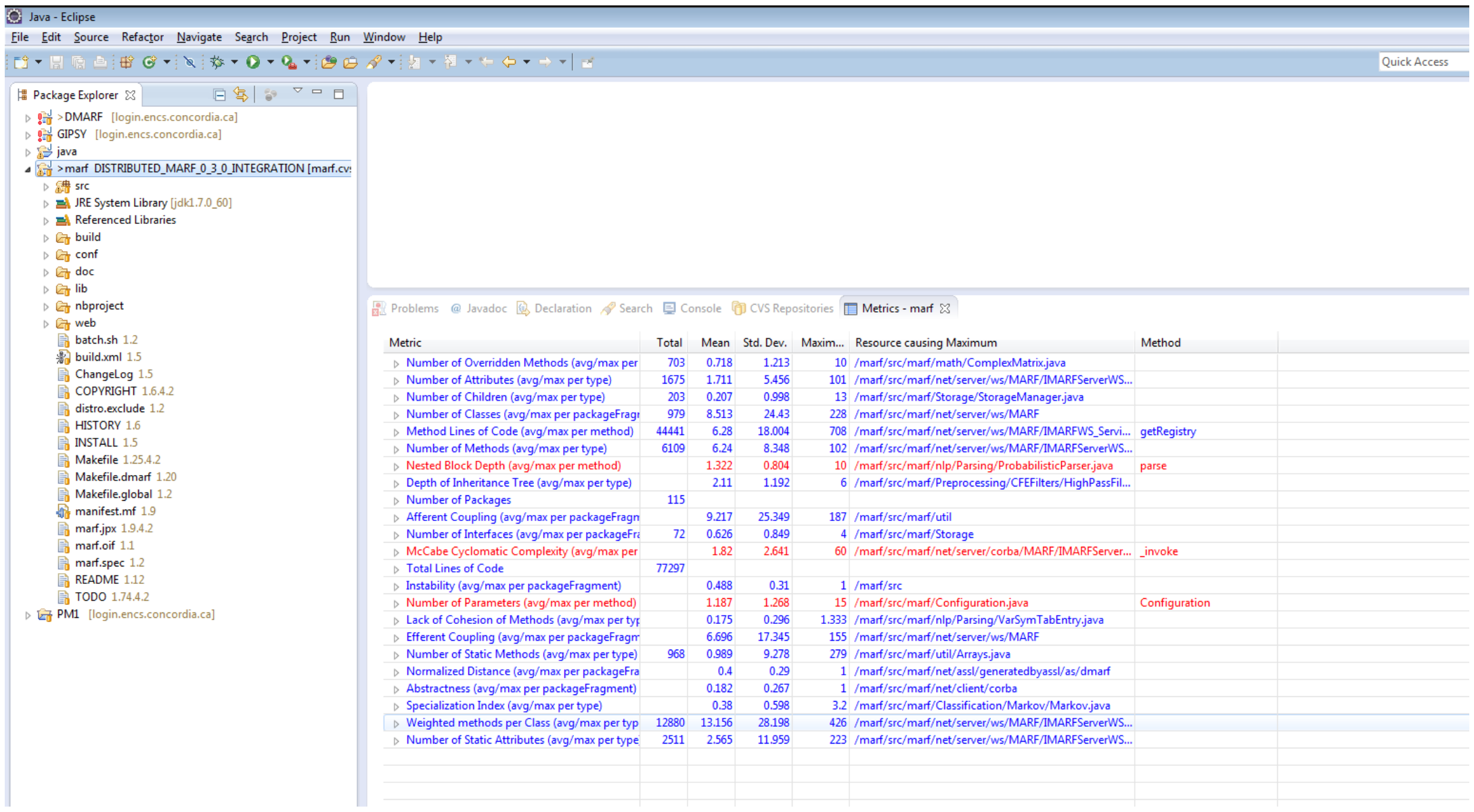

**Figure 49- Screenshot of metric plugin**

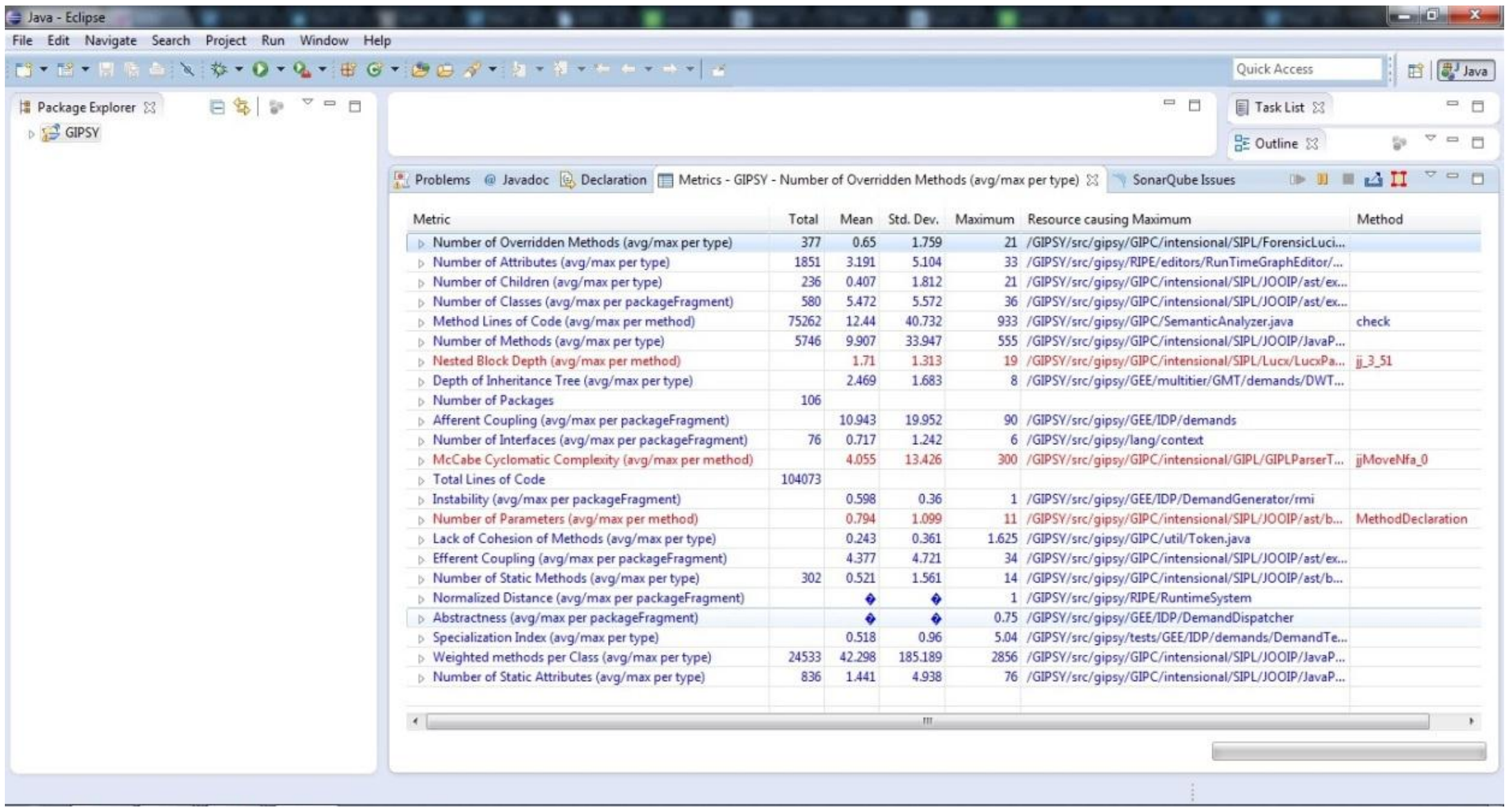

**Figure 50 - Screenshot for metrics plugin for GIPSY**

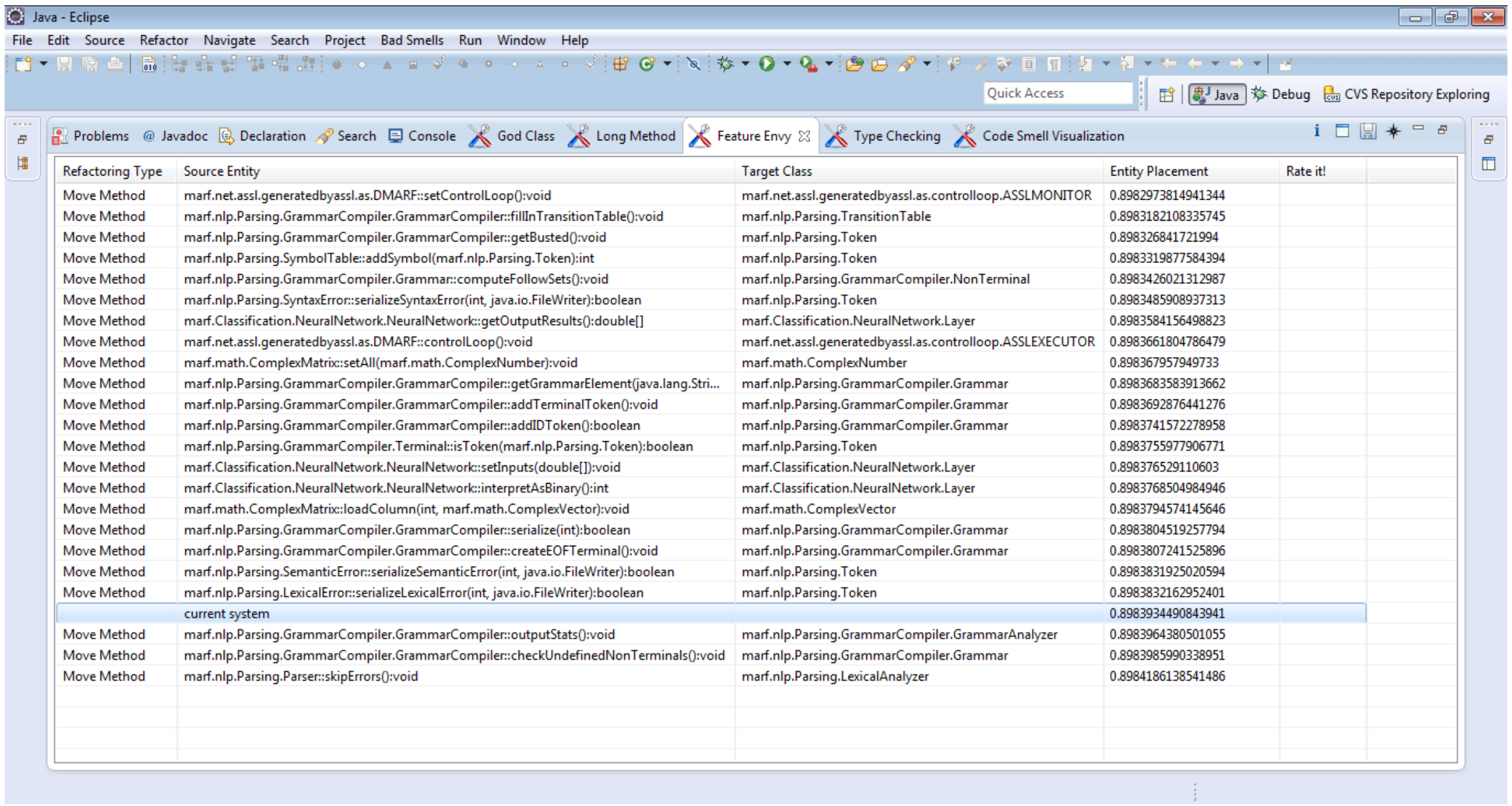

**Figure 51 - A snapshot showing JDeodorant's Feature Envy feature. Each listing indicates the position in the code of the code smell, as well as the proposed refactoring method.**

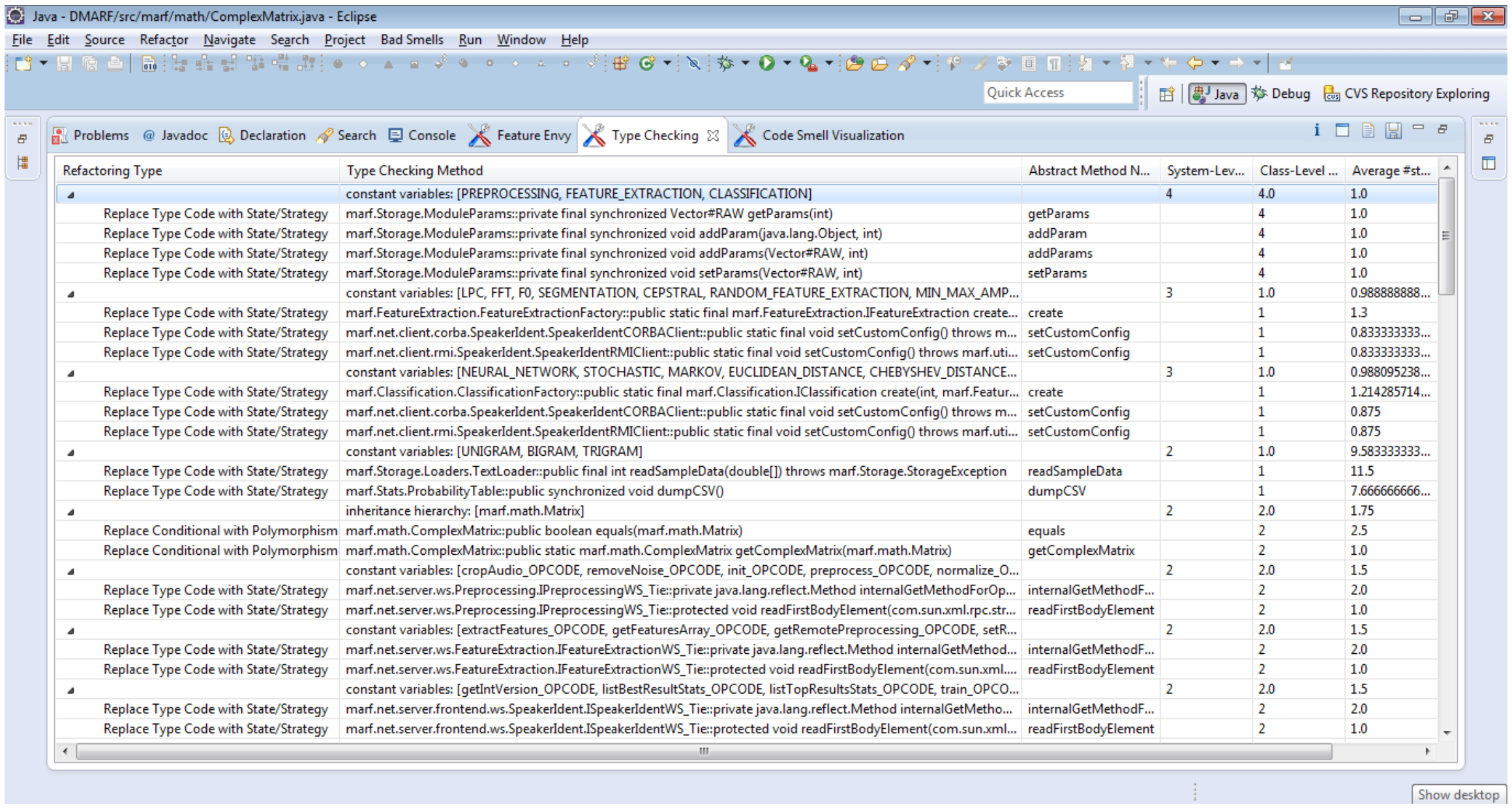

**Figure 52 - A snapshot showing JDeodorant's Type Checking feature. Each listing indicates the position in the code of the code smell, as well as the proposed refactoring method.**

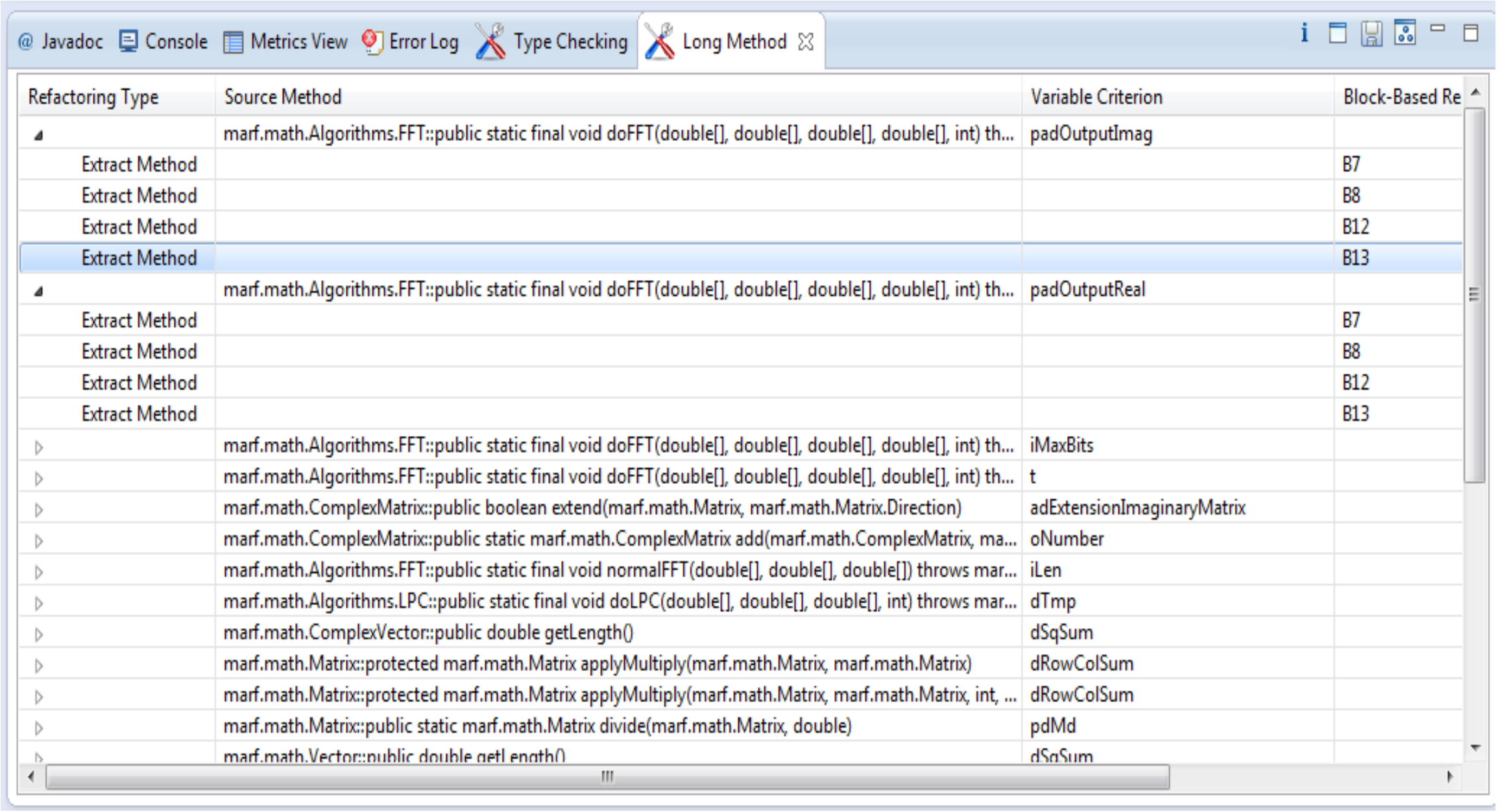

**Figure 53 - A snapshot showing JDeodorant's Long Method feature. Each listing indicates the position in the code of the code smell, as well as the proposed refactoring method.**

2. The LINUX commands used to measure the Number of Java files, Number of Classes and Java Lines of Code.

- Number of java files: find . –name "*.java" | wc –l
- Number of classes: find . –name "*.java" | xargs grep class | grep –v ".class"  | wc –l
- Number of lines of code: cat `find . –name "*.java" ` | wc –l

3. Fused DMARF and GIPSY Use case:

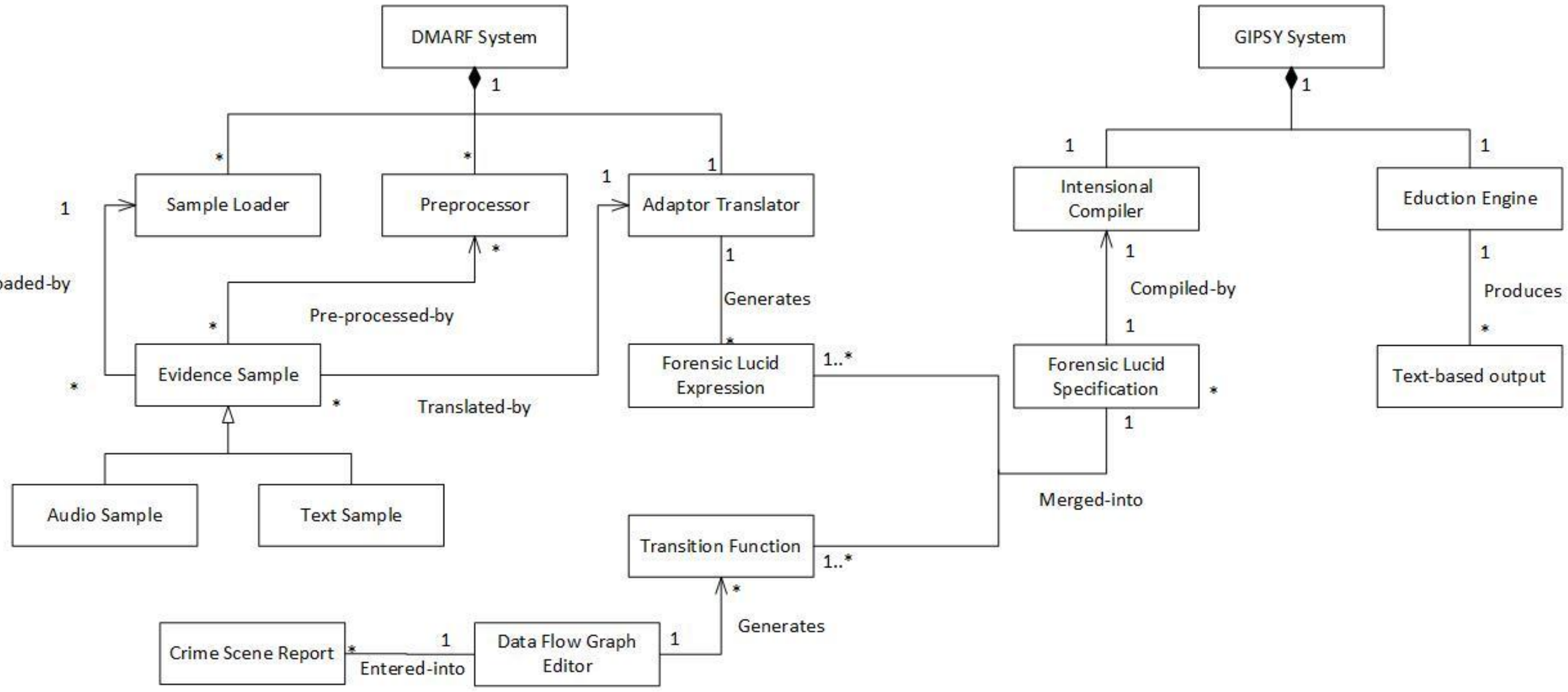

**Figure 54. Enlarged Fused DMARF and GIPSY use case**

The full Diff of Catch () block refactoring in section

```
### Eclipse Workspace Patch 1.0
#P GIPSY
Index: src/gipsy/GEE/IDP/DemandGenerator/jini/rmi/JINITransportAgent.java
===================================================================
RCS file:
/groups/m/me_soen6471_1/cvs_repository/GIPSY/src/gipsy/GEE/IDP/DemandGenerator/jini/rmi/JINITran
sportAgent.java,v
retrieving revision 1.2
diff -u -r1.2 JINITransportAgent.java
--- src/gipsy/GEE/IDP/DemandGenerator/jini/rmi/JINITransportAgent.java    22 Aug 2014 19:57:46 -
0000   1.2
+++ src/gipsy/GEE/IDP/DemandGenerator/jini/rmi/JINITransportAgent.java    24 Aug 2014 04:05:46 -
0000
@@ -159,6 +159,7 @@
                          }
                          catch (InterruptedException ex)
                          {
+                               ex.printStackTrace();
                          }
                   }
            }
@@ -750,6 +751,7 @@
             }
                   catch(InterruptedException ex)
                   {
+                        ex.printStackTrace();
             }
         }
     }
Index: src/gipsy/GIPC/DFG/DFGAnalyzer/LucidCodeGenerator.java
===================================================================
```

```
RCS file:
/groups/m/me_soen6471_1/cvs_repository/GIPSY/src/gipsy/GIPC/DFG/DFGAnalyzer/LucidCodeGenerator.j
ava,v
retrieving revision 1.2
diff -u -r1.2 LucidCodeGenerator.java
--- src/gipsy/GIPC/DFG/DFGAnalyzer/LucidCodeGenerator.java  22 Aug 2014 19:57:46 -0000 1.2
+++ src/gipsy/GIPC/DFG/DFGAnalyzer/LucidCodeGenerator.java  24 Aug 2014 04:05:46 -0000
@@ -396,7 +396,7 @@
        try {
          rf1.writeBytes(LCnode.Scope_label.substring(9)+": ");
        }
-       catch(Exception e) {}
+       catch(Exception e) {e.printStackTrace();}
      }


@@ -427,7 +427,7 @@
         try {
           rf1.writeBytes(prtspace2+"(");
         }
-        catch(Exception e) {}
+        catch(Exception e) {e.printStackTrace();}

       }

@@ -448,7 +448,7 @@
             try {
               rf1.writeBytes(prtspace2+ifstr);
             }
-            catch(Exception e) {}
+            catch(Exception e) {e.printStackTrace();}


           }
@@ -470,7 +470,7 @@
                 try {
                   rf1.writeBytes(prtspace2+opnode.pop());
                 }
-                catch(Exception e) {}
+                catch(Exception e) {e.printStackTrace();}
               }
             }
             gencode(n2, prtspace); // process each child
@@ -482,7 +482,7 @@
               try {
                 rf1.writeBytes(prtspace2+"fi");
               }
-              catch(Exception e) {}
+              catch(Exception e) {e.printStackTrace();}

             }
             if(i==1&&!opnode.peek().equals("START")) { //first node
@@ -493,7 +493,7 @@
                   rf1.writeBytes("\n"+prtspace2+opnode.pop()+"\n");
                   newline=1;
                 }
-                catch(Exception e) {}
+                catch(Exception e) {e.printStackTrace();}
               }
               else {

@@ -507,7 +507,7 @@
                 try {
                   rf1.writeBytes(prtspace2+opnode.pop());
                 }
```

```
-                 catch(Exception e) {}
+                 catch(Exception e) {e.printStackTrace();}

               }
             }
@@ -532,7 +532,7 @@
         try {
           rf1.writeBytes(prtspace2+")");
         }
-        catch(Exception e) {}
+        catch(Exception e) {e.printStackTrace();}

       }

@@ -553,7 +553,7 @@
             rf1.writeBytes(";\n");
             newline=1;
           }
-          catch(Exception e) {}
+          catch(Exception e) {e.printStackTrace();}

         }
       }
@@ -569,7 +569,7 @@
         try {
           rf1.writeBytes(prtspace2+"end");
         }
-        catch(Exception e) {}
+        catch(Exception e) {e.printStackTrace();}
       }
     }

@@ -599,7 +599,7 @@
         }

       }
-      catch(Exception e) {}
+      catch(Exception e) {e.printStackTrace();}

     }

@@ -906,7 +906,7 @@
       rf0.close();
       rf1=new RandomAccessFile(outfilename, "rw");
     }
-    catch(Exception e) {}
+    catch(Exception e) {e.printStackTrace();}

     genTable();
     genTableclose();
Index: src/gipsy/GIPC/DFG/DFGGenerator/DFGTranCodeGenerator.java
===================================================================
RCS file:
/groups/m/me_soen6471_1/cvs_repository/GIPSY/src/gipsy/GIPC/DFG/DFGGenerator/DFGTranCodeGenerato
r.java,v
retrieving revision 1.2
diff -u -r1.2 DFGTranCodeGenerator.java
--- src/gipsy/GIPC/DFG/DFGGenerator/DFGTranCodeGenerator.java      22 Aug 2014 19:57:48 -0000 1.2
+++ src/gipsy/GIPC/DFG/DFGGenerator/DFGTranCodeGenerator.java      24 Aug 2014 04:05:46 -0000
@@ -88,7 +88,7 @@
       SKrankmax.push(toname);
       toname = "w" + toname;
     }
-    catch (Exception e) {}
+    catch (Exception e) {e.printStackTrace();}
```

```
      drawRankgroup();

      try {
@@ -104,7 +104,7 @@
        rf1.writeBytes("color = hot_pink;\n");
        rf1.writeBytes("fontcolor = hot_pink;\n\n");
      }
-     catch (Exception e) {}
+     catch (Exception e) {e.printStackTrace();}
    }

    public void drawWhere() {
@@ -121,7 +121,7 @@
        this.rf1.writeBytes("\nsubgraph cluster" + itos(ncluster) + "{\n");
        this.rf1.writeBytes("node [shape=box];\n\n");
      }
-     catch (Exception e) {}
+     catch (Exception e) {e.printStackTrace();}
    }

    public void drawAssignLable(String image, String outpoint) {
@@ -140,7 +140,7 @@
        }
      }

-     catch (Exception e) {}
+     catch (Exception e) {e.printStackTrace();}
    }

    public void drawOperator(SimpleNode n, String oper, String filename1) {
@@ -165,7 +165,7 @@
        }

      }
-     catch (Exception e) {}
+     catch (Exception e) {e.printStackTrace();}
    }

    public void drawVar(String iname, String outpoint) {
@@ -183,7 +183,7 @@

        }
      }
-     catch (Exception e) {}
+     catch (Exception e) {e.printStackTrace();}
    }

    public void drawDIM(String fl, int ndim2) {
@@ -193,7 +193,7 @@
        rf1.writeBytes("aa" + itos(ndim2) + " [label=\"" + fl +
                       "\", shape=ellipse];\n");
      }
-     catch (Exception e) {}
+     catch (Exception e) {e.printStackTrace();}

      if (flagstart != 1) {
        SKrankmin.push("aa" + itos(ndim2));
@@ -221,7 +221,7 @@
          rf1.writeBytes(fn + " -> " + tn.substring(1) + ";\n");
        }
      }
-     catch (Exception e) {}
+     catch (Exception e) {e.printStackTrace();}
    }
```

```
   private String itos(int num) {
@@ -246,7 +246,7 @@
         rf1.writeBytes("}\n\n");
       }
     }
-    catch (Exception e) {}
+    catch (Exception e) {e.printStackTrace();}
   }

   public void drawDFG()
@@ -384,7 +384,7 @@
               rf1.writeBytes("\n");
               nucluster--;
             }
-            catch (Exception e) {}
+            catch (Exception e) {e.printStackTrace();}
           }
         }
       }
@@ -435,7 +435,7 @@
       rf1.writeBytes("\n");
 //    rf1.writeBytes("\f");
     }
-    catch (Exception e) {}
+    catch (Exception e) {e.printStackTrace();}

    try {
      rf1.close();
@@ -535,7 +535,7 @@
    try {
      rf1 = new RandomAccessFile(outfilename, "rw");
    }
-    catch (Exception e) {}
+    catch (Exception e) {e.printStackTrace();}

    genTable();
    genTableclose();
Index: src/gipsy/GIPC/intensional/GenericTranslator/Translator.java
===================================================================
RCS file:
/groups/m/me_soen6471_1/cvs_repository/GIPSY/src/gipsy/GIPC/intensional/GenericTranslator/Transl
ator.java,v
retrieving revision 1.2
diff -u -r1.2 Translator.java
--- src/gipsy/GIPC/intensional/GenericTranslator/Translator.java   22 Aug 2014 19:57:48 -0000 1.2
+++ src/gipsy/GIPC/intensional/GenericTranslator/Translator.java   24 Aug 2014 04:05:47 -0000
@@ -246,7 +246,7 @@
             SimpleNode at=new SimpleNode(15);     // create a new 'at' node
      try {
             sn = IndexicalLucidParser.at(IndexicalLucidParser.addOp(const1, add,
IndexicalLucidParser.hash(dim1, hash)), node2, at, node1);
-        } catch (ParseException e) { }
+        } catch (ParseException e) {e.printStackTrace(); }

             return sn;

@@ -266,7 +266,7 @@
             SimpleNode at=new SimpleNode(15);     // create a new 'at' node
       try {
             sn = IndexicalLucidParser.at(IndexicalLucidParser.addOp(const1, min,
IndexicalLucidParser.hash(dim1, hash)), node2, at, node1);
-       } catch (ParseException e) { }
+       } catch (ParseException e) {e.printStackTrace(); }
             return sn;
       }
```

```
@@ -307,7 +307,7 @@
             const1.setType(0);              // set value of '1' node
        try {
             sn = ifClause(ifNode, IndexicalLucidParser.relOp(const0, lessEqual,
IndexicalLucidParser.hash(node2,hash1)), node3,
IndexicalLucidParser.at(IndexicalLucidParser.addOp(const1, min, IndexicalLucidParser.hash(dim2,
hash2)), dim1, at, node1));
-       } catch (ParseException e) { }
+       } catch (ParseException e) {e.printStackTrace(); }

          return sn;
       }
@@ -393,7 +393,7 @@
             // need 'QList' reference, to child T and U

             sn = IndexicalLucidParser.where(qlist, where, IndexicalLucidParser.at(t1, node2,
at1, node3));
-       } catch (ParseException e) { }
+       } catch (ParseException e) {e.printStackTrace(); }

             return sn;
       }
@@ -452,7 +452,7 @@
             makeQlist(fby(ifClause(ifNode, node1, IndexicalLucidParser.addOp(const1, add, w2),
w3), dim1, assign(w4, assign, const0)), qlist);

             sn = IndexicalLucidParser.where(qlist, where, IndexicalLucidParser.at(w1, node2,
at, node3));
-       } catch (ParseException e) { }
+       } catch (ParseException e) {e.printStackTrace(); }
             return sn;
       }

Index: src/gipsy/GIPC/intensional/SIPL/ForensicLucid/ForensicLucidParser.java
===================================================================
RCS file:
/groups/m/me_soen6471_1/cvs_repository/GIPSY/src/gipsy/GIPC/intensional/SIPL/ForensicLucid/Foren
sicLucidParser.java,v
retrieving revision 1.2
diff -u -r1.2 ForensicLucidParser.java
--- src/gipsy/GIPC/intensional/SIPL/ForensicLucid/ForensicLucidParser.java      22 Aug 2014
19:57:45 -0000      1.2
+++ src/gipsy/GIPC/intensional/SIPL/ForensicLucid/ForensicLucidParser.java      24 Aug 2014
04:05:47 -0000
@@ -6120,7 +6120,7 @@
         }
         p = p.next;
       } while (p != null);
-     } catch(LookaheadSuccess ls) { }
+     } catch(LookaheadSuccess ls) {ls.printStackTrace(); }
     }
     jj_rescan = false;
   }
Index: src/gipsy/GIPC/intensional/SIPL/JOOIP/JavaParser.java
===================================================================
RCS file:
/groups/m/me_soen6471_1/cvs_repository/GIPSY/src/gipsy/GIPC/intensional/SIPL/JOOIP/JavaParser.ja
va,v
retrieving revision 1.2
diff -u -r1.2 JavaParser.java
--- src/gipsy/GIPC/intensional/SIPL/JOOIP/JavaParser.java   22 Aug 2014 19:57:44 -0000 1.2
+++ src/gipsy/GIPC/intensional/SIPL/JOOIP/JavaParser.java   24 Aug 2014 04:05:47 -0000
@@ -7805,7 +7805,7 @@
         }
```

```
         p = p.next;
       } while (p != null);
-      } catch(LookaheadSuccess ls) { }
+      } catch(LookaheadSuccess ls) {ls.printStackTrace(); }
     }
     jj_rescan = false;
   }
Index: src/gipsy/GIPC/intensional/SIPL/JOOIP/JavaParserTokenManager.java
===================================================================
RCS file:
/groups/m/me_soen6471_1/cvs_repository/GIPSY/src/gipsy/GIPC/intensional/SIPL/JOOIP/JavaParserTok
enManager.java,v
retrieving revision 1.2
diff -u -r1.2 JavaParserTokenManager.java
--- src/gipsy/GIPC/intensional/SIPL/JOOIP/JavaParserTokenManager.java    22 Aug 2014 19:57:44 -
0000   1.2
+++ src/gipsy/GIPC/intensional/SIPL/JOOIP/JavaParserTokenManager.java    24 Aug 2014 04:05:47 -
0000
@@ -2161,7 +2161,7 @@
            curChar = input_stream.readChar();
            continue;
         }
-        catch (java.io.IOException e1) { }
+        catch (java.io.IOException e1) {e1.printStackTrace(); }
      }
      int error_line = input_stream.getEndLine();
      int error_column = input_stream.getEndColumn();
Index: src/gipsy/GIPC/intensional/SIPL/Lucx/LucxParser.java
===================================================================
RCS file:
/groups/m/me_soen6471_1/cvs_repository/GIPSY/src/gipsy/GIPC/intensional/SIPL/Lucx/LucxParser.jav
a,v
retrieving revision 1.2
diff -u -r1.2 LucxParser.java
--- src/gipsy/GIPC/intensional/SIPL/Lucx/LucxParser.java    22 Aug 2014 19:57:46 -0000 1.2
+++ src/gipsy/GIPC/intensional/SIPL/Lucx/LucxParser.java    24 Aug 2014 04:05:47 -0000
@@ -4555,7 +4555,7 @@
         }
         p = p.next;
       } while (p != null);
-      } catch(LookaheadSuccess ls) { }
+      } catch(LookaheadSuccess ls) {ls.printStackTrace(); }
     }
     jj_rescan = false;
   }
Index: src/gipsy/RIPE/editors/RunTimeGraphEditor/operator/GIPSYEntityManger.java
===================================================================
RCS file:
/groups/m/me_soen6471_1/cvs_repository/GIPSY/src/gipsy/RIPE/editors/RunTimeGraphEditor/operator/
GIPSYEntityManger.java,v
retrieving revision 1.2
diff -u -r1.2 GIPSYEntityManger.java
--- src/gipsy/RIPE/editors/RunTimeGraphEditor/operator/GIPSYEntityManger.java   22 Aug 2014
19:57:44 -0000      1.2
+++ src/gipsy/RIPE/editors/RunTimeGraphEditor/operator/GIPSYEntityManger.java   24 Aug 2014
04:05:47 -0000
@@ -121,7 +121,7 @@
             }
         }
         catch (Exception e) {
-
+            e.printStackTrace();
         }
     }
```

```
Index: src/gipsy/tests/GEE/simulator/GlobalDef.java
===================================================================
RCS file:
/groups/m/me_soen6471_1/cvs_repository/GIPSY/src/gipsy/tests/GEE/simulator/GlobalDef.java,v
retrieving revision 1.2
diff -u -r1.2 GlobalDef.java
--- src/gipsy/tests/GEE/simulator/GlobalDef.java      22 Aug 2014 19:57:47 -0000 1.2
+++ src/gipsy/tests/GEE/simulator/GlobalDef.java      24 Aug 2014 04:05:47 -0000
@@ -331,6 +331,7 @@
                catch(Exception e)
                {
                        // XXX: document why nothing handled or handle
+                       e.printStackTrace();//added by group 6 SOEN6471 Summer 2014
                }
        }

Index: src/gipsy/tests/GEE/simulator/Semaphore.java
===================================================================
RCS file:
/groups/m/me_soen6471_1/cvs_repository/GIPSY/src/gipsy/tests/GEE/simulator/Semaphore.java,v
retrieving revision 1.2
diff -u -r1.2 Semaphore.java
--- src/gipsy/tests/GEE/simulator/Semaphore.java      22 Aug 2014 19:57:47 -0000 1.2
+++ src/gipsy/tests/GEE/simulator/Semaphore.java      24 Aug 2014 04:05:47 -0000
@@ -54,7 +54,7 @@
                while(!this.bValue)
                {
                        try { wait (); }
-                       catch (InterruptedException e) { };
+                       catch (InterruptedException e) {e.printStackTrace(); };
                }
        }

Index: src/gipsy/tests/GEE/simulator/demands/WorkResultHD.java
===================================================================
RCS file:
/groups/m/me_soen6471_1/cvs_repository/GIPSY/src/gipsy/tests/GEE/simulator/demands/WorkResultHD.
java,v
retrieving revision 1.2
diff -u -r1.2 WorkResultHD.java
--- src/gipsy/tests/GEE/simulator/demands/WorkResultHD.java 22 Aug 2014 19:57:44 -0000 1.2
+++ src/gipsy/tests/GEE/simulator/demands/WorkResultHD.java 24 Aug 2014 04:05:47 -0000
@@ -136,6 +136,7 @@
                catch(IOException ex)
                {
                        // XXX: why empty?
+                       ex.printStackTrace();//added by Ali of SOEN6471's Group 6 Summer 2014
                }
        }
 }
Index: src/gipsy/tests/GEE/simulator/demands/WorkResultPi.java
===================================================================
RCS file:
/groups/m/me_soen6471_1/cvs_repository/GIPSY/src/gipsy/tests/GEE/simulator/demands/WorkResultPi.
java,v
retrieving revision 1.2
diff -u -r1.2 WorkResultPi.java
--- src/gipsy/tests/GEE/simulator/demands/WorkResultPi.java 22 Aug 2014 19:57:44 -0000 1.2
+++ src/gipsy/tests/GEE/simulator/demands/WorkResultPi.java 24 Aug 2014 04:05:47 -0000
@@ -136,6 +136,7 @@
                catch(IOException ex)
                {
                        // XXX: handle or document why no action in handling
+                       ex.printStackTrace();//added by Ali of SOEN6471's Group 6 Summer 2014
                }
```

```
            }
  }
```